\documentclass[useAMS,usenatbib]{mn2e}
\usepackage{graphicx,amsmath,multirow,amssymb}
\usepackage{natbib}
\newcommand{\comment}[1]{}

\def\simgt{\lower.5ex\hbox{$\; \buildrel > \over \sim \;$}}
\def\simlt{\lower.5ex\hbox{$\; \buildrel < \over \sim \;$}}

\title[Chemical patterns in Globular Clusters]{A view of the H-band light-element chemical patterns in Globular Clusters under the AGB self-enrichment scenario}
\author[Dell'Agli et al.]{F. Dell'Agli$^{1,2}$, D. A. Garc{\'{\i}}a-Hern{\'a}ndez$^{1,2}$, P. Ventura$^{3}$, S. M{\'e}sz{\'a}ros$^{4,5}$,
\newauthor 
T. Masseron$^{1,2}$, J. G. Fern{\'a}ndez-Trincado$^{6,7}$, B. Tang$^{6}$, M. Shetrone$^{8}$,
\newauthor
 O. Zamora$^{1,2}$, S. Lucatello$^{9}$  \\
$^1$Instituto de Astrof\'{\i}sica de Canarias, V\'{\i}a L\'actea s/n, E-38205 La Laguna, Tenerife, Spain \\
$^2$Departamento de Astrof\'{\i}sica, Universidad de La Laguna (ULL), E-38206 La Laguna, Spain\\
$^3$INAF -- Osservatorio Astronomico di Roma, Via Frascati 33, 00040, Monte Porzio Catone (RM), Italy \\
$^4$ELTE E\"otv\"os Lor{\'a}nd University, Gothard Astrophysical Observatory, Szombathely, Hungary\\
$^5$Premium Postdoctoral Fellow of the Hungarian Academy of Sciences\\
$^6$Departamento de Astronom\'{\i}a, Casilla 160-C, Universidad de Concepci{\'o}n, Concepci{\'o}n, Chile\\
$^7$Institut Utinam, CNRS UMR6213, Univ. Bourgogne Franche-Comt\'e, OSU THETA , \\
Observatoire de Besan\c{c}on, BP 1615, 25010 Besan\c{c}on Cedex, France \\
$^8$University of Texas at Austin, McDonald Observatory, Fort Davis, TX 79734, USA\\
$^9$INAF -- Osservatorio Astronomico di Padova, vicolo dell'Osservatorio 5, I-35122 Padova, Italy \\
}

\begin{document}

\date{Accepted, Received; in original form }

\pagerange{\pageref{firstpage}--\pageref{lastpage}} \pubyear{2012}

\maketitle

\label{firstpage}

\begin{abstract}
We discuss the self-enrichment scenario by AGB stars for the formation of multiple
populations in globular clusters (GCs) by analyzing data set of giant stars
observed in 9 Galactic GCs, covering a wide range of metallicities and for which
the simultaneous measurements of C, N, O, Mg, Al, Si are available. To this aim
we calculated 6 sets of AGB models, with the same chemical composition as the
stars belonging to the first generation of each GC.
We find that the AGB yields can reproduce the set of observations available,
not only in terms of the degree of contamination shown by stars in each GC
but, more important, also the observed trend with metallicity, which agrees well with the predictions from AGB evolution modelling.
While further observational evidences are required to definitively fix the main
actors in the pollution of the interstellar medium from which new generation of
stars formed in GCs, the present results confirm that the gas ejected by 
stars of mass in the range $4~\rm M_{\odot} \leq \rm M \leq 8~\rm M_{\odot}$ during the AGB phase
share the same chemical patterns traced by stars in GCs.
 
\end{abstract}


\section{Introduction}
In the last decades the results from high-resolution spectroscopy and 
space photometry have challenged the traditional paradigma that stars
in globular clusters (GCs) are the best example of a simple stellar
population \citep[e.g.][]{gratton12,piotto07}.

Several studies have outlined that the chemical composition of GC stars,
far from being homogeneous, shows well defined abundance patterns
involving the chemical species up to silicon, the most confirmed and
investigated being the O-Na anticorrelations \citep[see][and references therein]{gratton12}. The existence of
these chemical patterns was observed in all the Galactic GCs studied to date 
(with the exception of Ruprecht 106; see Villanova et al. 2013), though 
with clear differences from cluster to cluster \citep[see e.g.][]{carretta09a}. 
The presence of these anomalies in unevolved stars in the main sequence 
and in the sub giant branch \citep{gratton01} indicates that these 
chemical signatures were present in the gas
from which the stars formed, ruling out the action of any in situ mechanism.

On the photometric side, the discovery of multiple main sequences in the
colour-magnitude diagram (CMD) of some GCs \citep{bedin04, piotto07}, combined with the analysis of the 
morphology of the horizontal branch (HB) \citep{dantona04, caloi05, caloi07, caloi08},
lead to the conclusion that part of the stars in GCs were born enriched
in helium.

All the afore mentioned observational pieces of evidence indicate that star formation
in GCs was far from being fully homogeneous, and that after the formation
of the first generation (FG) of stars, a second generation (SG) 
formed\footnote{Here we refer generally with SG to all the stars formed after
the FG, despite in some GCs it was confirmed that more than two groups of stars
exist, differing in their chemical composition}, from gas which was exposed
to proton-capture nucleosynthesis. This opened the way to the search of possible
polluters to be active in a self-enrichment mechanism. Because no significant 
spread is observed in the turn-off region of the CMD of all
the GCs observed, the gas from which SG stars formed, processed by internal 
nucleosynthesis, must have been expelled by FG stars evolving in a time scale 
shorter than $\sim 1$ Gyr.

So far four candidates have been proposed: super massive main sequence stars 
\citep{denissenkov14}; fast rotating massive stars \citep{krause13},
massive binaries \citep{demink09} and massive AGB stars \citep{dercole10}. 
The problems arising when trying to interpret the observed multiple populations 
on the basis of the afore mentioned scenarios are thoroughly described and 
discussed in \citet{renzini15}.

On the way to understand how SG stars formed, an important contribution
will be given by data from the Apache Point Observatory Galactic Evolution 
Experiment \citep[APOGEE; e.g.][]{majewski17}, a high-resolution, near-IR, 
large spectroscopic survey. The APOGEE data allow a good study of the Mg-Al anticorrelation, which is extremely important to discriminate among the possible polluters of the intra-cluster medium
from which SG stars formed. This is because the activation of Mg-Al nucleosynthesis is much more sensitive to the temperature than CNO cycling and the Ne-Na chain \citep{ventura13}. Moreover, the Mg and Al abundances observed in giant stars definitively reflect their 
initial chemistry, as there is no way that deep mixing can reach stellar 
regions touched by magnesium burning.

Meszaros et al. (2015, hereafter ME15) presented APOGEE data for 10 GCs, 
spanning a range of metallicity from $[\rm Fe/H]=-2.37$ to $[\rm Fe/H]=-0.78$. For 
the stars in all 10 clusters the surface abundances, when possible, of the elements
involved in CNO cycling and Mg-Al-Si nucleosynthesis were measured.

A first attempt to interpret ME15 data was made by \citet{ventura16}, focusing on only 5 of the GCs in the ME15 sample: the main conclusion
reached by \citet{ventura16} is that the extent of the Mg-Al and
Mg-Si anticorrelations observed, particularly the variation with 
the metallicity, could be explained by hypothesizing that SG stars
formed from gas lost by stars of mass in the range $4-8~\rm M_{\odot}$
during the AGB phase.

In this paper we make a step forward, by considering 9 out of 10 of the GCs studied
by ME15 (we skipped NGC 5466 where only 8 stars were observed), attempting to infer their star formation history after the formation
of the FG. While the analysis by \citet{ventura16} is limited to the Mg, Al, Si and O
abundances, here we consider also the carbon and nitrogen data, being aware
that the surface mass fractions of these elements are subject to a significant
alteration during the RGB ascending. Furthermore, while \citet{ventura16} used
pre-existing AGB models with a chemistry similar to the stars in the ME15 sample,
here we use updated AGB models, calculated specifically for the present
analysis, where the initial chemical composition is assumed ad hoc
to be the same as derived from ME15 for the FG stars of the different clusters.
This is extremely important, as the interpretation of the results does not
rely on any artificial scaling of the abundances, which may partly alter
the results obtained.

In the present study we focus on the interpretation of the
high-resolution spectroscopic data based on the yields from massive AGB stars, 
without entering into the dynamical details of the self-enrichment process.
All the arguments related to the modality with which SG stars formed \citep{dercole08}, 
the difference in the spatial distributions of FG and SG stars, which makes the 
former more exposed to escape from the cluster \citep{vesperini13}, the need for 
dilution with pristine gas \citep{dercole11}, the possibility that part of the pristine 
gas survives to the epoch of type II SNe explosions \citep{dercole16}, are thoroughly
documented in the literature and will not be addressed here. Our aim is to
answer the following question: can the large data set presented by ME15 be explained
on the chemical side within the context of the self-enrichment scenario by AGB stars?

To this goal, we study the ME15 GCs individually, doing a detailed comparison 
between the
data set and the chemistry of the AGB ejecta. The main scope is to deduce the
origin of SG stars and whether the formation of SG occurred from the ``pure"
AGB ejecta or if some dilution with pristine gas in the cluster is required.
Whenever possible, we confront our conclusions with results from photometry,
as the latter provides valuable information on the helium spread of the clusters stars, a key indicator of the existence and of the degree of contamination
of SG stars compared to their FG counterparts.

The paper is organized as follows: section 2 contains an overview of the physical input for the AGB models adopted and the main chemical 
points regarding their ejecta; section 3 is devoted to 
the analysis of the data for each GC, via a detailed comparison with the yields
from the AGB stars with the same chemical composition; in section 4 we extend our comparison to the O-Na plane; a general discussion of the results obtained is presented in section 5, whereas the conclusions are given in
section 6.

\section{Massive AGBs models: nucleosynthesis and yields}
The AGB models used here have been calculated specifically for the present investigation.
They were computed by means of the ATON code for the stellar evolution. The interested
reader can find in \citet{ventura98} a detailed description of the numerical structure of
the code; the most recent updates are given, e.g., in \citet{ventura09}.

\subsection{The initial chemistry}
To choose the O, Mg, Al and Si initial chemical composition of the models used to study each individual cluster
we relied on the assumed chemistry of FG stars, reported in Table 7 of ME15, where they summarise the average FG abundances of each cluster. Clusters similar in
metallicity and in the O, Mg, Al and Si abundances were grouped together, and studied by means 
of a unique set of models. The mass fractions of the other species were scaled according to the 
overall metallicity and the solar distribution by Lodders et al. (2003). This choice 
also includes C and N, for which we assumed $[\rm{C/Fe}]=[\rm{N/Fe}]=0$.
Note that, because of the effects of the first dredge-up (FDU) and ``non conventional" mixing \citep[see e.g.][]{corinne10}, the surface abundance of C and N in RGB stars is not expected to reflect the composition of the gas from which the stars formed.

The overall theoretical framework consist in 6 sets of AGB models, with mass in the range
$4-8~\rm M_{\odot}$, whose initial chemical composition is reported in Table \ref{input}. 

\begin{table*}
\begin{center}
\caption{The chemical composition assumed to calculate the AGB models used to study 
the GCs in ME15 sample. The metallicity and the mass fraction of the individual species
have been taken from ME15. The only exceptions are C and N, for which a solar-scaled
abundance was adopted in all cases.} 
\label{input}
\begin{tabular}{cccccccccc}
\hline
 cluster & [Fe/H] & Z & Y & [C/Fe] & [N/Fe] & [O/Fe] & [Mg/Fe] & [Al/Fe] & [Si/Fe] \\
\hline
\\
 M92, M15 & -2.3 & $2\times 10^{-4}$ & 0.25 & 0.0 & 0.0 & 0.65 & 0.30 & -0.20 & 0.40 \\
 M53 & -2.0 & $4\times 10^{-4}$ & 0.25 & 0.0 & 0.0 & 0.60 & 0.40 & -0.10 & 0.40 \\
 M13, M2, M3 & -1.5 & $10^{-3}$ & 0.25 & 0.0 & 0.0 & 0.55 & 0.20 & 0.00 & 0.40 \\
 M5 & -1.3 & $1.5\times 10^{-3}$ & 0.25 & 0.0 & 0.0 & 0.40 & 0.25 & 0.00 & 0.35 \\
 M107 & -1.0 & $2.5\times 10^{-3}$ & 0.25 & 0.0 & 0.0 & 0.40 & 0.35 & 0.50 & 0.50 \\
 M71 & -0.8 & $6\times 10^{-3}$ & 0.26 & 0.0 & 0.0 & 0.50 & 0.45 & 0.45 & 0.40 \\
\\
\hline
\end{tabular}
\end{center}
\end{table*}

\subsection{Hot bottom burning in massive AGB stars}
\label{hbb}
The evolution of stars during the AGB phase is thoroughly documented in the
literature. The reviews by \citet{herwig05} and \citet{karakas14} present an
exhaustive description of the physical and chemical evolution of these stars,
and a discussion of how the uncertainties regarding some of the physical
ingredients adopted affect the results obtained.

For what attains the stars of interest here, we know that the AGB 
evolution of $\rm M \geq 4~\rm M_{\odot}$ stars is characterized by the ignition of the so
called hot bottom burning (hereianfter HBB). HBB consists in the ignition of a series
of proton capture reactions at the base of the convective mantle, taking place
when the temperature in those regions of the stars, $\rm T_{bce}$, exceeds 
$\sim 30$MK \citep{renzini81, blocker91}. The activation of HBB, due to a partial 
overlapping of the convective envelope with the CNO burning shell, requires core masses 
above $\sim 0.75~\rm M_{\odot}$, corresponding to initial masses above $ \sim 4~\rm M_{\odot}$
\citep{ventura13}.

When HBB is active, the gas ejected by the stars will show the imprinting of the
nucleosynthesis experienced at the base of the convective envelope, because convective
currents are extremely efficient in homogenizing the whole external mantle, up to
the outermost layers, which are gradually lost via stellar winds.

Common features of all the stars experiencing HBB are the destruction of the surface
C and the synthesis of great amounts of N, a process which requires
$\rm T_{bce} \sim 30$MK. The possibility that more advanced proton captures are activated
depends primarily on the values reached by $\rm T_{bce}$.

\begin{figure*}
\begin{minipage}{0.48\textwidth}
\resizebox{1.\hsize}{!}{\includegraphics{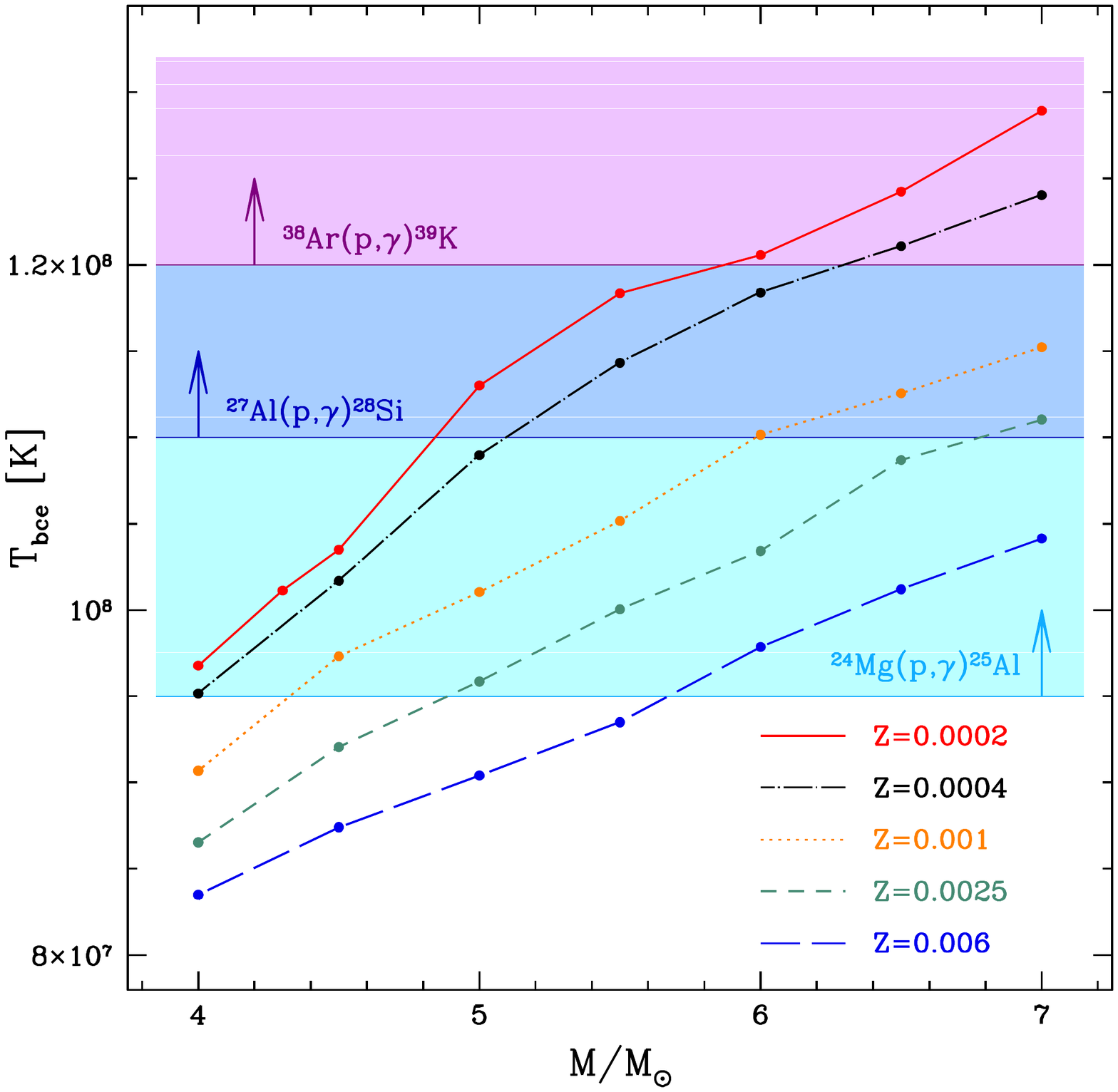}}
\end{minipage}
\begin{minipage}{0.48\textwidth}
\resizebox{1.\hsize}{!}{\includegraphics{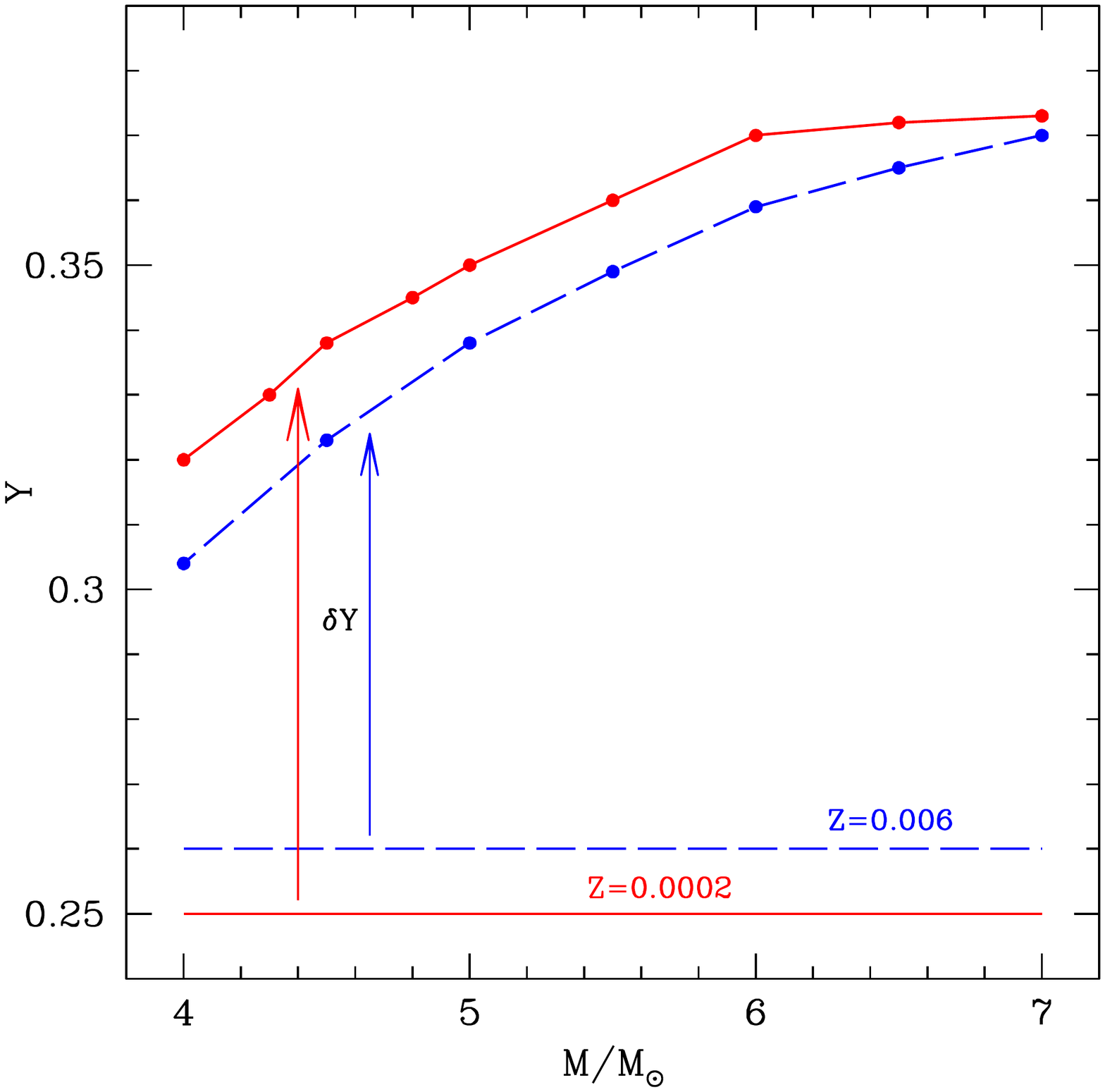}}
\end{minipage}
\vskip-50pt
\caption{Left: typical temperature at the base of the convective envelope during the AGB phase,  
as function of the initial mass of the star at different metallicities. The temperatures 
above which some important nuclear reactions (labelled) are efficient are shown as 
horizontal lines and correspondent shaded area. Right: helium yields as function of stellar 
masses for the highest (Z$=6\times 10^{-3}$, solid-red line) and lowest (Z$=2\times10^{-4}$, 
dashed-blue line) metallicity considered in this work. The horizontal lines at Y=0.26 and 
Y=0.25 represent the initial helium content considered respectively.}
\label{Tbce}
\end{figure*}

The left panel of Fig. \ref{Tbce} shows the values of the temperatures at the base of the envelope of
the AGB models used in the present analysis\footnote{The temperature at the base of the
surface convection region increases in the early AGB phases and decreases in the final
stages, after most of the envelope was lost by stellar winds. However, for most of the
time during which mass loss occurs, the temperature keeps approximately constant, thus
allowing us to provide a typical value.}, for different masses and metallicities.
Two general trends are evident from the figure: a) for a given chemical composition,
higher masses evolve at larger temperatures, due to their higher core masses; b)
lower metallicity AGB stars achieve higher $\rm T_{bce}$. 

In the left panel of Fig. \ref{Tbce} we also show the threshold temperatures above which different
proton-capture channels are activated. This information is crucial to understand 
the kind of pollution expected from these objects and how advanced the proton-capture
nucleosynthesis experienced will be. While we will discuss in details the various 
yields in the following sections, a clear result shown in Fig. \ref{Tbce} is that the most
advanced nuclear channels, in particular the ignition of Mg-Al-Si nucleosynthesis,
can be achieved only in the most metal-poor AGB stars. 

We believe important to mention at this point that while the trend of the HBB
temperature with mass and metallicity, shown in the left panel of Fig. 1, can be considered
general, i.e. independent of the details of AGB modelling, the strength of HBB experienced,
hence the temperatures reached by the base of the convective mantle, depend on the
treatment of convection, particularly on the way that the temperature gradient is
calculated in regions unstable to convective motions. In the analysis by Ventura \&
D'Antona (2005a) it is shown that use of the Full Spectrum of Turbulence (FST, 
Canuto \& Mazzitelli 1991) to model convection leads to strong HBB, with
large temperatures at the base of the envelope. Conversely, use of the classic Mixing
Length Theory (MLT), particularly when the standard calibration used to fit the evolution of the 
Sun is adopted, determines much weaker HBB conditions.
The AGB models used in the present work are based on the FST description, which has the
advantage that the results obtained in this context are independent of any calibration
of free-parameters. When the MLT model is used the situation is more complex, because
the results are sensitive to the choice of the free parameter $\alpha$, giving the 
mixing length. When the solar-calibrated MLT $\alpha$ is adopted, the degree of the 
nucleosynthesis experienced is much lower than in the FST case, because the temperatures 
are below $10^8$ K, even in low-metallicity massive AGB stars \citep[e.g.][]{cristallo15}. 

Thus, in the MLT case there is no way that the AGB yields can reproduce the
observed O-Mg-Al-Si chemical patterns. The use of a higher $\alpha$ leads to a more
advanced HBB nucleosynthesis; although only in the most metal-poor stars with
mass close to the threshold for the core collapse, the temperatures become high
enough to efficiently activate magnesium burning \citep[e.g.][]{fishlock14}.

\subsection{AGB stars as helium manufacturers}
\label{helium}
An important point in the pollution by massive AGB stars is that the gas expelled
by these objects is enriched in helium. This is shown in the right panel of Fig. \ref{Tbce}, 
where we show the average helium abundance of the gas ejected as function of stellar masses 
for the highest (Z$=6\times 10^{-3}$) and lowest (Z$=2\times10^{-4}$) metallicity considered 
in this work.
While the temperatures at the base of the envelope, that reflect into the degree of 
the nucleosynthesis experienced, display an extreme sensitivity to the metallicity,
the helium in the ejecta is mainly determined by the initial mass of the star.
This is because most of the surface helium enrichment occurs during the
second dredge-up following the core He-burning phase, and is primarily determined by the 
extent of the inwards penetration of the convective envelope down to regions previously 
touched by CNO activity: the efficiency of this process is related to the mass of the 
envelope and is scarcely affected by the chemical composition of the star
\citep{ventura10, karakas14}. The second dredge-up provokes a rise in the surface helium, 
which increases with the mass of the star, ranging from $\delta \rm Y \sim 0.02$, for 
$\rm M=4~\rm M_{\odot}$, to $\delta \rm Y \sim 0.12$, for $\rm M=8~\rm M_{\odot}$. 

Based on the results shown in right panel of Figure \ref{Tbce}, we understand that if the formation of 
SG stars in GCs is determined by the AGB ejecta, these stars must be enriched in
helium. The extent of such enrichment depends on the range of masses involved
in the pollution of the interstellar medium and on the degree of dilution of
the AGB gas with pristine matter in the cluster. If no dilution occurred, SG stars
will definitively show evidences of helium enrichment, which leaves an
important signature in the distribution of the stars on the MS \citep{piotto07} and in the
morphology of the HB \citep{dantona02}. The most extreme case is when the formations of SG stars
occurs within $\sim 10-20$Myr after the type II SNe explosions epoch,
directly from the winds of the most massive ($\rm M\sim 7-8~\rm M_{\odot}$) AGB stars,
without dilution with pristine gas. In this case, a high-helium stellar component
would be present in the cluster, which would define a blue-hook structure on the
HB. This is the scenario invoked by \citet{dantona04} to explain the bluest 
stars in the HB of NGC 2808 and by \citet{marcella11} to account for the ``blue hook"
component of the HB of NGC 2419. 

\section{Results}
 
\subsection{The low metallicity clusters: M15 and M92}

M15 and M92 are the most metal poor clusters in the ME15 sample, with $[\rm Fe/H] \sim -2.3$, which
corresponds to a metallicity Z$ = 2\times 10^{-4}$. Fig.~\ref{Tbce} shows that a very
advanced HBB nucleosynthesis is expected to occur in AGB models of this chemical
composition, because all the stars with mass M$>4.3$M$_{\odot}$ reach temperatures at the base
of the envelope exceeding 100MK. 

\begin{figure*}
\begin{minipage}{0.48\textwidth}
\resizebox{1.\hsize}{!}{\includegraphics{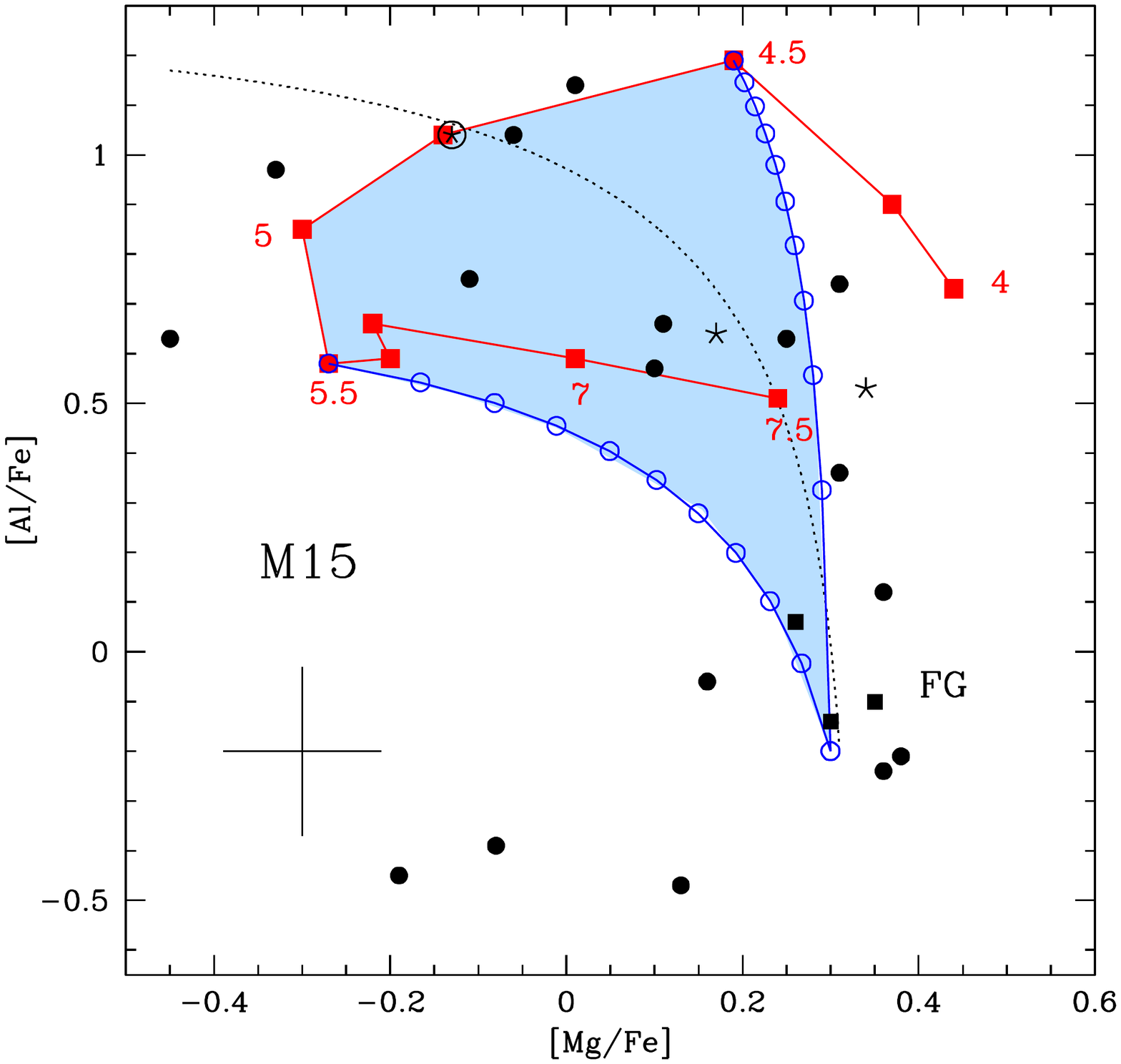}}
\end{minipage}
\begin{minipage}{0.48\textwidth}
\resizebox{1.\hsize}{!}{\includegraphics{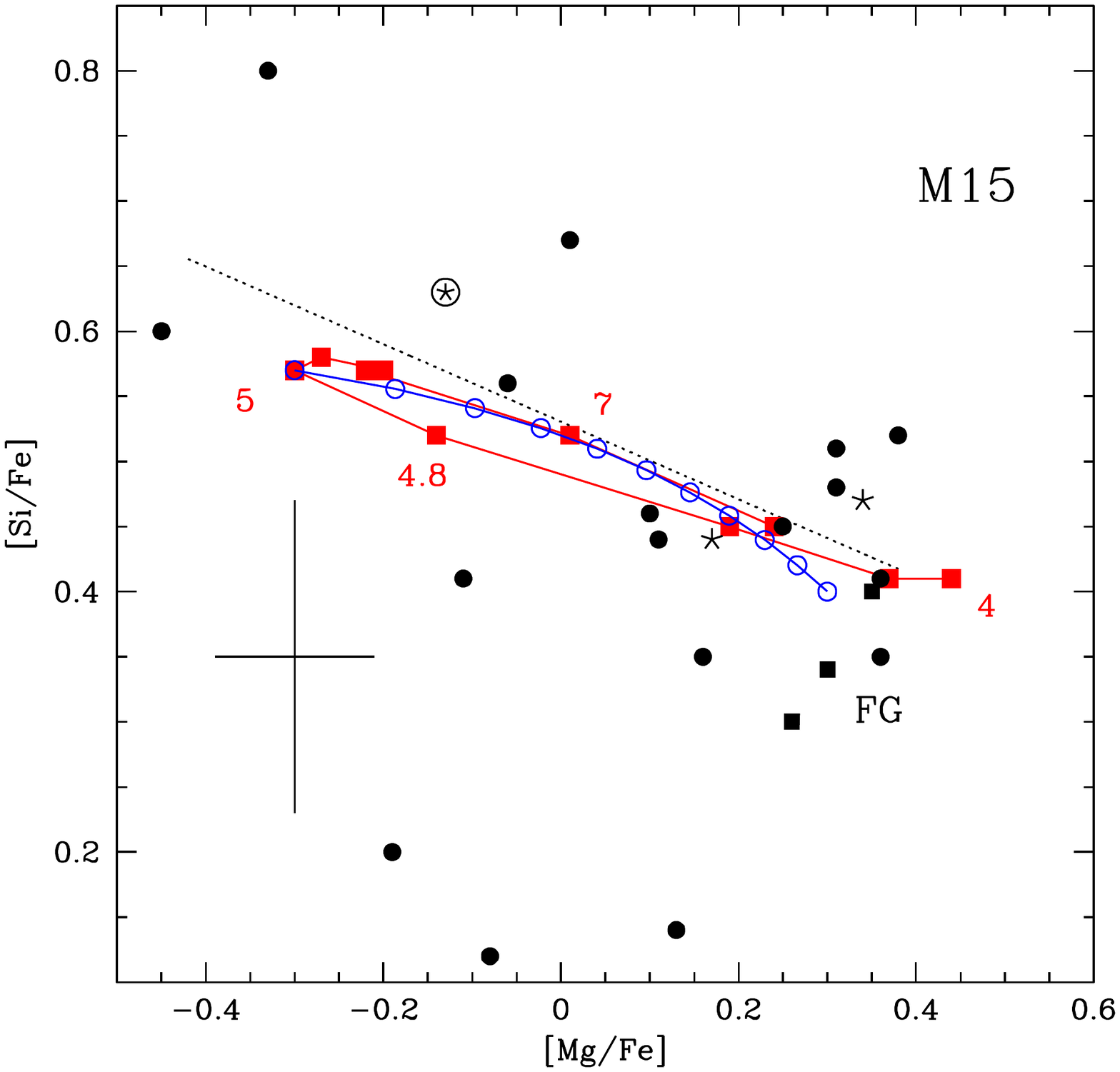}}
\end{minipage}
\vskip-70pt
\begin{minipage}{0.48\textwidth}
\resizebox{1.\hsize}{!}{\includegraphics{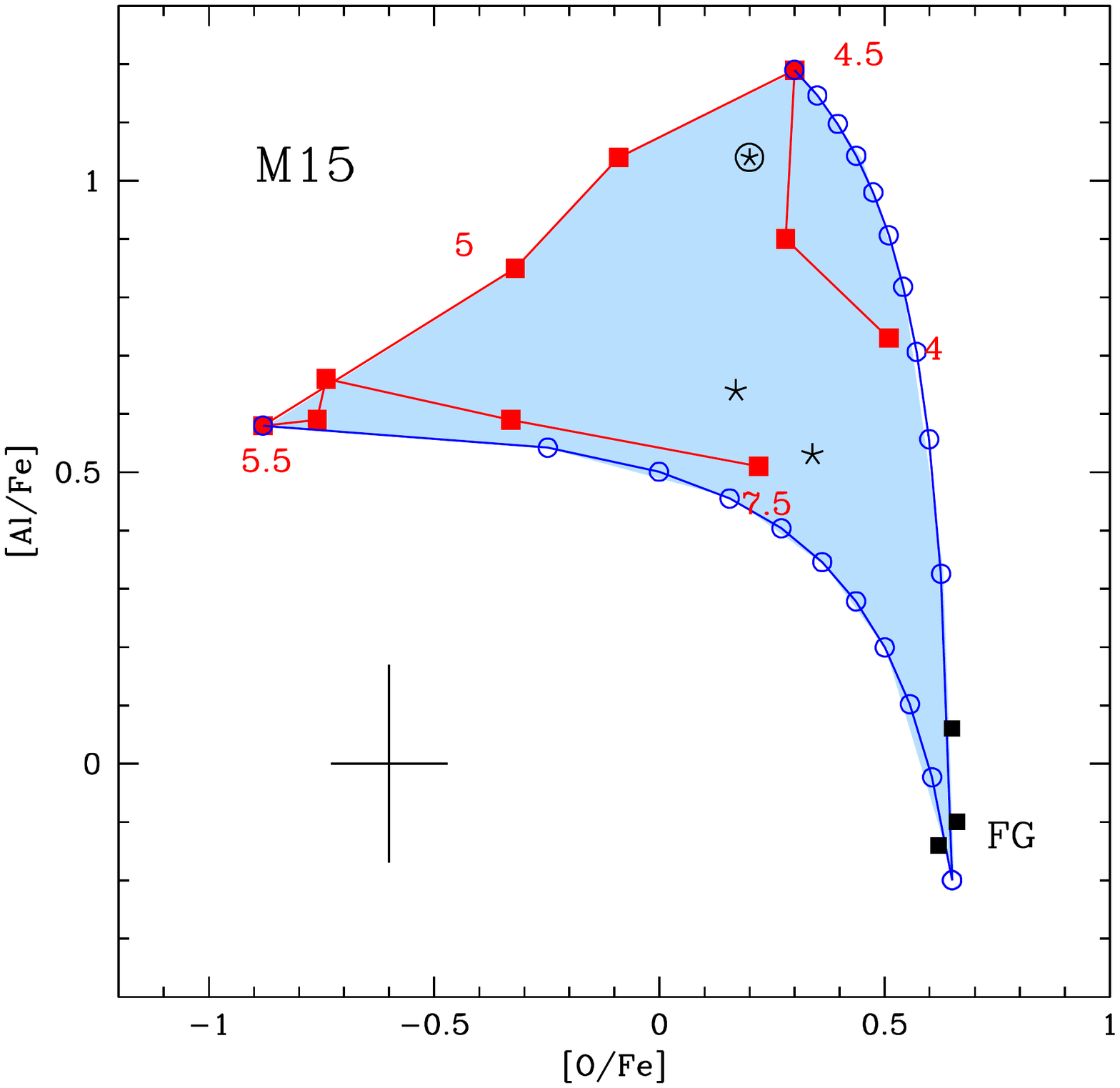}}
\end{minipage}
\begin{minipage}{0.48\textwidth}
\resizebox{1.\hsize}{!}{\includegraphics{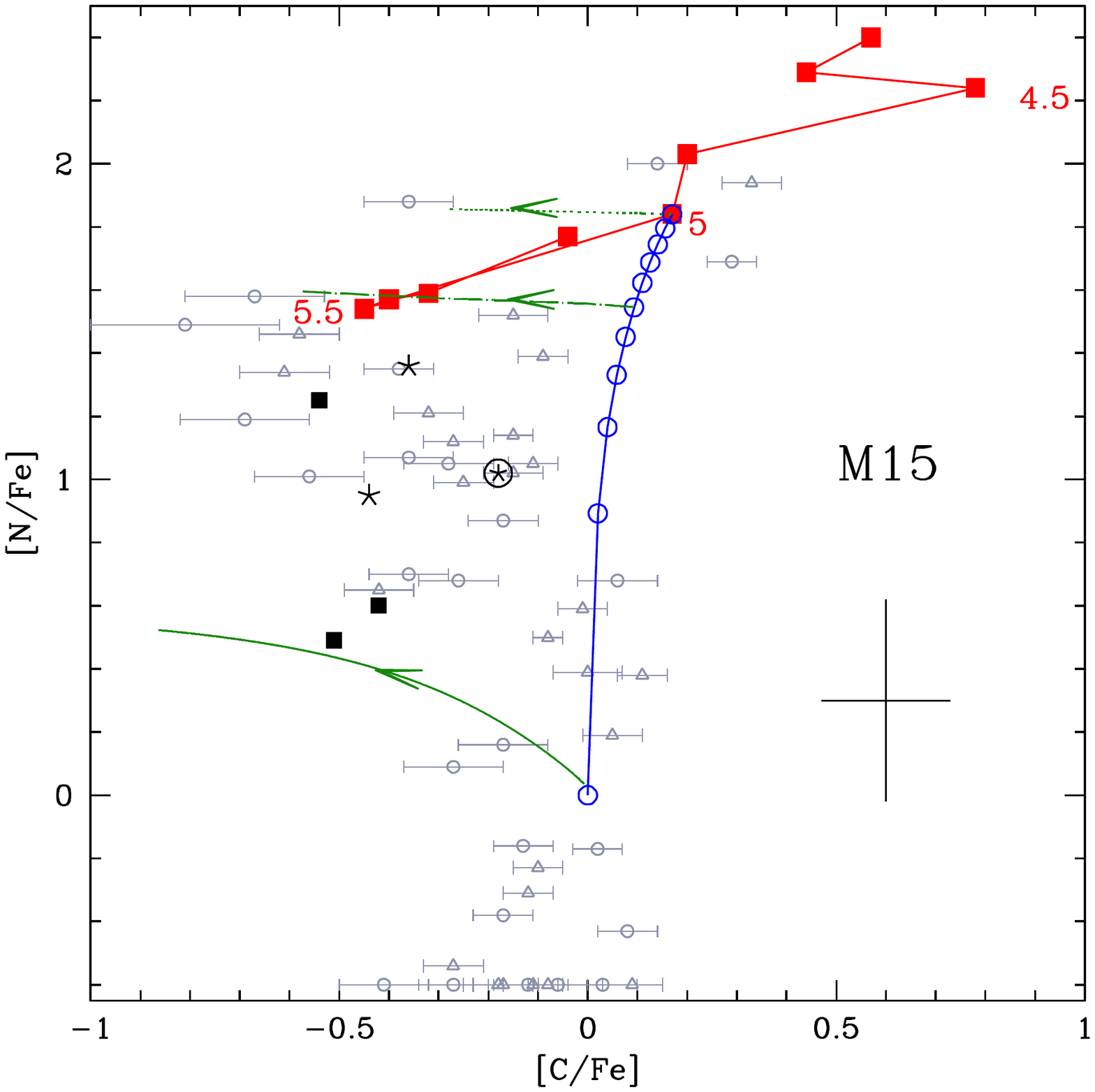}}
\end{minipage}
\vskip-50pt
\caption{M15 stars from ME15 are shown as full dots in the Mg-Al (top-left) and Mg-Si 
(top-right). CNO abundances are measured for part of the sample. They are identified by 
black squares, stars and stars surrounded by open circles according to the anticorrelation 
pattern (low, intemediate and high, respectively) shown in the O-Al (bottom-left) and C-N 
(bottom-right) planes. The black crosses indicate typical APOGEE error bars. 
Results from the fits are shown with dotted lines. Dataset from \citet{cohen05a} are 
shown with gray, open triangles ($\rm T_{\rm eff}<5300$ K) and circles ($\rm T_{\rm eff}>5300$ K), in the 
C-N plane. The yields of AGB models of metallicity Z$ = 2\times10^{-4}$ are indicated with red squares; the numbers close to the squares indicate the initial masses. Blue, open circles indicate the chemistry expected from mixing of the AGB ejecta with pristine gas, with percentages of the latter ranging from 0 to $100\%$, in steps of $10\%$. The points corresponding to $100\%$ dilution are close to the FG label. Green curves represent the evolution of a $0.9~\rm M_{\odot}$ star, currently evolving through the RGB. Three initial chemical compositions are considered: the chemistry of FG stars (solid), the chemistry of pure AGB ejecta (dotted) and of a $50\%$ mix of 
pristine matter with AGB gas (dashed-dotted).}
\label{M15}
\end{figure*}

\begin{figure*}
\begin{minipage}{0.48\textwidth}
\resizebox{1.\hsize}{!}{\includegraphics{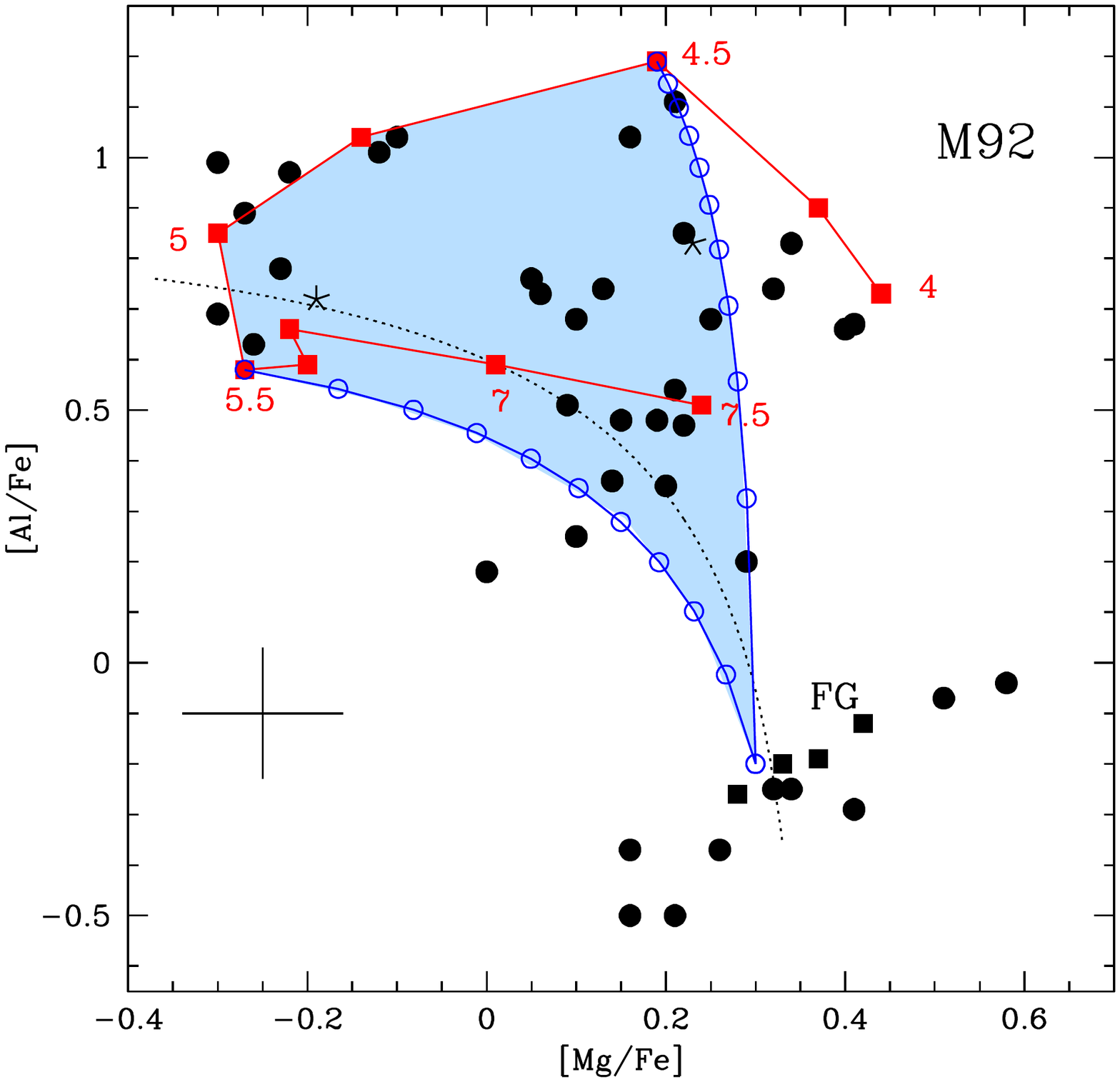}}
\end{minipage}
\begin{minipage}{0.48\textwidth}
\resizebox{1.\hsize}{!}{\includegraphics{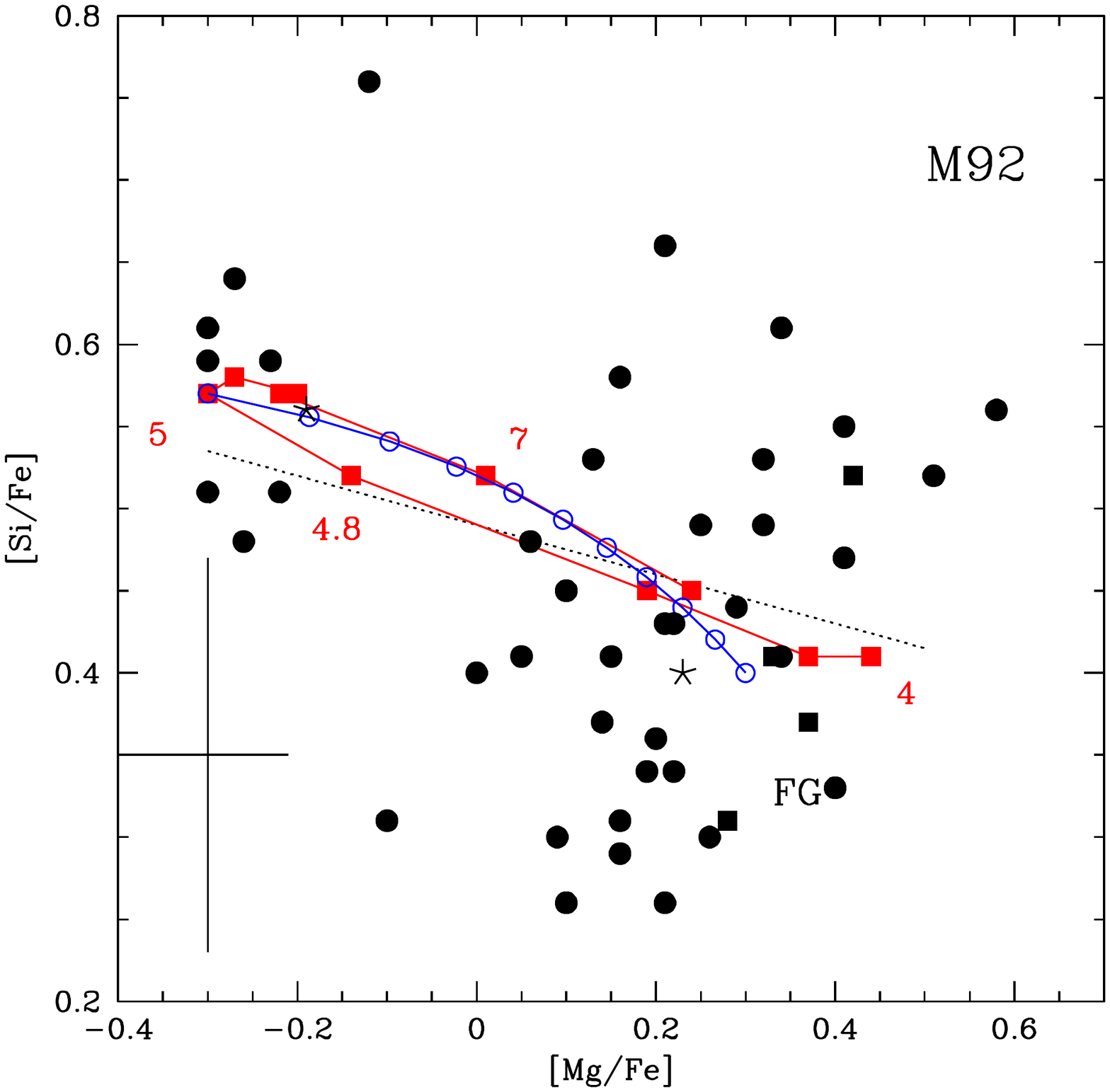}}
\end{minipage}
\vskip-70pt
\begin{minipage}{0.48\textwidth}
\resizebox{1.\hsize}{!}{\includegraphics{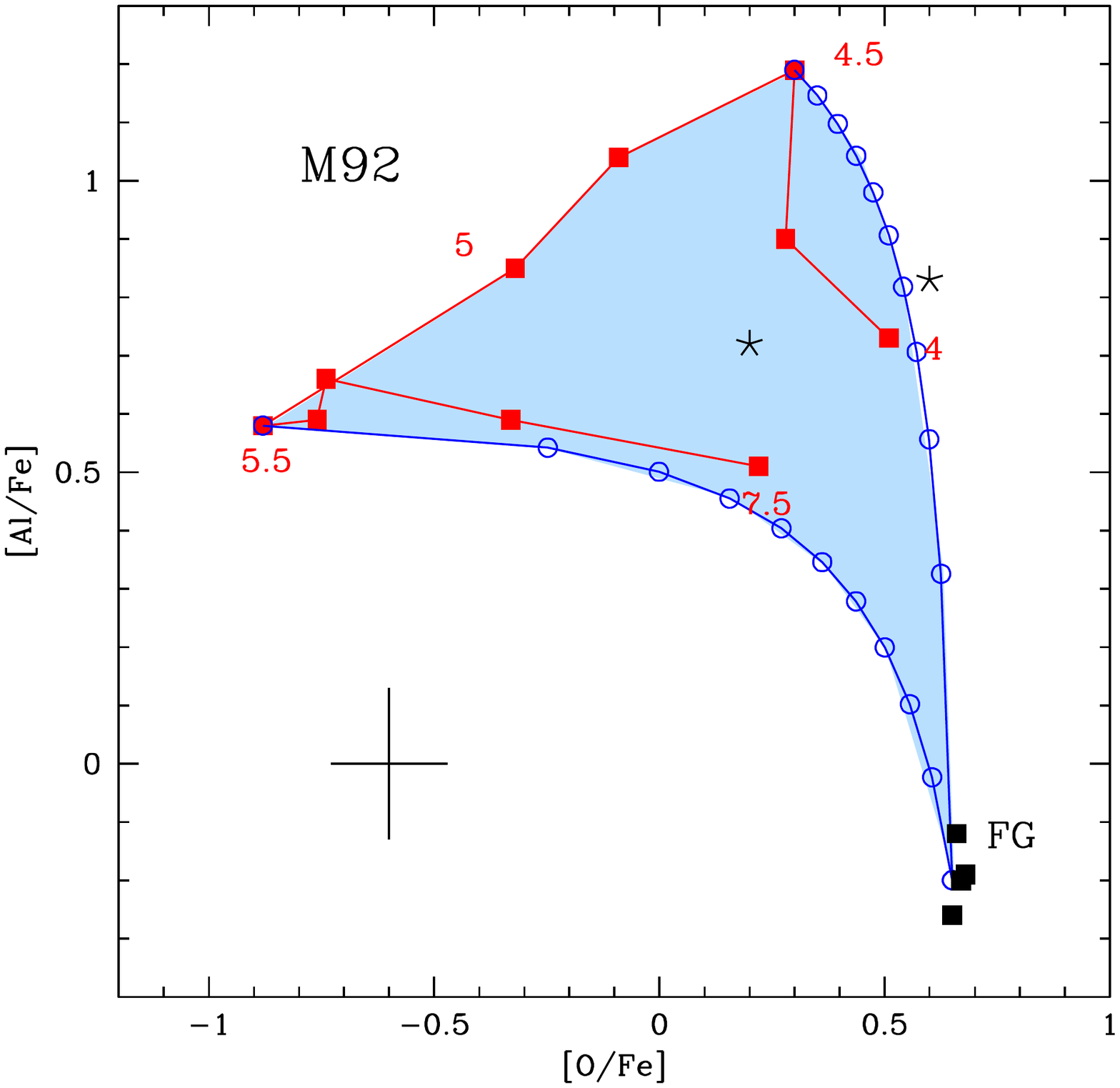}}
\end{minipage}
\begin{minipage}{0.48\textwidth}
\resizebox{1.\hsize}{!}{\includegraphics{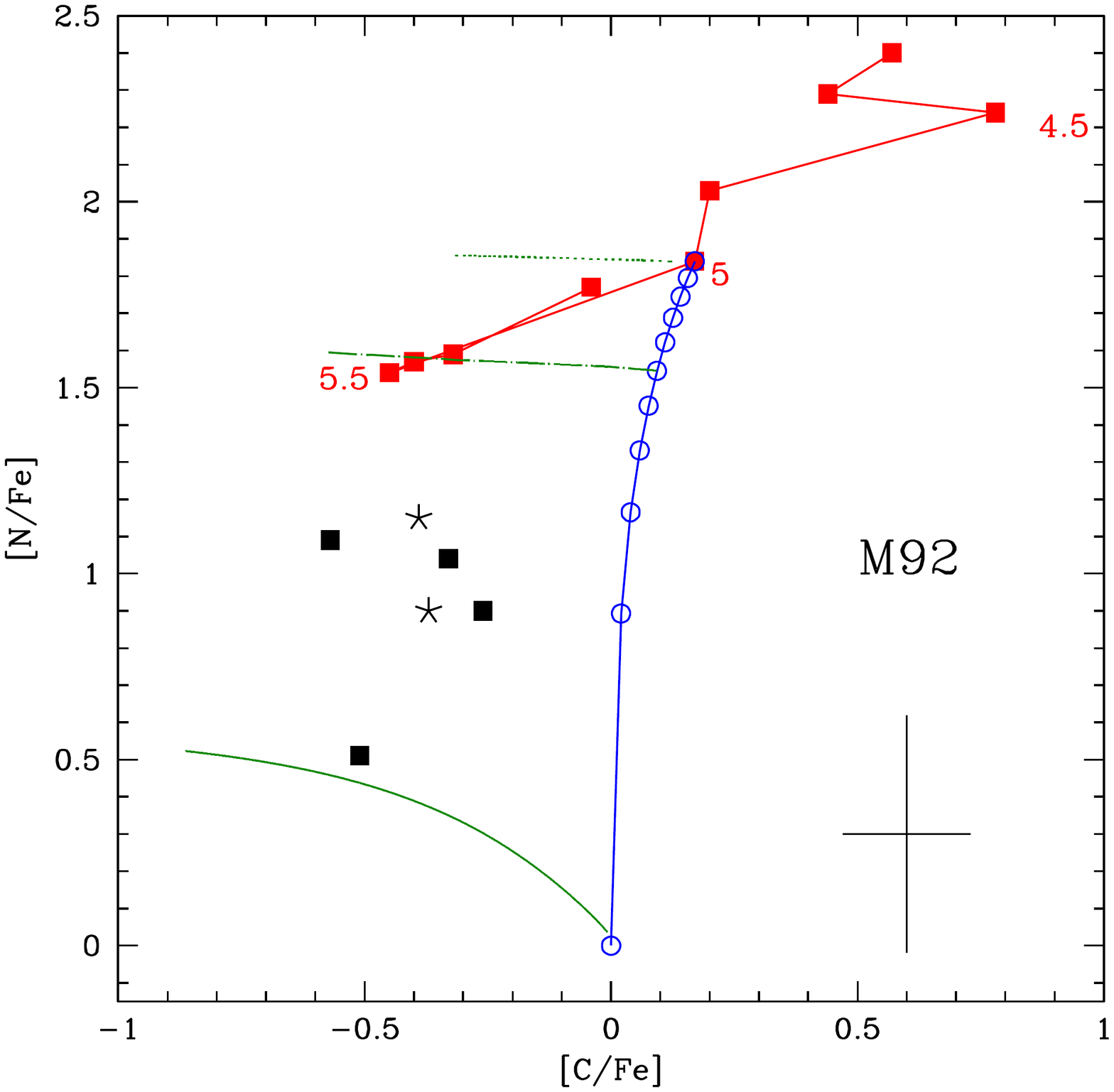}}
\end{minipage}
\vskip-50pt
\caption{M92 stars from ME15 are shown in the same planes and with the same symbols as in Figure \ref{M15}. They are compared with the yields of AGB models of the same metallicity (Z$=2\times 10^{-4}$) as in the case of M15.}
\label{M92}
\end{figure*}

Figs. \ref{M15} and \ref{M92} shows the abundances of the ejecta of Z$=2\times 10^{-4}$ AGB models (red squares) in different planes.
The high HBB temperatures given above trigger a significant depletion of Mg into Al. From the figures we see that Mg depletion reaches $\delta [\rm{Mg/Fe}] \sim -0.6$, whereas the increase 
in aluminium is $\delta [\rm{Al/Fe}] \sim 1$. The HBB temperatures at these low metallicities are
so hot that part of the aluminium produced is converted into silicon (see Fig.~\ref{Tbce}). The reason for the negative trend with mass
of the aluminium content of the ejecta is that higher mass models are exposed to a more
efficient Al burning, as confirmed by the increase in the silicon content, shown in
the top, right panels of Figs.~\ref{M15} and \ref{M92}. 

These extremely high temperatures certainly imply also a very advanced 
CNO cycling. Indeed,
Figs.~\ref{M15} and \ref{M92} show that the AGB ejecta are oxygen-poor ($\delta [\rm{O/Fe}] \sim -1.1$). At the same time the gas 
from these stars is strongly enriched in N, which has an enhancement factor\footnote{We
use the standard definition of ``enhancement factor", as the ratio between the average
abundance of a given chemical species in the ejecta and the initial mass fraction.}
in the range $50-200$. The ejecta are not
carbon poor, despite the high rates of carbon burning reactions.
This is because at these extremely low metallicities even a small number of third dredge-up 
events is sufficient to increase significantly the surface carbon. Since part of this
carbon is converted into nitrogen by HBB, this same reason explains the extremely large
nitrogen abundances.

Figs.~\ref{M15} and \ref{M92} show the observations of M15 and M92 by 
ME15 (black dots) overlapped to the models. The blue regions identify the area between 
the dilution curves for AGB stars of 4.5 and 5.5 solar masses, obtained 
by mixing different fractions of the ejecta of AGB stars with pristine gas, sharing 
the same chemical composition of FG stars (reported in Table \ref{input}). 
When a sufficient number of data are available, to compare the theoretical dilution pattern with the one traced by the data, we have used the following equation:

\begin{equation}
[\rm X]=log[(1-\textit{dil})10^{[\rm X]_{FG}}+\textit{dil }10^{[\rm X]_{SG}}]
\label{dilution}
\end {equation}

where [X] is the logarithmic abundance of a given element. [X]$_{FG}$ and [X]$_{SG}$ are, respectively, the abundances of the pristine gas and 
processed material; \textit{dil} represents the fraction of pristine gas in the gas where second generation formed. 
In order to obtain the data pattern, for [X]$_{FG}$ we assume the average values of [Mg/Fe] and [Al/Fe] for the FG reported in Table 7 by ME15; for [X]$_{SG}$ we assume the minimum value of magnesium observed and we derive the SG aluminum value by minimising the root mean square (rms) of the residulas along the fitting relations. In the case of M15 in Fig.~\ref{M15} we obtain [Al/Fe]$_{SG}=1.18\pm 0.05$ with rms=0.46, while for M92 we find [Al/Fe]$_{SG}=0.85\pm 0.06$ with rms=0.47. The same procedure is adopted in the O-Al plane, for all the GCs where a sufficient number of stars are available.

We caution the reader since the beginning regarding the fact that the interpretation of the observations in the C-N plane is trickier than others.
This is because the abundances of C and N are altered during 
the RGB ascending, due to deep mixing. While ``standard" evolution 
theories\footnote{We indicate with ``standard" the results obtained when the border of the 
convective regions is fixed via the Schwartzschild criterion.} for the FDU predict a C depletion of 
$\sim 20-30 \%$, several observational evidences \citep[e.g.][]{gratton04} suggest a deeper mixing in the post-bump 
phases, likely associated to thermohaline mixing \citep{corinne10}. Therefore, the C and N 
abundances of the giant stars do not reflect the initial chemistry. 

To quantify this effect in the comparison with the observations, we calculated stellar models evolving through the RGB in which deep
mixing is applied after the luminosity bump, with the same extent required to reproduce results
from thermohaline mixing models. We assumed a typical mass currently evolving through 
the RGB ($0.9~\rm M_{\odot}$). Starting from the chemical composition discussed in this section, we compute the evolution adopting three chemical compositions. They correspond to 
the chemistry of FG stars (pure pristine gas), of a $50\%$ mix of 
pristine matter with AGB gas and of a pure AGB ejecta. The results are indicated with green curves, respectively, with solid, dotted-dashed 
and dotted lines in Figs. \ref{M15} and \ref{M92}. We note that in all cases the surface carbon 
decreases during the RGB evolution, because of mixing surface matter with C-poor matter
exposed to CNO processing. On the other hand, the surface N of the models with chemistry
polluted by AGB gas remains unchanged, because the initial N is higher than the
equilibrium abundances expected in regions exposed to CNO cycling. We did the same computation at higher metallicity (Z$=2.5\times10^{-3}$) and we found that similar spread are predicted: $\delta(\rm[C/Fe])$ varies from $\sim 0.7$ dex in the case of a pure gas pristine initial chemistry, to $\sim 0.5$ dex in the 100\% AGB ejecta. Spread in N is found only in the first case ($\delta(\rm[N/Fe]\sim 0.6)$), for the same reasons previously stated.
Therefore, we generally conclude that detailed N measurements can be a valuable indicators of the degree of dilution of the contaminating gas
with pristine matter in the metallicity range here considered. 
Due to known errors in the ASPCAP determination of the main stellar parameters (e.g., Teff, logg), which affect
the derived abundances (including [C,N,O/Fe]) of SG stars in GCs (Jonsson et al. 2018, in prep.), ME15 chose to derive abundances independent
of ASPCAP, using photometric temperatures and gravities. The abudances derived by ME15 do not exhibit the systematic
errors that ASPCAP do. Nevertheless, [N/Fe], together with [C/Fe], are still affected by high level of uncertainty
due to the fact that CN and CO lines are often blended with other molecular or atomic lines. This affect low-metallicity measurements in particular ([Fe/H]$\leq-2$), as the CO, OH and CN lines are extremely weak in this parameter region; therefore it is not possible to give more information than upper and lower limits in this range of metallicities. This is clearly reported by ME15 where [N/Fe] and [C/Fe] errors are relatively high ($\pm 0.1-0.32$)\footnote{Note that these errors do not include possible systematic/random effects due to the methodology employed in the chemical abundances derivation (the use of spectral windows vs. the entire spectrum, model atmospheres, linelists, etc.)}. We take into account all these points in the following discussion and to draw any conclusion from this analysis.

\subsubsection{M15}
For this cluster we mainly base 
our analysis on the Mg-Al-Si abundances, because the abundances of the CNO elements are
available for only a few stars. In the comparisons we do not consider the three
outlier stars\footnote{These stars show a peculiar chemical composition very different from the rest of cluster stars, while they do not show stellar parameters and radial velocities so different to the rest of the cluster stars. We believe that the ME15 methodology for the derivation of their chemical abundances, or the DR10 data reduction pipeline has failed with these stars for some unknown technical reasons.}, showing very low Al ($[\rm{Al/Fe}]\sim-0.5$) and Si ($[\rm{Si/Fe}]\leq0.2$) abundances, 
with [Mg/Fe] in the range $-0.2 < [\rm{Mg/Fe}] < +0.15$. 

In the Mg-Al plane, shown in the top-left panel of Fig. \ref{M15}, we can clearly
distinguish an anticorrelation pattern, the stars with most extreme chemical composition (Mg-poor/ Al-rich),
in the left, upper region of the plane, showing up an overall magnesium depletion
with respect to the FG of $\sim 0.6$ dex, and an Al enhancement of $\sim +1.4$ dex. 
As shown in the figure, the yields of AGB stars show the same Mg and Al spread observed. These
results suggest that these stars formed from the ejecta of 4.5-6 solar masses AGBs, with only a
$\sim 10\%$ contribution from pristine gas.

Some stars, which fall within the blue region, are characterized by a chemical composition
somewhat intermediate between the FG and the stars with the most extreme chemical composition:
these are SG stars, formed from AGB ejecta diluted with $\sim 40-50\%$ of pristine 
gas of the cluster.

In the Mg--Si plane, the data indicate the presence of a silicon spread\footnote{This is confirmed by the application of the least square fit method to the data, from which we obtain a straight line with a steepness of -0.3 and a scatter of 0.09.}, $\delta\rm[Si/Fe]\sim +0.2$ dex, which indicates that the material from which the Mg-poor and Si-enhanced stars
formed was exposed to Al-burning. The observed silicon spread is consistent with what is expected from the yields as shown in the top-right panel of Fig. \ref{M15}.
While the distinction between FG and SG stars is more clear in the Mg-Al panel
discussed previously, here we stress the importance of the existence of a Si spread, because Si production requires hotter temperatures than those for Al, above $\sim 100$MK, thus
providing important hints on the class of polluters which contaminated the intra-cluster
medium.

The CNO abundances are available only for 6 stars in the sample. Three of them do not 
show any Al enhancement nor O and Mg depletion, which suggests that they belong to the FG
of the cluster; these stars have been indicated as full squares in Fig. \ref{M15}.
The three stars left show the imprinting of Mg and O depletion and Al enhancement, thus
they belong to the SG. Two of these stars, indicated with asterisks, show a chemical
composition intermediate beween FG and SG stars, thus they formed from dilution of AGB
gas with pristine matter. One out of the 6 stars (2M21291235+1210498), indicated 
with an asterisk surrounded by a circle, shows up a significant enhancement of aluminium 
(above 1 dex) and a magnesium content $\sim 3$ times smaller than FG stars, a trace of 
a more extreme chemical composition. Interestingly, the combination of the Mg, Al and O 
abundances for this star reflect the chemical composition of the ejecta from AGB stars 
of mass $4.5-5~\rm M_{\odot}$. 

As stated above, any further conclusion 
based on the comparison between the present dataset and the models in the C-N plane is 
not straightforward, in particular in the lower metallicity cases and with such a small 
sample of stars. In the case of M15, nonetheless, we may take advantage of the comparison with 
the results by \citet{cohen05a}, who measured the C-N abundances of a wide number of RGB 
stars, shown in the bottom-right panel of Fig. \ref{M15}. The wide spread 
of $\sim 1$ dex observed in [N/Fe] is in agreement with the prediction 
of the models. \citep{cohen05a} sampled the entire RGB, including stars below the bump, 
thus not exposed to any deep-mixing. We indicate these sources according to their estimated 
temperature in order to distinguish stars below (circles, $\rm T_{\rm eff}>5300$ K) and above 
(triangle, $\rm T_{\rm eff}<5300$ K) the location of the FDU. It is interesting to discuss
the first 
group of stars: part of them show [N/Fe]$\sim 0$, without carbon depletion, clearly 
indicating the pure initial chemistry of the FG, not affected by any mixing episodes. 
On the other hand, in the same class of stars we can identify the SG, on the basis of 
their higher nitrogen abundances, associated with a progressively increasing carbon 
depletion. This reflects the imprinting of the medium where they formed, in remarkable 
agreement with the predictions from the AGB yields.
 
To summarise, ME15 data indicate that M15 harbours SG stars, formed from nuclearly
processed gas, which is consistent with the AGB self-enrichment expectations. The presence of SG stars
in M15 is not new, as evidences for the presence of SG stars in this cluster
dates back to the work by \citet{cohen05a}. The UVES data by \citet{carretta09b} 
also suggest that SG stars exist in M15: the observations outline the presence
of a O-Na anticorrelation trend (see Section \ref{ONacarretta}), despite a solid interpretation of this result is
partly disturbed by the fact that for most stars only upper limits exist for the
oxygen abundances. More recently, \citet{milone17}, 
based on high-precision photometry of RGB stars, estimated that $60\%$ of the stars in 
M15 belong to the SG. 
The presence of SG stars could also explain the helium spread invoked to account for the
morphology of the HB: \citet{vdb16} claims that helium mass fractions $Y > 0.29$ are
required to reproduce the blue part of the HB of M15, whereas \citet{milone14}
conclude that a helium spread $\delta \rm Y \sim 0.07$ would be compatible with the
distribution of M15 stars on the HB. Interestingly, such large spread would imply
the existence of stars formed with $Y\sim 0.32$, the helium obtained when diluting pure
AGB ejecta with $\sim 20\%$ of pristine gas; this is in agreement with our interpretation
of M15 SG stars with the most extreme chemistry.

\subsubsection{M92}
On the chemical side, M92 is extremely similar to M15, in terms of metallicity and
of the Mg-Al-Si content of FG stars. For these elements a deeper analysis is possible
compared to M15, because more data are available, while the CNO
elements are available for 6 stars only.
The comparison between the observations and Z$=2\times 10^{-4}$ models are shown in 
Fig. \ref{M92}, with the dilution curves obtained by mixing gas from AGB stars with
variable percentage of pristine matter.
 
Similarly to M15, the Mg-Al plane proves the best locus to draw information on the
formation of SG stars, owing to the small increase expected for silicon and the paucity
of data available for CNO elements. The most extreme chemistries of M92 stars
correspond to an overall Mg decrease of $\sim 0.6$ dex and a $\sim 1.3$ dex 
of increase in Al, with respect to the average abundances of the FG stars; same abundances are expected from the AGB ejecta.

These results confirm on qualitative grounds the conclusions drawn by \citet{ventura16}
regarding the distribution of M92 giants in the Mg-Al plane. Compared to \citet{ventura16}, here
we obtain a better agreement between data and theoretical predictions,
because in this case we used the same Mg abundance of FG stars, namely $[\rm{Mg/Fe}]=+0.3$, $0.1$ dex smaller than in 
\citet{ventura16}: this is the reason for the smaller Mg and Al contents in the ejecta
of AGBs, which nicely fit the data shown in the left, bottom panel
of Fig. \ref{M92}.

Compared to M15, we note that the data cover the Mg-Al plane more uniformly, suggesting
a formation of the SG with different degrees of dilution with pristine gas of the
cluster. However, we cannot draw conclusions on any difference in the formation of the SG in the
two clusters, as the dissimilarity between the two clusters could be due to the smaller 
dataset available for M15.

The Mg-Si anticorrelation is more evident in M15 than in M92\footnote{The least square fitting of the data provide a steepness of -0.15 and a scatter of 0.1.} due to the high Si dispersion of the data around [Mg/Fe]$\sim$0.2. 
On the other hand, stars with the lowest magnesium ([Mg/Fe]$<-0.1$) show up a silicon content 
[Si/Fe]$>0.45$; this is in good agreement with the prediction from yields of AGB stars, 
suggesting that these stars possibly formed from gas with a very limited (if any) degree of dilution
with pristine matter.

Turning to the O-Al plane (see left, bottom panel of Fig. \ref{M92}), we find that 4 out 
of the 6 stars for which oxygen data are available belong to the FG. These stars, indicated
with full squares, exhibit a rather homogenous Al content; their position in the two
planes discussed previously confirm this interpretation. 
The other two stars (asterisks) for which CNO data are available, are more Al-rich, one of them with a
Mg and O contents significantly smaller than FG stars: from their surface chemistry 
we interpret them as belonging to the SG of the cluster, formed from AGB gas diluted
with $\sim 20\%$ (the one with the smallest Mg, 2M17163748+4306155) and $\sim 50\%$ 
(2M17171307+4309483) of pristine material.
While the same reasons stated for M15 (limits in the determination of the abundances in the low metallicity range, limited number of stars, etc.) prevents a clear interpretation of the C-N plane, in the
right, bottom panel of Fig. \ref{M92}, we display the results of the AGB ejecta and
stellar evolution calculations with deep mixing.

In conclusion, the analysis of the wide Mg-Al anticorrelation measured provide a strong indication of the presence of a SG in M92,
formed from gas which was exposed to advanced proton-capture nucleosynthesis. \citet{milone17} conclude that
the FG stars account for $\sim 30\%$ of the whole population of M92, a result
consistent with our understanding of the Mg-Al data (see left-upper panel of 
Fig.~\ref{M92}). The comparison of the data points and the chemistry of AGB 
ejecta in the Mg-Al and Mg-Si planes suggests that M92 hosts a few stars with a very
extreme chemical composition, formed by pure AGB gas, possibly mixed with
a very tiny percentage of pristine matter. These stars, which would have a helium
content $Y \geq 0.30$ (see right panel of Figure \ref{Tbce}), apparently have no counterparts on the HB of 
M92, according to the analysis by \citet{vdb16}, who claim that a very small helium
spread is required to reproduce the morphology of the HB of this cluster.

\begin{figure*}
\begin{minipage}{0.48\textwidth}
\resizebox{1.\hsize}{!}{\includegraphics{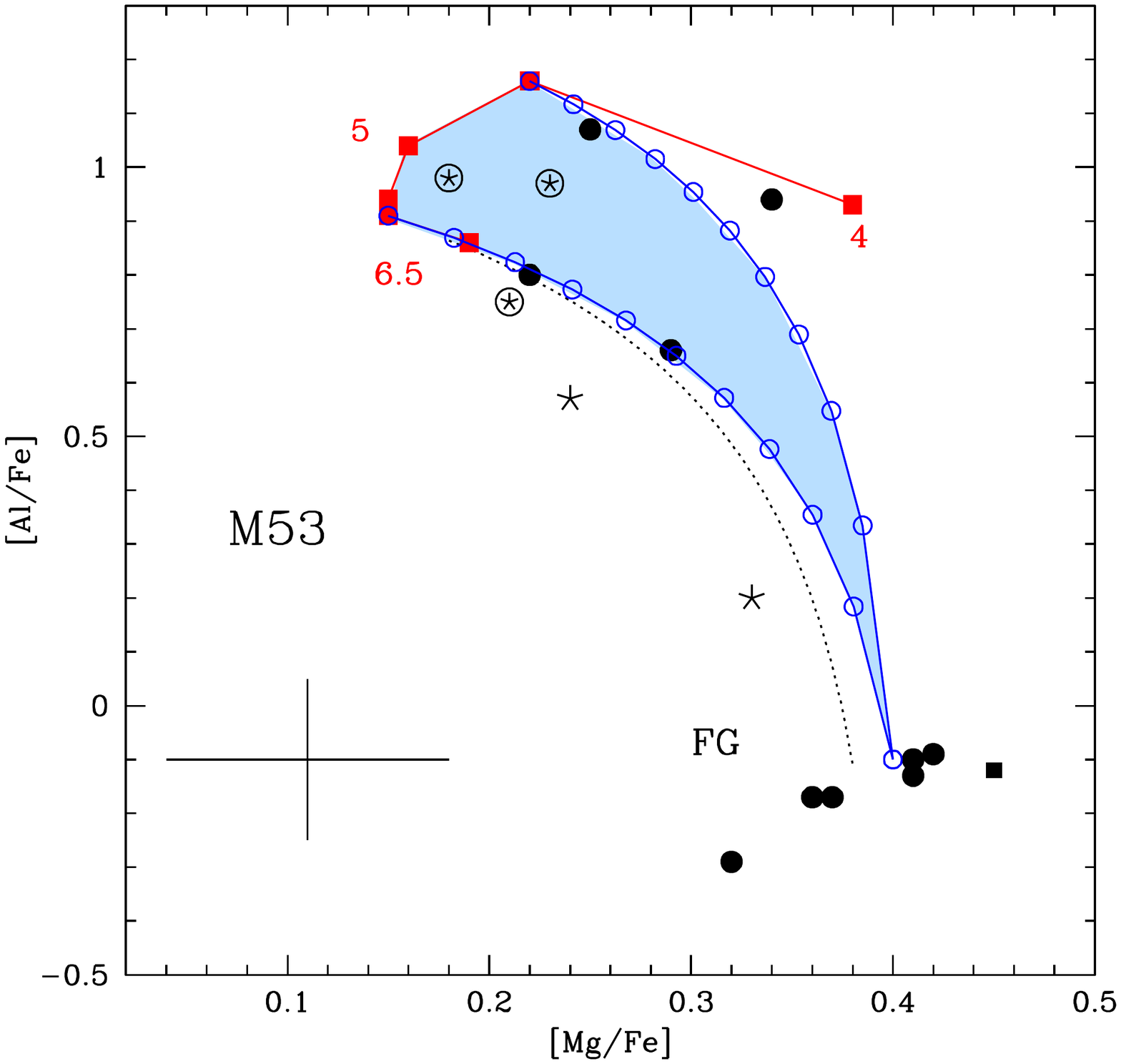}}
\end{minipage}
\begin{minipage}{0.48\textwidth}
\resizebox{1.\hsize}{!}{\includegraphics{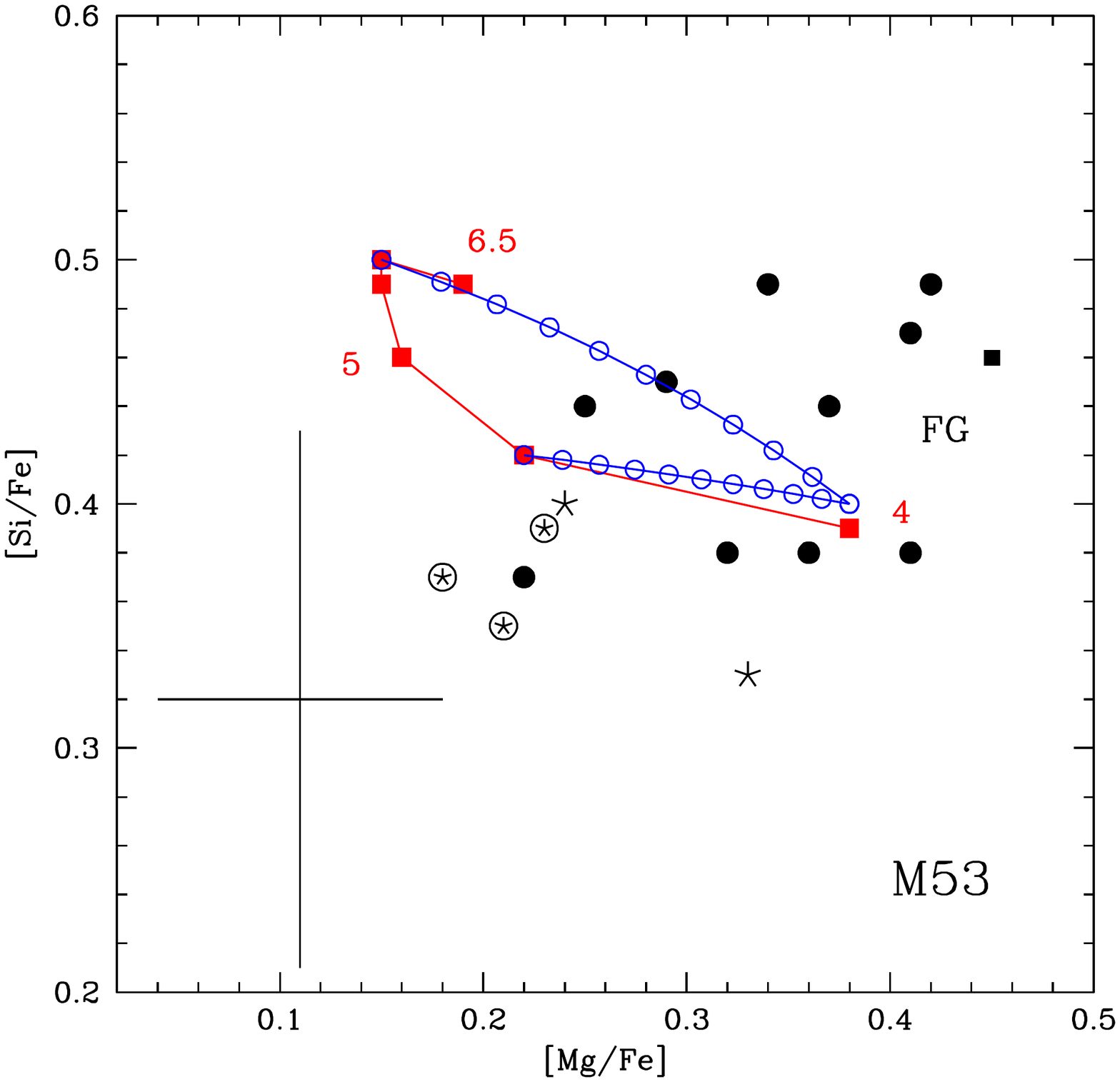}}
\end{minipage}
\vskip-70pt
\begin{minipage}{0.48\textwidth}
\resizebox{1.\hsize}{!}{\includegraphics{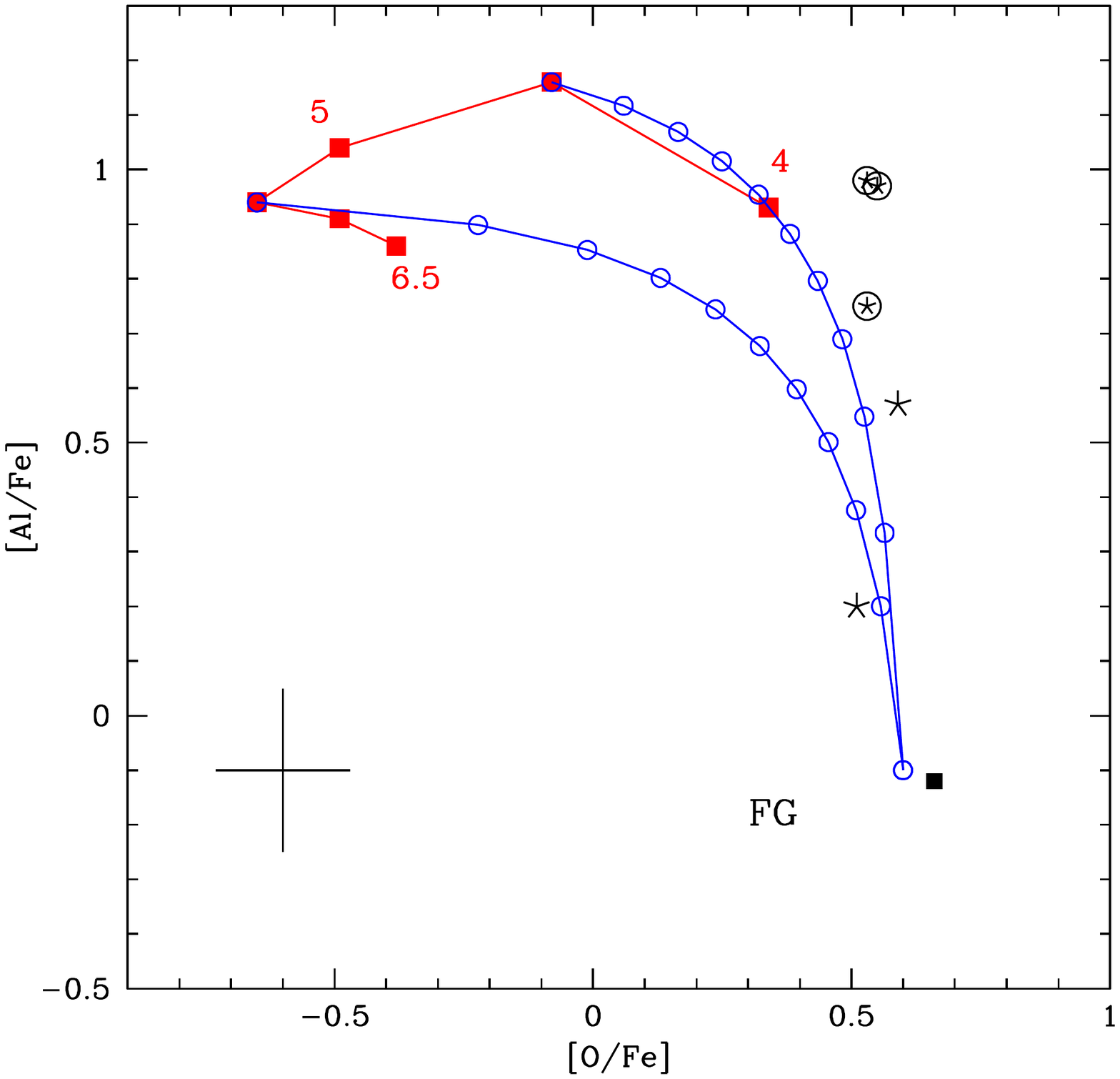}}
\end{minipage}
\begin{minipage}{0.48\textwidth}
\resizebox{1.\hsize}{!}{\includegraphics{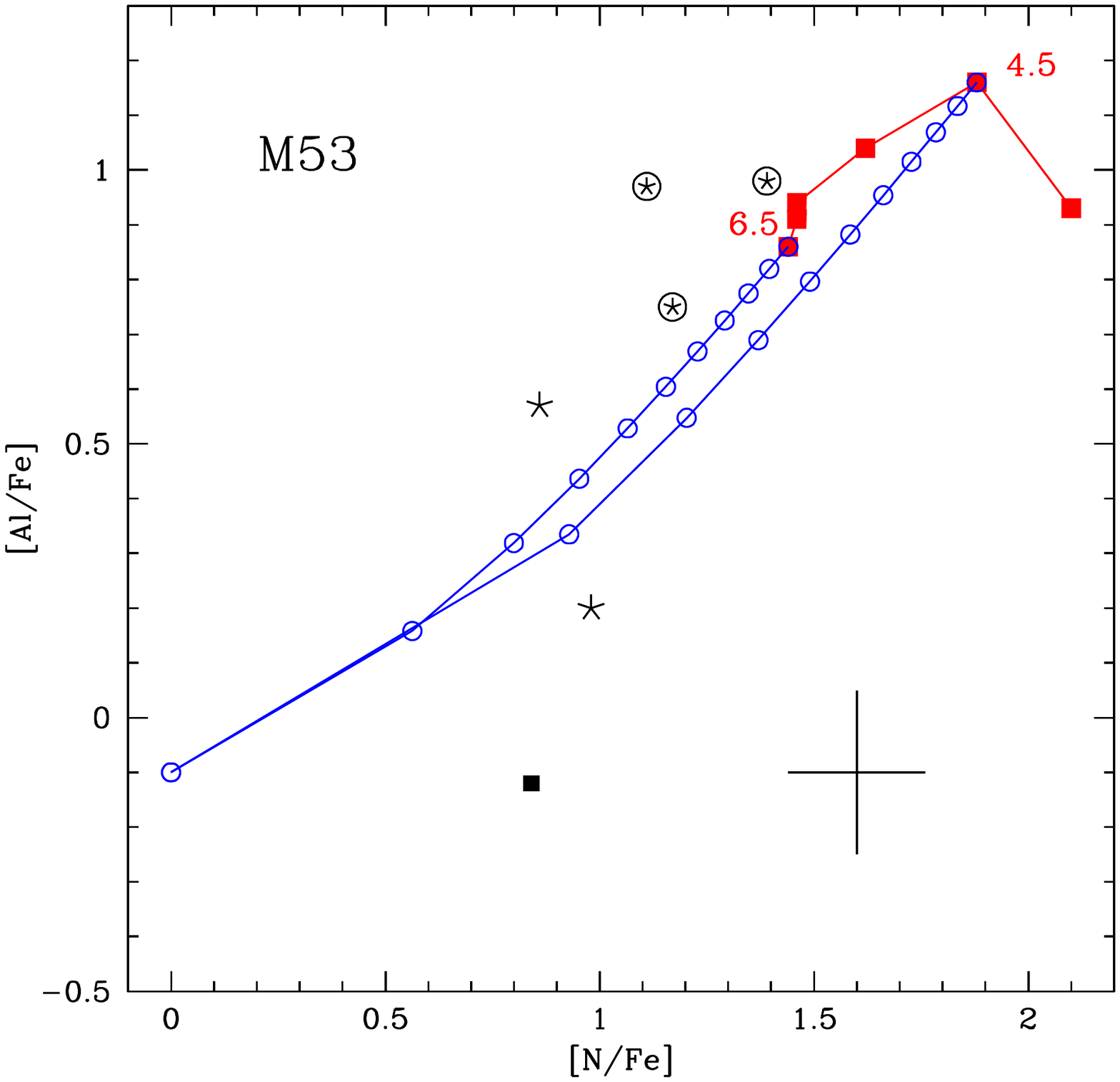}}
\end{minipage}
\vskip-50pt
\caption{M53 stars from ME15 are shown as full dots in the Mg-Al (top--left) and Mg-Si (top--right). CNO abundances are measured for part of the sample. They are identified by black squares, stars and stars surrounded by open circles according to the anticorrelation pattern (low, intemediate and high, respectively) shown in the O-Al (bottom--left) and N-Al (bottom--right) planes. Results from the fits are shown with dotted lines and the black crosses indicate typical error bars. The yields of AGB models of metallicity Z$ = 4\times10^{-4}$ are indicated with red squares; the numbers close to the squares indicate the initial masses. Blue, open circles indicate the chemistry expected from mixing of the AGB ejecta with pristine gas, with percentages of the latter ranging from 0 to 100\%, in steps of 10\%. The points corresponding to 100\% dilution are close to the FG label.}
\label{M53}
\end{figure*}

\subsection{M53: a cluster at the edge}
The metallicity of this cluster is higher ($[\rm Fe/H]=-1.96$) than M15 and M92. 
Magnesium, aluminium and silicon abundances have been measured for 16 stars,
for 6 of which the CNO values are available. The models used to study M53
have been calculated with metallicity Z$=4\times10^{-4}$. The comparison between the data and
the models is reported in Fig.~\ref{M53}.
 
The extent of Mg-depletion and Al-enhancement observed in this cluster and shown in the top--right panel of 
Fig.~\ref{M53}, are $\delta [\rm{Mg/Fe}] \sim -0.2$ dex and 
$\delta [\rm Al/Fe] \sim +1.2$ dex, respectively. These numbers are in good agreement with the AGB 
models shown here. In the same panel of this Figure, the result of the fitting procedure for M53 is shown, obtaining  [Al/Fe]$_{SG}=0.86\pm 0.06$ with rms=0.34.

Compared to M15 and M92, the Mg-depletion is 
significantly reduced, whereas the abundances of the most Al-rich stars are comparable. 
These differences can be understood on the 
basis of the lower HBB temperatures of the intermediate mass stars that evolved in M53, 
compared to M15 and M92. Based on the discussion in Section \ref{hbb} and on the 
results shown in Fig.~\ref{Tbce}, we know that the HBB experienced at 
the base of the convective envelope is stronger the lower is the metallicity. 
The Al-enhancement, as shown by the data and the models, is rather independent of Z 
in this metallicity domain, unlike the depletion of magnesium. This reflects the difference in the 
nucleosynthesis experienced by the two chemical species under the effects of HBB. 
Magnesium is destroyed during the whole AGB phase, the depletion rate being larger 
the higher is T$_{\rm bce}$. Al is subject to both a creation path and a destruction 
channel, the latter leading to the production of 
silicon. It is not surprising that the amount of Al produced is almost independent of Z:
as far as the HBB temperatures exceed $\sim 110$ MK 
it is expected that silicon production occurs (see Fig.~\ref{Tbce}), with a partial
depletion of the Al previously synthesized by Mg burning (an exhaustive discussion on the 
Mg-Al-Si nucleosynthesis can be found in Arnould et al. 1999).

The maximum silicon spread expected from the AGB ejecta is $\delta \rm[Si/Fe]<0.1$ dex in 
the case of M53. This is because the metallicity of this cluster is sufficiently 
low for the FG stars of intermediate mass to have produced Mg-poor and Al-rich gas, 
but at the same time it is high enough to avoid a more advanced HBB nucleosynthesis, 
required to eject gas enriched in silicon. For this reason we do not expect 
any Si-enhancement for higher metallicities GCs and we consider M53 as an edge 
point in this excursion. The silicon spread expected is smaller than the dispersion and 
the typical error of the data is shown in the right, top panel of Fig.~\ref{M53}. It makes 
really hard to confirm or disproof the presence of Mg-Si anticorrelation in this cluster. 
More precise measurements and larger sample of data would be useful to draw any conclusion 
on this regard and on the duration of the formation of the SG observed in the Mg-Al plane.

In the O-Al plane we base our analysis on the 6 stars for which the oxygen is available.
One star (2M13123617+1807320), indicated with a full square, shows a typical
FG chemistry, based on the O, Mg and Al abundances. The 5 stars left show the presence
of some Al-enrichment, suggesting pollution from gas exposed to nuclear activity.
For two out of the 5 stars the aluminium abundance observed is $[\rm{Al/Fe}]<+0.5$ (shown 
with asterisks in Fig.~\ref{M53}), which suggests significant ($>50\%$) dilution with pristine 
matter; the positions in the Mg-Al plane supports this interpretation. For the 3 stars left,
indicated with asterisks surrounded by circles, the interpretation of their position in
the O-Al and Mg-Al planes leads to different conclusions. Looking at the O-Al plane,
the comparison of data with the AGB chemistry suggests that the Al abundances are slightly 
overestimated (still within 
the error bars) and that they formed from a $\sim 50\%$ mixing of AGB ejecta and pristine 
matter. On the other hand, their surface magnesium reflects the chemistry of the AGB
gas, with practically no dilution; in this case we should invoke an overestimation of
their oxygen to allow consistency with the models. Our favourite option is the former,
which is compatible with the large error bars of magnesium, relative to the overall Mg 
spread observed. This understanding is in agreement with the studies by \citet{boberg16}, 
who suggest that SG stars in M53 formed from gas diluted with pristine matter, and by
\citet{caloi11}, that, based on the analysis of the HB, concluded that only a small helium spread, 
$\delta \rm Y \sim 0.04$, is present, thus ruling out the existence of a purely AGB 
contaminated stellar component. 
Concerning the bottom-right panel of Fig.~\ref{M53}, from the AGB ejecta we expect a clear N-Al correlation, shown by SG stars of this cluster.

\subsection{The intermediate metallicity GCs}
We discuss M13, M2 and M3 together, owing to their close metallicities and
chemical compositions. The pattern traced by the multiple populations of the GCs 
considered in this section will be discussed on the basis of their intermediate metallicity,
i.e. Z$=10^{-3}$. These models were calculated ad hoc for the distribuiton of Mg, Al and O shown in these clusters (see Table 1), slightely enhanced in Mg with respect to those used in \citet{ventura16}. As shown in Fig.~\ref{Tbce} the HBB temperatures of
stars of this metallicity are not sufficient to allow any silicon production,
if not in the stars of mass close to the threshold for core collapse. This is 
confirmed by the results for these three clusters, indicating that no silicon
spread is observed. Therefore, we will focus our analysis on the Mg-Al
plane and on the abundances of the elements involved in CNO cycling.

\begin{figure*}
\begin{minipage}{0.33\textwidth}
\resizebox{1.\hsize}{!}{\includegraphics{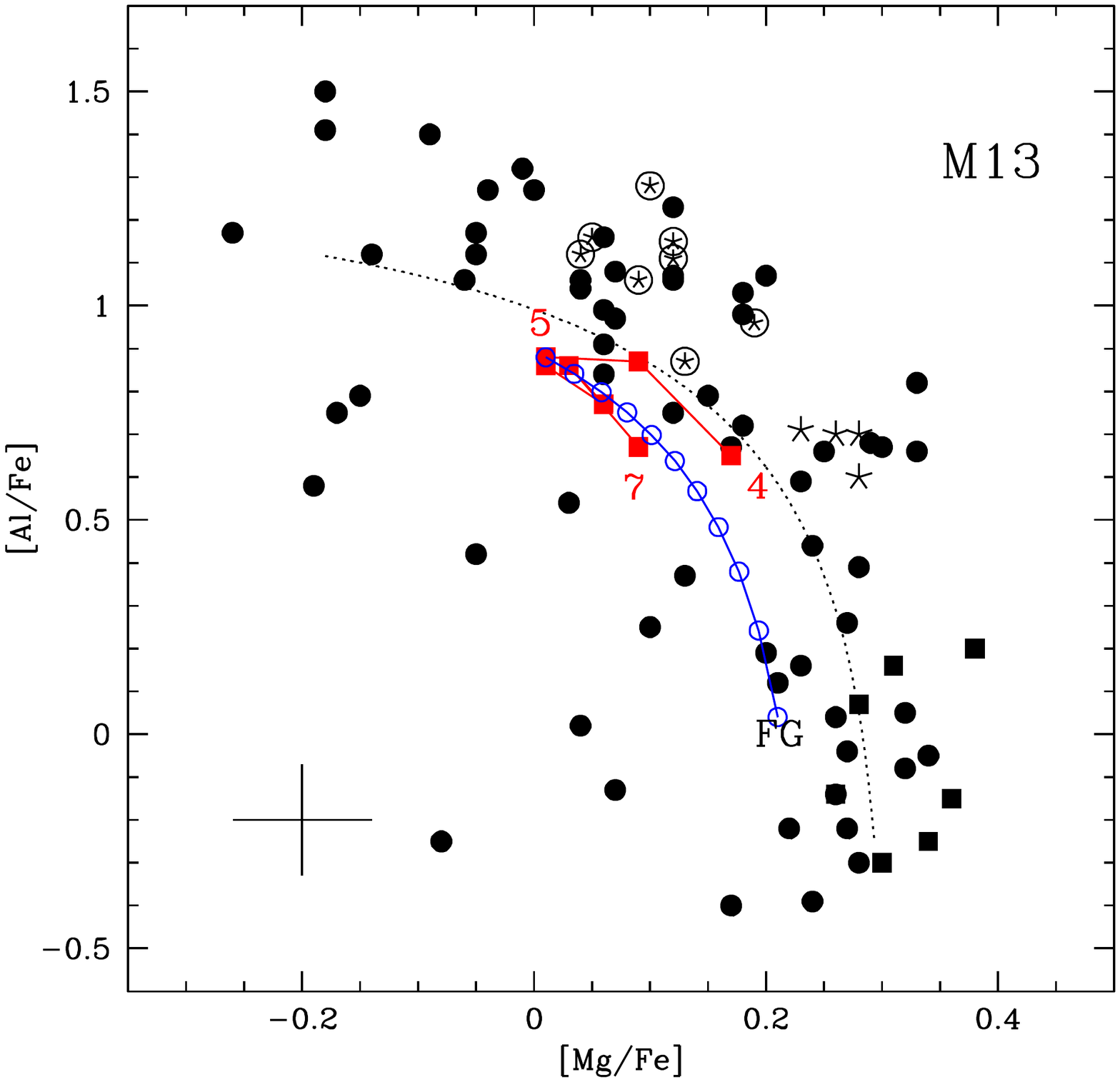}}
\end{minipage}
\begin{minipage}{0.33\textwidth}
\resizebox{1.\hsize}{!}{\includegraphics{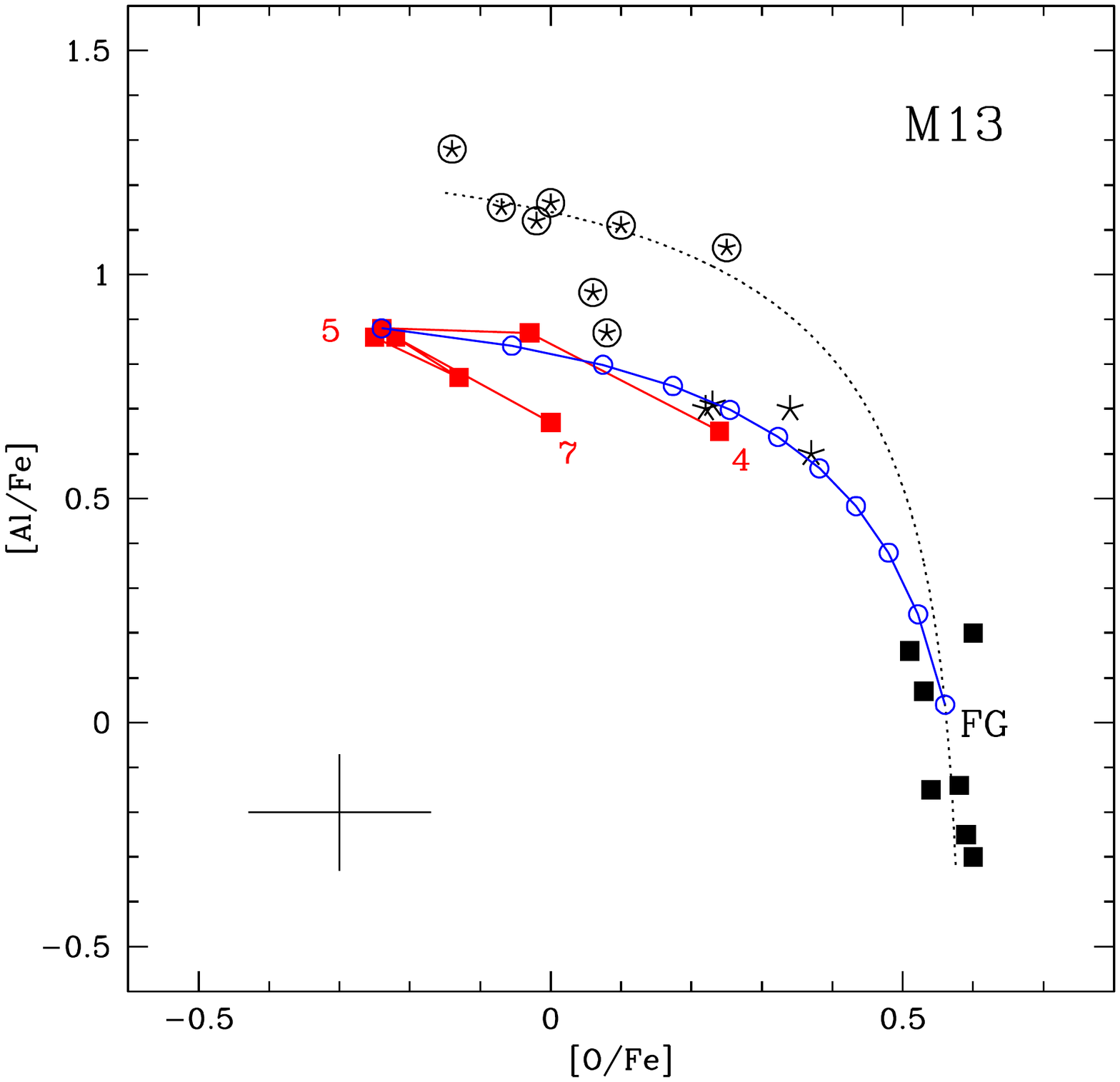}}
\end{minipage}
\begin{minipage}{0.33\textwidth}
\resizebox{1.\hsize}{!}{\includegraphics{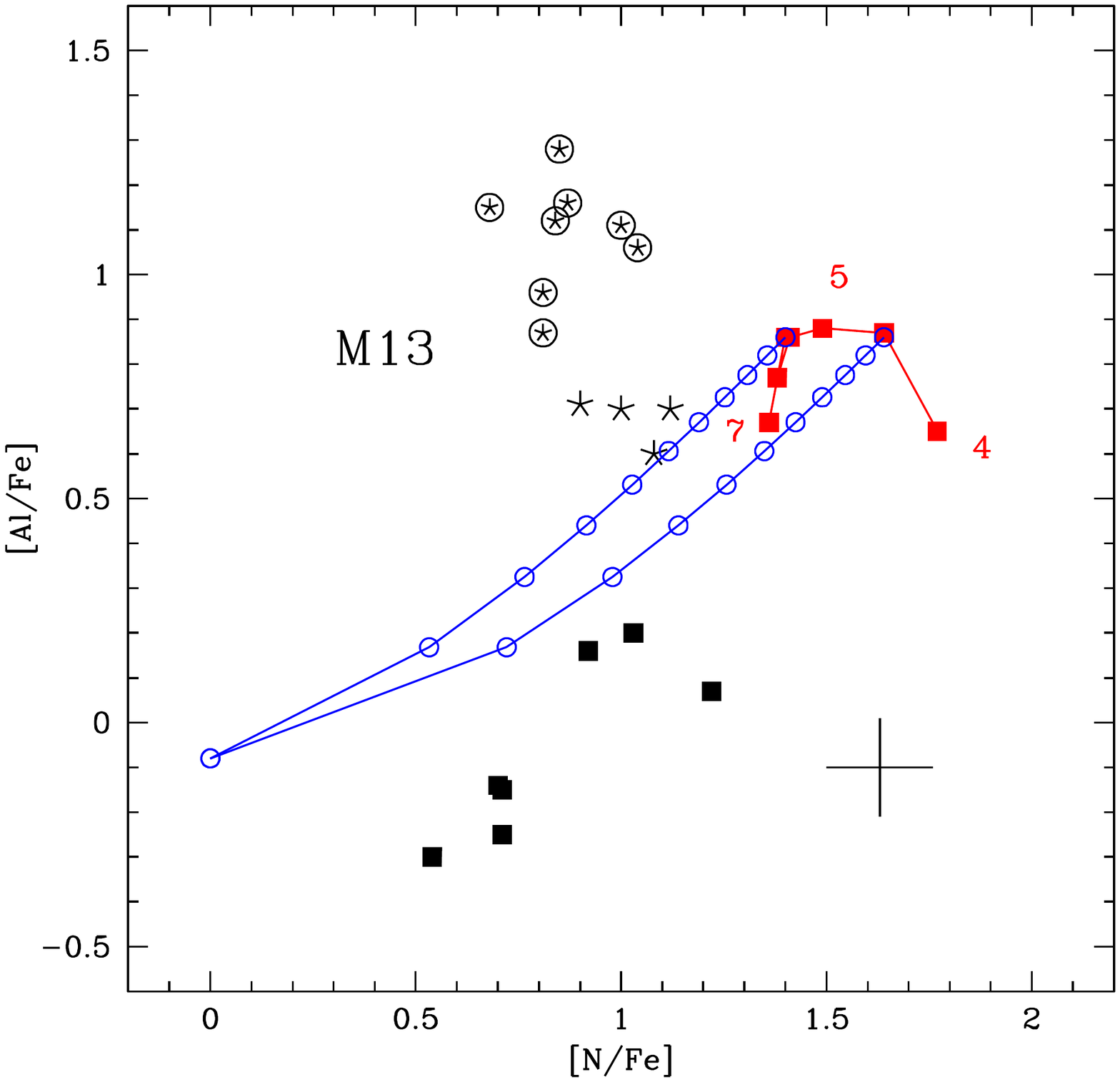}}
\end{minipage}
\vskip-10pt
\caption{The Mg-Al (left), O-Al (center) and N-Al (right) distribution of stars in M13. The observations are compared with AGB models of metallicity Z$ = 10^{-3}$. The meaning of the symbols is the same as in Figure \ref{M53}.}
\label{M13}
\end{figure*}

\subsubsection{M13}
\label{secM13}
The M13 sample is composed by 81 stars, for which
the Mg and Al abundances were calculated. For 18 of these sources the CNO abundances
are also available. The distribution of the data points on the Mg-Al, O-Al and N-Al
planes is shown in Fig.~\ref{M13}; overimposed to the data we show the yields from
massive AGBs.

In the Mg-Al plane (see left panel of Fig.~\ref{M13}) we report the reults of the fit, with [Al/Fe]$_{SG}=1.29\pm 0.04$ with rms=0.56. Form this panel we note that the 
yields from AGB stars show up a depletion of magnesium of $\sim -0.2$ dex, which is 
sufficient to reproduce the majority of the stars in the sample, with
the exception of $\sim 10$ stars, in which the measured Mg is $\sim 0.4$ dex smaller than 
the average of FG stars, as given by ME15. These stars are located in the left, upper 
zone of the Mg-Al plane, with $[\rm{Mg/Fe}]<0$; their Al abundances are $\sim +0.3$ dex larger 
than the Al content of the yields. The data errors seem not sufficiently large to explain
this difference.

In the O-Al plane\footnote{[Al/Fe]$_{SG}=1.18\pm 0.03$ with rms=0.28.} we see that the observed O spread, $\delta [\rm{O/Fe}] \sim -0.7$ dex, 
is well reproduced by the AGB models. This comparison suggests the presence of SG
stars in M13, some of which were likely formed from the AGB ejecta, with a small
(if any) degree of dilution with pristine gas. However, this result must be
taken with some caution. The smallest Mg abundances measured for the 10 stars mentioned above would indicate that the AGB nucleosynthesis at the 
intermediate metallicities, while correctly predicting the depletion of oxygen, slightly 
underestimates the strength of Mg burning.  We expect that these stars have $\delta [\rm{O/Fe}] > -0.8$; unfortunately for none of the $\sim 10$ Mg-poor stars
discussed above the CNO data is available to draw firm conclusions on this regard.

As shown for M53, an correlation pattern is expected from the AGB ejecta in the N-Al plane, as displayed in the right panel of Fig. \ref{M13}. This is also confirmed by previous studies on these two elements \citep[e.g.][]{bragaglia10,gratton12}. Nevertheless, no clear distinction between FG and SG stars is detected in the distribution of the N abundances of this cluster. This suggests that the N abundances should be revisited and avoid us to draw any firm conclusion on this regard.

M13 has traditionally played an important role in the studies focused on GCs,
particularly for the complex morphology of the HB, which displays an extended blue
tail. On the side of the chemical patterns traced by GC stars, early results
presented by \citet{kraft97} showed an extended 
O-Na anticorrelation, with an oxygen spread of $\delta [\rm{O/Fe}] \sim -0.8$ dex,
very similar to that shown in Fig.~\ref{M13} (see middle panel); a less defined,
albeit statistically significant Mg-Al trend is found in the same work, with an overall
 Mg and Al spreads $\delta [\rm{Mg/Fe}] \sim -0.4$ dex and $\delta([\rm Al/Fe]) \sim +1.2$ dex, again in agreement with the results from
ME15. Important results from high resolution spectroscopy of M13 giant stars were
published by \citet{sneden04} and \citet{cohen05b}, that confirmed the existence of
a O-Na anticorrelation pattern, with the same oxygen and Na spreads given in
\citet{kraft97}. In both studies the stars with the largest sodium are magnesium
poor, with the same Mg spread, $\delta [\rm{Mg/Fe}] \sim -0.4$ dex, found by ME15.
A slightly lower Mg spread was detected by \citet{johnson05}. 

In all the afore mentioned investigations the largest
abundances of magnesium are $[\rm{Mg/Fe}] \sim +0.4$, higher than the Mg of FG stars
given by ME15, on which we based our computations; while this difference has practically
no effects on the Mg spread, Al production is affected by this choice, because a higher 
Al is expected when the initial Mg is larger. This could reconcile the largest Al
given in ME15 with the AGB yields.

These studies, although less homogeneous and with a lower statistics than ME15,
confirm the oxygen and magnesium spreads given in ME15. As stated previously, while the
oxygen spread is nicely reproduced by the AGB models of the M13 metallicity, the 
expected Mg spread is too small to reproduce the stars with the most extreme chemical
composition; this is indeed not new, as the same problem was addressed by \citet{ventura11},
when comparing the AGB ejecta with the most Mg-poor stars in M13 and NGC2808.
The analysis by \citet{ventura11} showed that the key quantity relevant for the
determination of the overall Mg depletion in the gas from massive AGB stars is the
cross-section of the $^{25}\rm Mg(p,\gamma)^{26}\rm Al$ reaction; this is because a significant
depletion of the total magnesium is made possible only via an efficient destruction of
the  $^{25}\rm Mg$ produced by  $^{24}\rm Mg$ burning. An analysis of the enhancement of the
$^{25}\rm Mg$ destruction rate required to reproduce the observed Mg spread is in 
progress (Ventura et al., submitted).

Taken as a whole, the results by ME15, in agreement with the previous works cited above,
indicate that M13 harbours a significant fraction of SG stars; this is in agreement with 
the recent analysis by \citet{milone17}, who suggest that only $\sim 20\%$ of the stars
in M13 belong to the FG of the cluster. The wide oxygen and magnesium spreads observed
indicate that part of SG stars formed from almost pure AGB ejecta, thus suggesting the
present of an extremely rich population. This is in agreement with the study by
\citet{caloi05}, who claimed the presence of stars significantly enriched in helium
to account for the complex morphology of the HB of M13. A later analysis by
\citet{dalessandro13}, aimed at interpreting UV, HST data, confirmed the presence of
stars enriched in helium, although the largest helium proposed, $Y_{max}=0.30$, is 
slightly smaller than the helium expected in stars formed from pure AGB ejecta. Our interpretation is much more consistent with
the study by \citet{dantona08}, that suggest the presence of a fraction of M13 stars
with helium mass fractions in the range $0.27 < \rm Y < 0.35$.

\begin{figure*}
\begin{minipage}{0.33\textwidth}
\resizebox{1.\hsize}{!}{\includegraphics{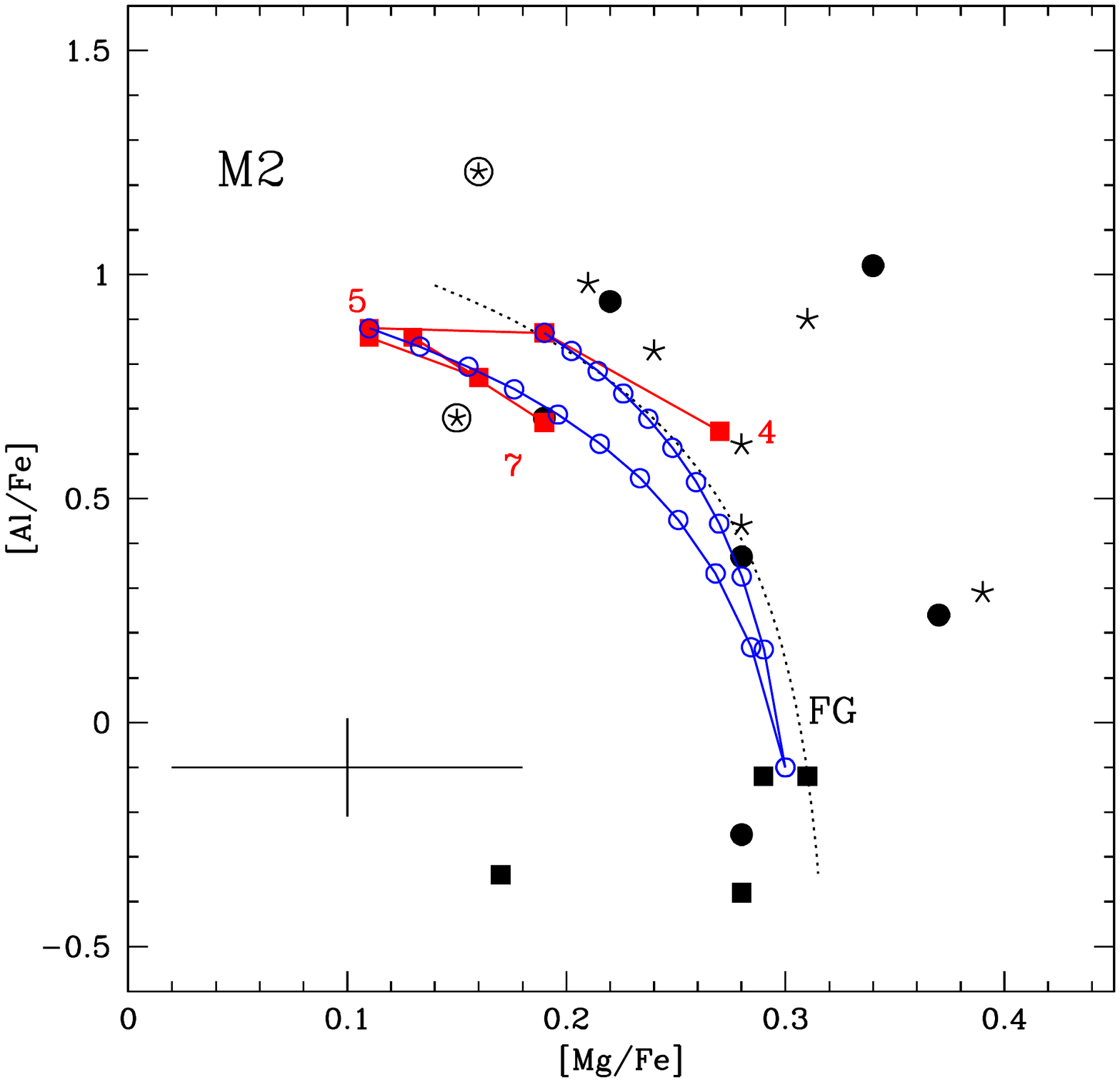}}
\end{minipage}
\begin{minipage}{0.33\textwidth}
\resizebox{1.\hsize}{!}{\includegraphics{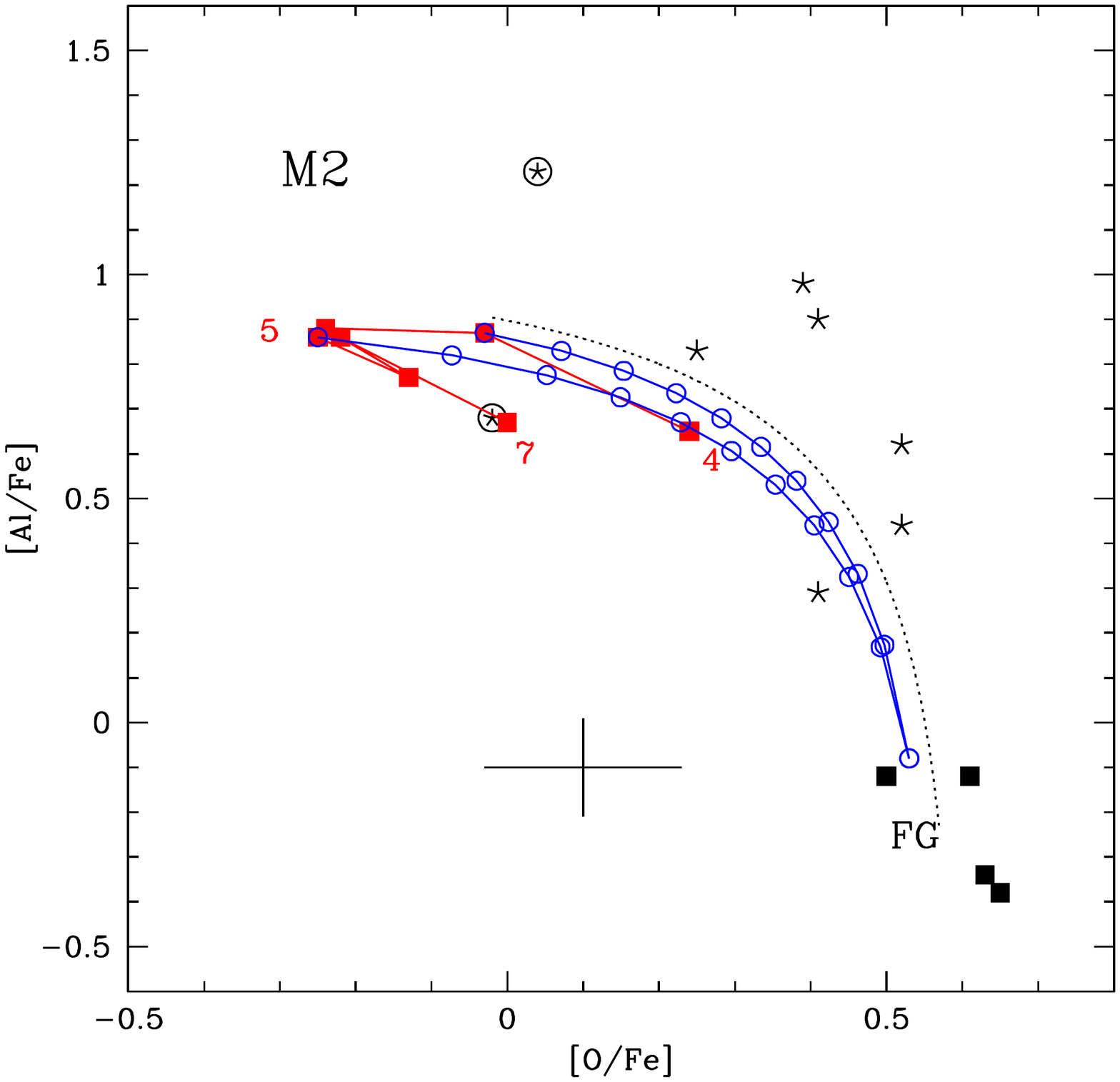}}
\end{minipage}
\begin{minipage}{0.33\textwidth}
\resizebox{1.\hsize}{!}{\includegraphics{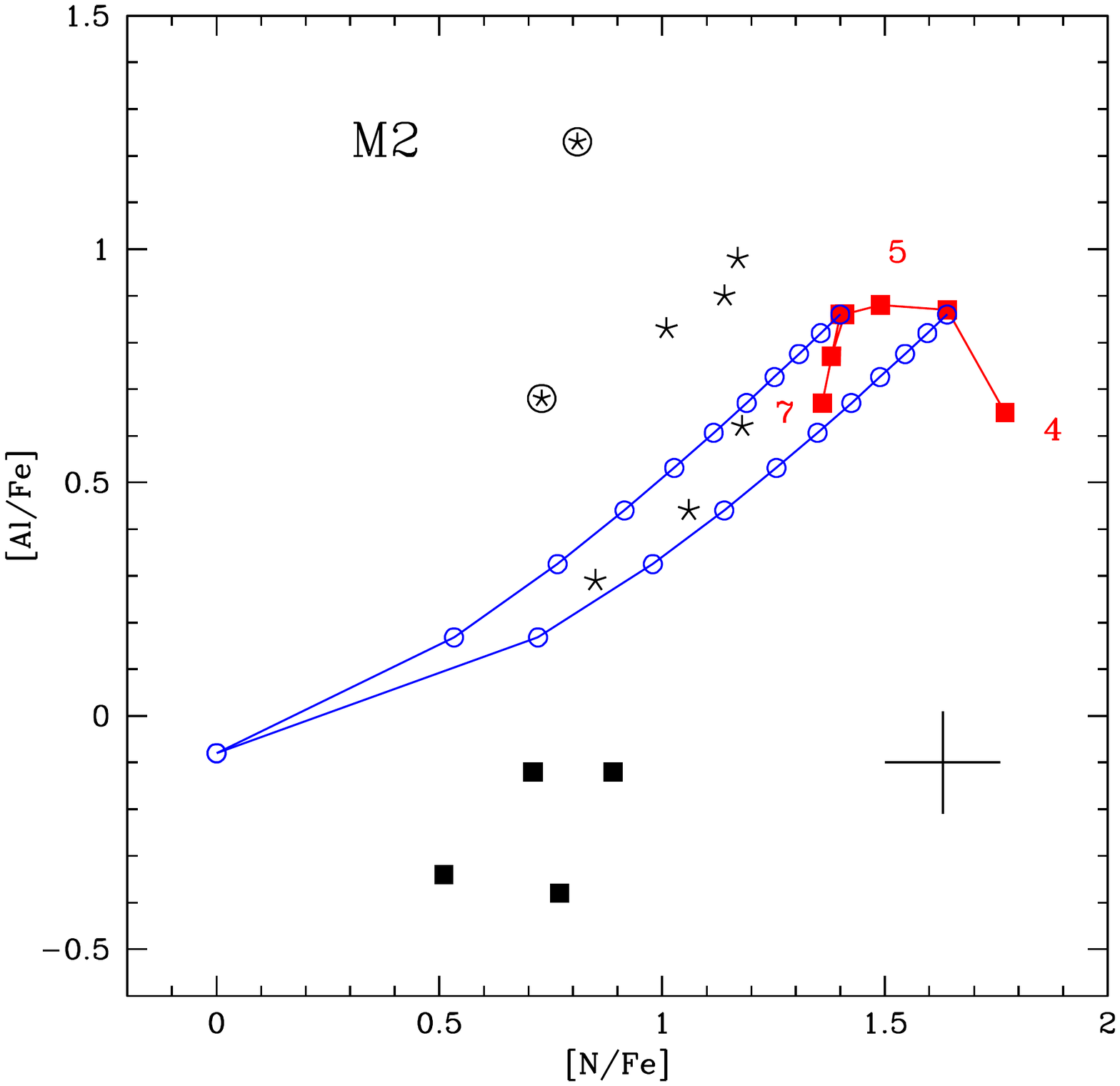}}
\end{minipage}
\vskip-10pt
\caption{The Mg-Al (left), O-Al (center) and N-Al (right) distribution of stars in M2. The observations are compared with AGB models of metallicity Z$ = 10^{-3}$. The meaning of the symbols is the same as in Figure \ref{M53}.}
\label{M2}
\end{figure*}

\subsubsection{M2}
Data for M2 are limited to 18 stars, 12 of which has the CNO abundances. 
This will allow us to undergo a less robust and complete analysis, compared
to M13 and M3.

The comparison with the chemical composition of the AGB ejecta is shown in Fig.~\ref{M2}.
To account for the slightly higher magnesium of FG stars in M2 compared to the 
other intermediate metallicity clusters, we shifted all the AGB magnesium by
$+0.1$ dex; we warn here that while this choice does not alter the results
regarding the Mg depletion, it is going to underestimate the Al content of the
ejecta, which is practically independent of the initial aluminium whereas it 
is extremely sensitive to the magnesium initially present in the star. 

The data in the Mg-Al and O-Al planes\footnote{[Al/Fe]$_{SG}=1.15\pm 0.11$ with rms=0.5 and [Al/Fe]$_{SG}=0.9\pm 0.07$ with rms=0.35, in the Mg-Al and O-Al planes respectively.} trace two well defined anticorrelation
patterns. Five stars (indicated with full squares in Fig.~\ref{M2}) belong to the FG, 
while the 13 left (asterisks) show evidences of 
proton-capture processing. The chemical composition of the two stars with the 
most extreme chemistry (2M21332216-0048247 and 
2M21332527-0049386), with $\delta [\rm{Mg/Fe}]=-0.15$ and 
$\delta [\rm{O/Fe}] = -0.5$, are compatible with dilution of gas from AGB with 
$\sim 30\%$ of pristine matter; these stars are indicated with asterisks+circles. 
The paucity of data in this case prevents any conclusion regarding a possible 
discreteness in the distributions of the data on the planes.

In the N-Al plane, while FG and asterisk stars show [N/Fe] values more compatible with the correlation pattern traced by the dilution curves, this is not true for two O-poor stars in particular (2M21332216-0048247 and 
2M21332527-0049386). This should indicate possible problems in the determination of N abundances for the most extreme chemistry stars.

\citet{lardo13} claimed evidences of multiple populations in M2, based on the
splitting of the RGB and a well define C-N anticorrelation.
High-resolution spectroscopy of M2 stars was presented by \citet{yong14}, who
outlined the presence of the classic O-Na anticorrelation and of a 
Na-Al correlation trend; the Mg-Al trend was not clear. In the same study 
\citet{yong14} identified a stellar component enriched in iron ($\delta [\rm Fe/H]\sim+0.7$ respect to the main peak, $[\rm Fe/H]=-1.7$), whose
numerical consistency is however below $5\%$; because of such a small percentage,
and taking into account that all the M2 stars in the ME15 sample share the same
iron content ($[\rm Fe/H] \sim -1.5 \pm 0.15$), we may 
safely neglect it in the present analysis. \citet{pancino17} have rencently confirmed the presence of a clear Mg-Al anticorrelation in this cluster, analysing the data from $Gaia$-ESO survey.
This is extremely important for our analysis, because in this metallicity range
the Mg-Al trend is a valuable indicator of the nucleosynthesis experienced by the
material from which SG stars formed. The Mg-Al anticorrelation traced by M2 stars
is nicely reproduced by the AGB gas of the proper metallicity; the same holds for
the O-Al pattern. The 
extent of the Mg and O depletion require mixing of the AGB ejecta with
$\sim 20-30\%$ of pristine gas: this mixing would produce SG stars
with a helium spread up to $Y \sim 0.32$, which is in excellent agreement
with the results found by \citet{milone15} for this cluster.

\begin{figure*}
\begin{minipage}{0.33\textwidth}
\resizebox{1.\hsize}{!}{\includegraphics{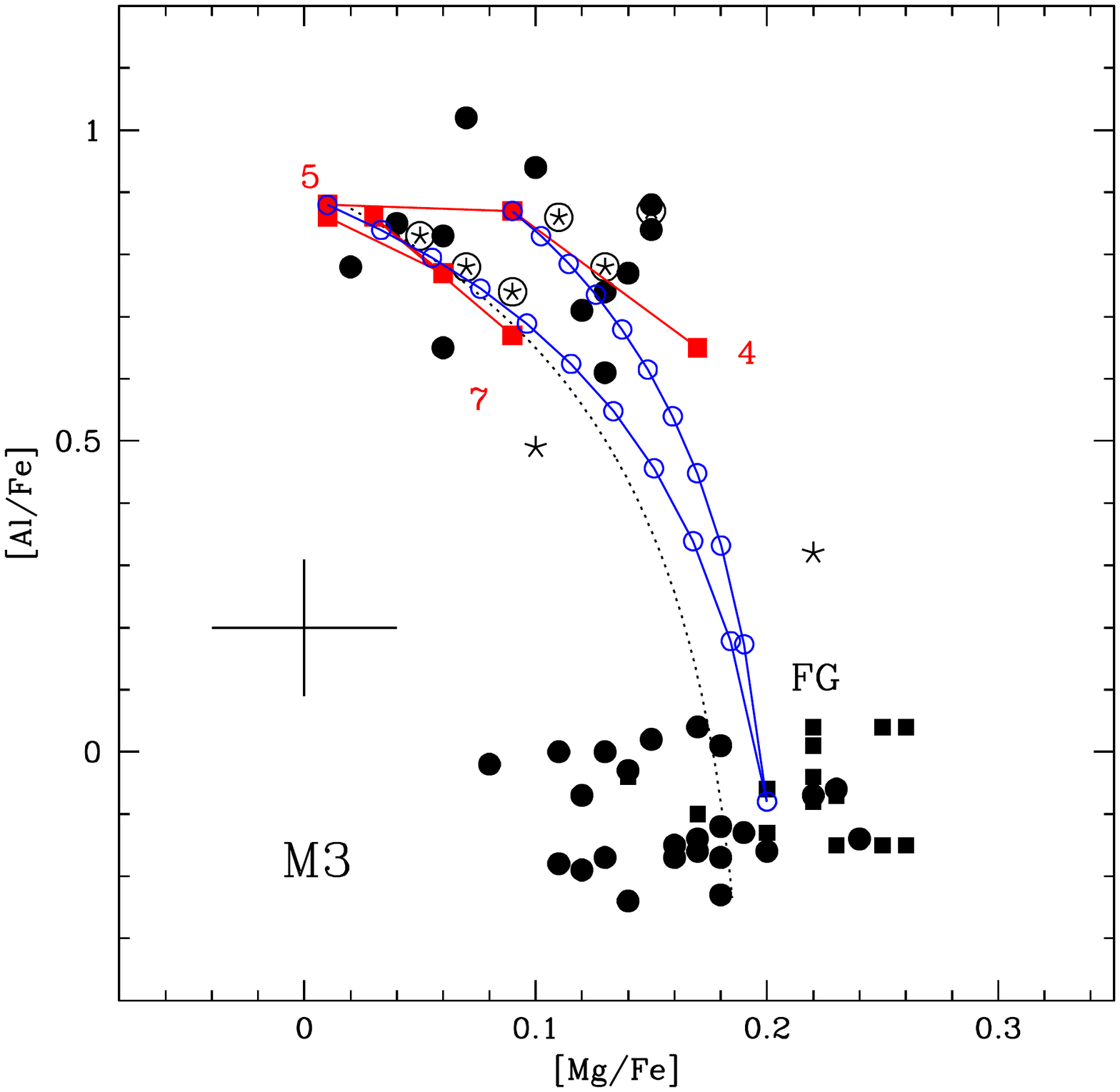}}
\end{minipage}
\begin{minipage}{0.33\textwidth}
\resizebox{1.\hsize}{!}{\includegraphics{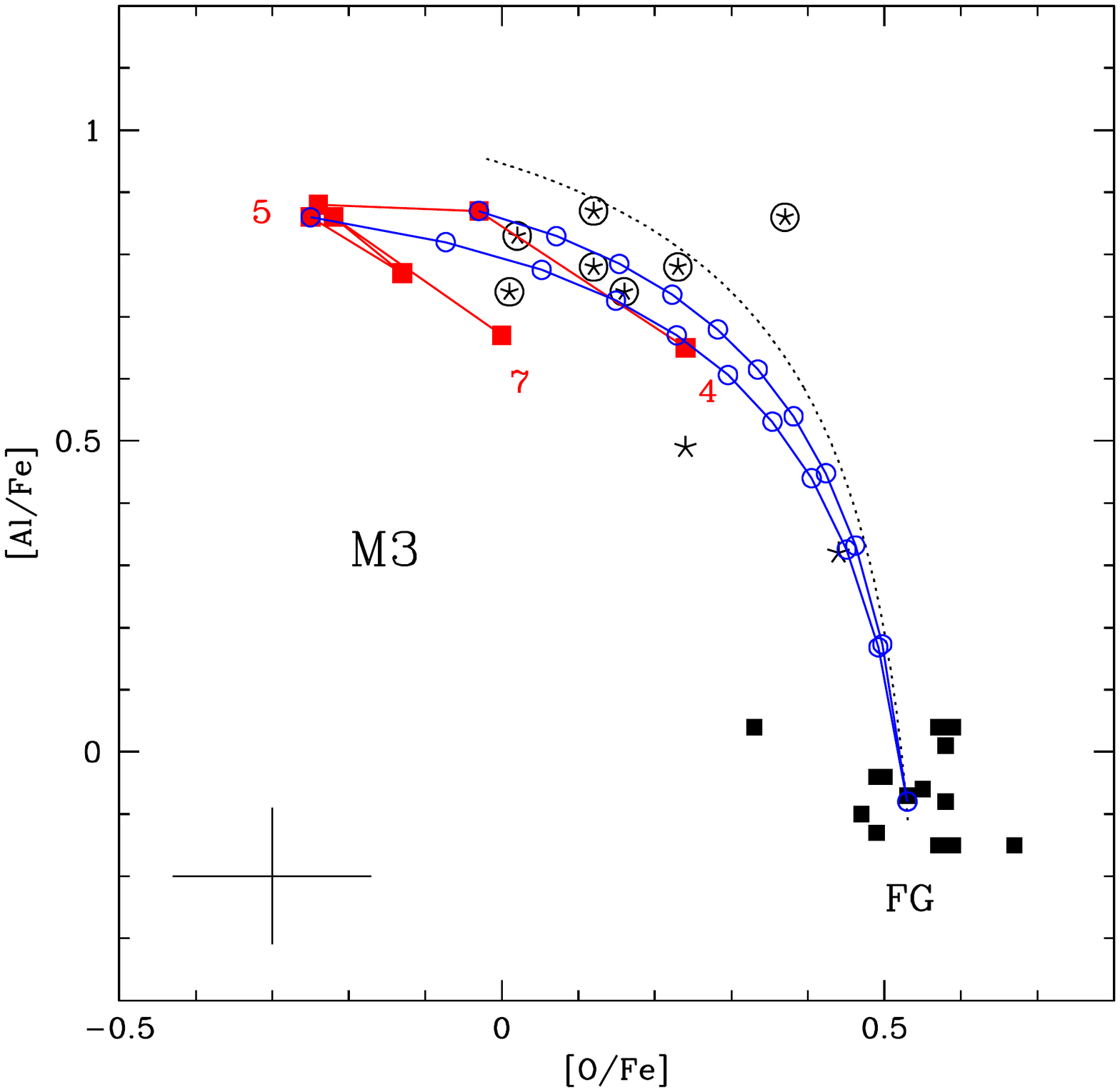}}
\end{minipage}
\begin{minipage}{0.33\textwidth}
\resizebox{1.\hsize}{!}{\includegraphics{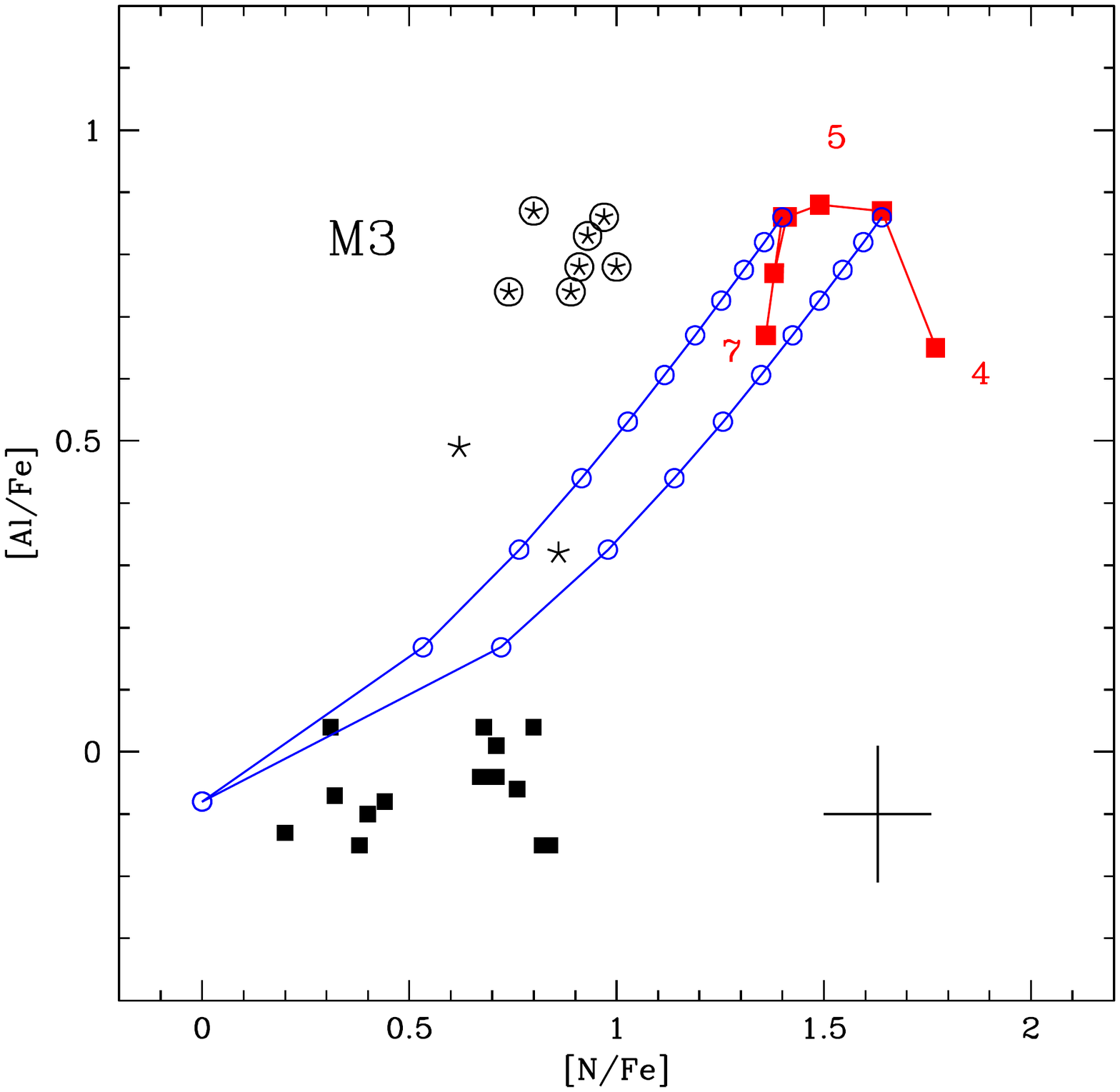}}
\end{minipage}
\vskip-10pt
\caption{The Mg-Al (left), O-Al (center) and N-Al (right) distribution of stars in M3. The observations are compared with AGB models of metallicity Z$ = 10^{-3}$. The meaning of the symbols is the same as in Figure \ref{M53}.}
\label{M3}
\end{figure*}

\subsubsection{M3}
\label{secM3}
ME15 measured the abundances of a large number of stars (59) in M3. The data for M3
stars and the comparison with the AGB models are shown in Fig.~\ref{M3}.

The left panel of Fig.~\ref{M3} shows a clear Mg-Al anticorrelation\footnote{[Al/Fe]$_{SG}=0.87\pm 0.03$ with rms=0.4}, with a
maximum magnesium spread $\delta [\rm{Mg/Fe}] \sim -0.2$ and an overall Al increase of 
$\delta [\rm{Al/Fe}] \sim 1$. This is an indication that the gas from which SG stars formed
was exposed to nuclear activity, with a less advanced nucleosynthesis compared to
the clusters analysed previously, that exhibited traces of a much stronger Mg depletion.
The stars with the most extreme chemistry are reproduced by mixing AGB gas with 
a small percentage (below $20\%$) of pristine matter. 
The distribution of the stars in the Mg-Al plane suggests a gap in the star 
formation history, which prevented star formation from dilution with large fraction 
of pristine gas; confirmation of this result requires more statistics, possibly accompained with a detailed analysis of the distribution of the stars in the CMD.

The ME15 data in the O-Al plane\footnote{[Al/Fe]$_{SG}=0.95\pm 0.04$ with rms=0.34}, shown in the middle panel of Fig.~\ref{M3}, trace
a clear anticorrelated pattern. The AGB models can nicely reproduce the maximum oxygen 
spread observed, $\delta[\rm{O/Fe}] \sim -0.4$ dex: in this case a higher fraction of pristine gas
is required to reproduce the O abundances of the stars with the highest Al, compared to
what we found in the Mg-Al plane. Even in this case we note the clear distinction 
between the FG and the group of stars with $[\rm{Al/Fe}]>+0.7$. 
Only 2 stars, marked with an asterisk, show an intermediate chemistry in the Mg-Al and O-Al planes, in agreement with a higher degree of dilution with the pristine gas. 
In the right panel of Fig. \ref{M3}, N-Al correlation is displayed by the data. FG stars show the lowest [N/Fe] values (in a range compatible with the effect of deep-mixing), while the Al-rich group of stars show a somewhat lower N respect to the expectation.

The presence of stars with a small content of Mg and O and with large abundances of
Al is a clue of the presence of a SG in M3. The Mg-Al anticorrelation in M3 is clearly bimodal, as M4 \citep[e.g.][]{villanova11},
at odds with all the others GCs presented in this work.. The data by ME15 confirm 
star-to-star differences, as in the study by \citet{cohen05b}, who found an O-Mg 
correlation and an O-Na anticorrelation in the chemical composition of giant stars 
in M3. Indications of the presence of stars formed from processed gas were also found
by \citet{johnson05}, who claimed the detection of a Mg-Al anticorrelation, although
the scatter in magnesium was not fully confirmed. On the photometric side, a recent
study by \citet{massari16} outlined the presence of at least two stellar populations
in M3. 

While the presence of SG stars in M3 is definitively confirmed by the investigations 
mentioned above and the ME15 data, the modality with which the SG formed is far from 
being clear. The analysis of the Mg-Al plane suggests that SG stars formed from 
basically pure AGB gas, whereas from the O-Al data we deduce that significant dilution 
is required. We are much more favourable to the second possibility, because the
formation of stars from undiluted AGB ejecta would imply the presence of a helium-rich
population, which is ruled out by the morphology of the HB of this cluster, which 
was shown to be well reproduced by use of a small helium dispersion. A $\sim 50\%$
dilution with pristine gas would imply a maximum helium $Y \sim 0.30$, the same quantity
invoked by \citet{caloi08} to explain the bluest stars in the HB of M3. The presence
of stars with this helium enrichment is at odds with the conclusions by 
\citet{vdb16, valcarce16, denissenkov17}, who claim a very small dispersion 
$\delta \rm Y < 0.02$ to explain the distibution of the stars along the HB.

\begin{figure*}
\begin{minipage}{0.33\textwidth}
\resizebox{1.\hsize}{!}{\includegraphics{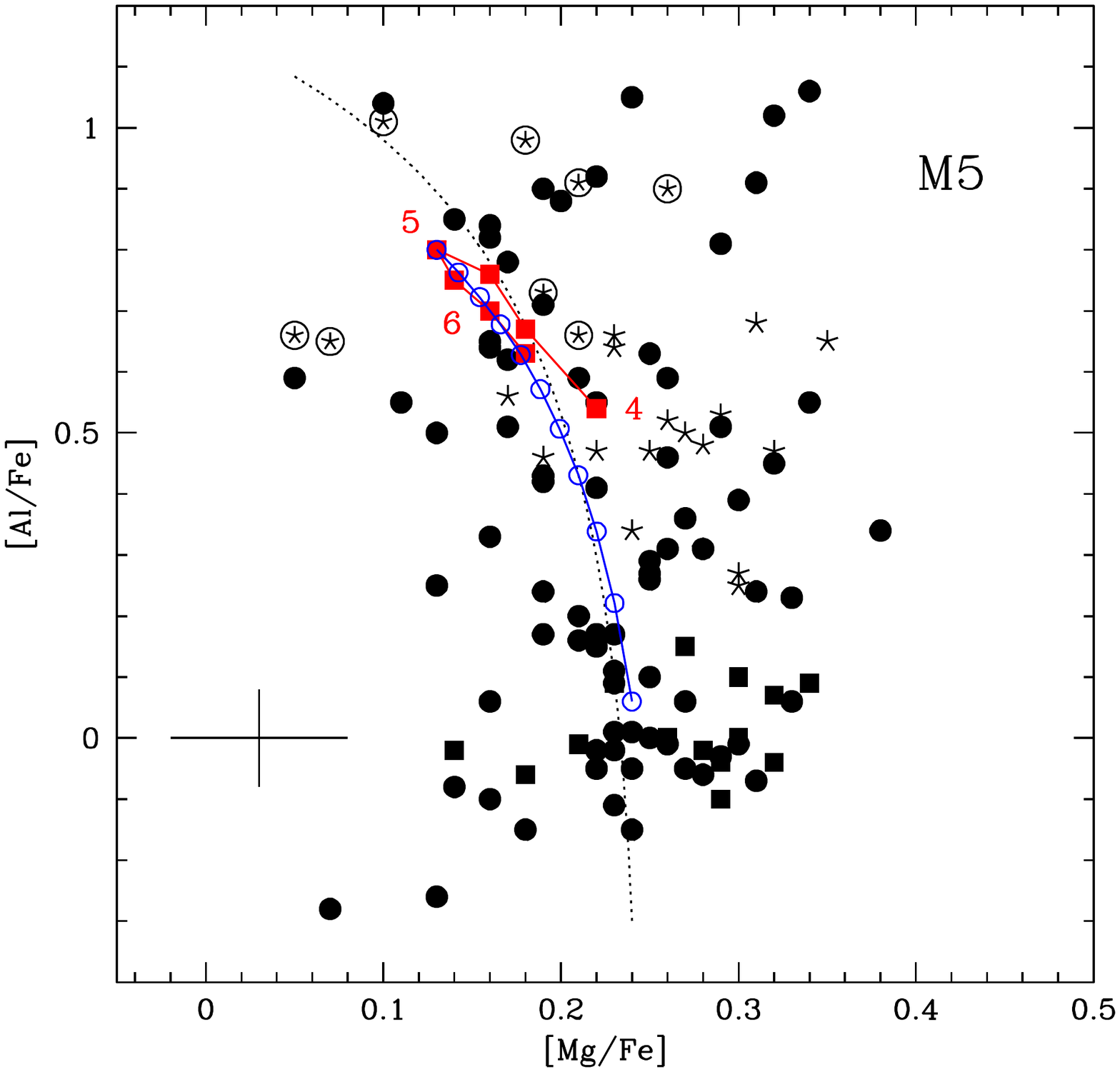}}
\end{minipage}
\begin{minipage}{0.33\textwidth}
\resizebox{1.\hsize}{!}{\includegraphics{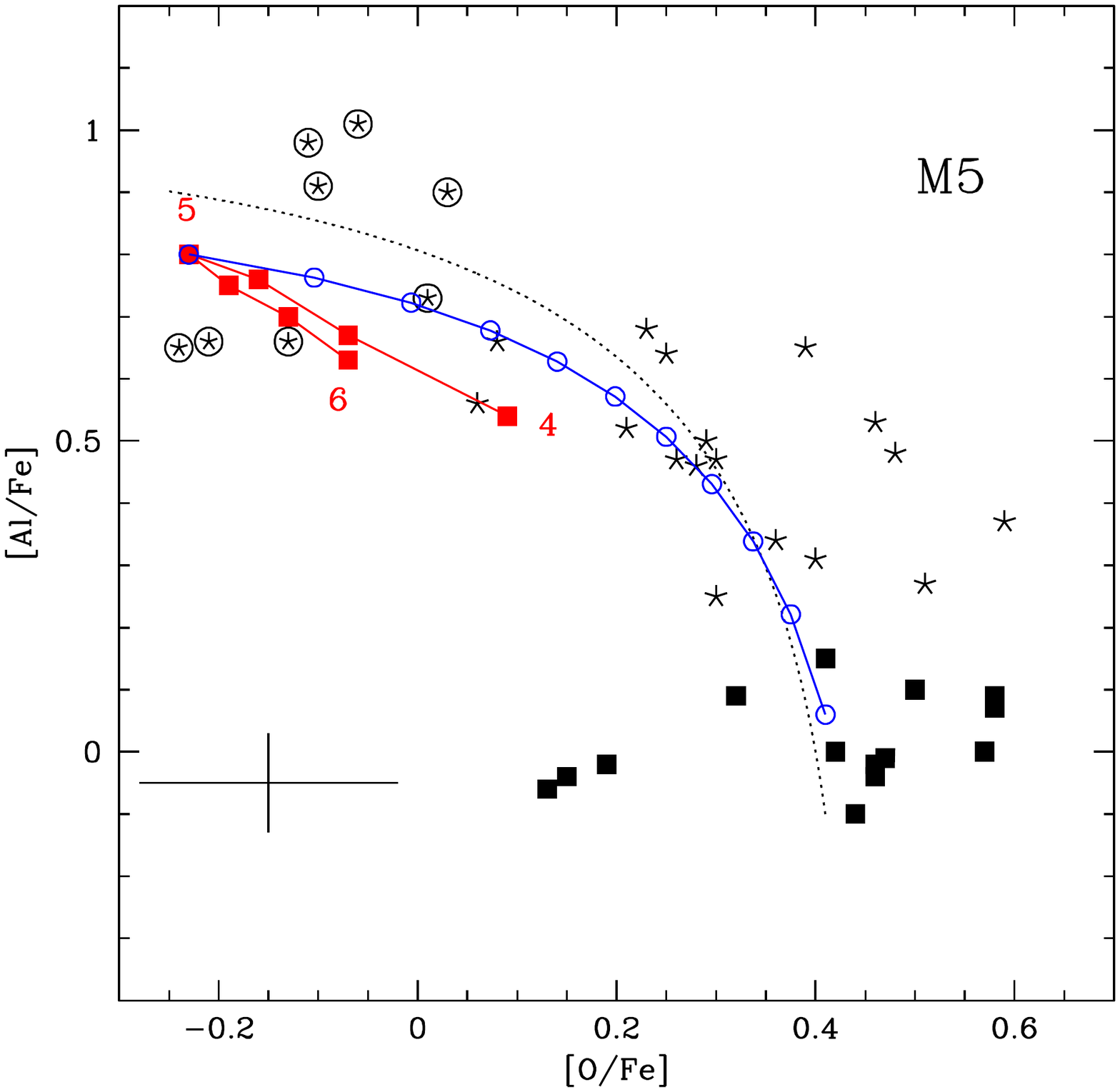}}
\end{minipage}
\begin{minipage}{0.33\textwidth}
\resizebox{1.\hsize}{!}{\includegraphics{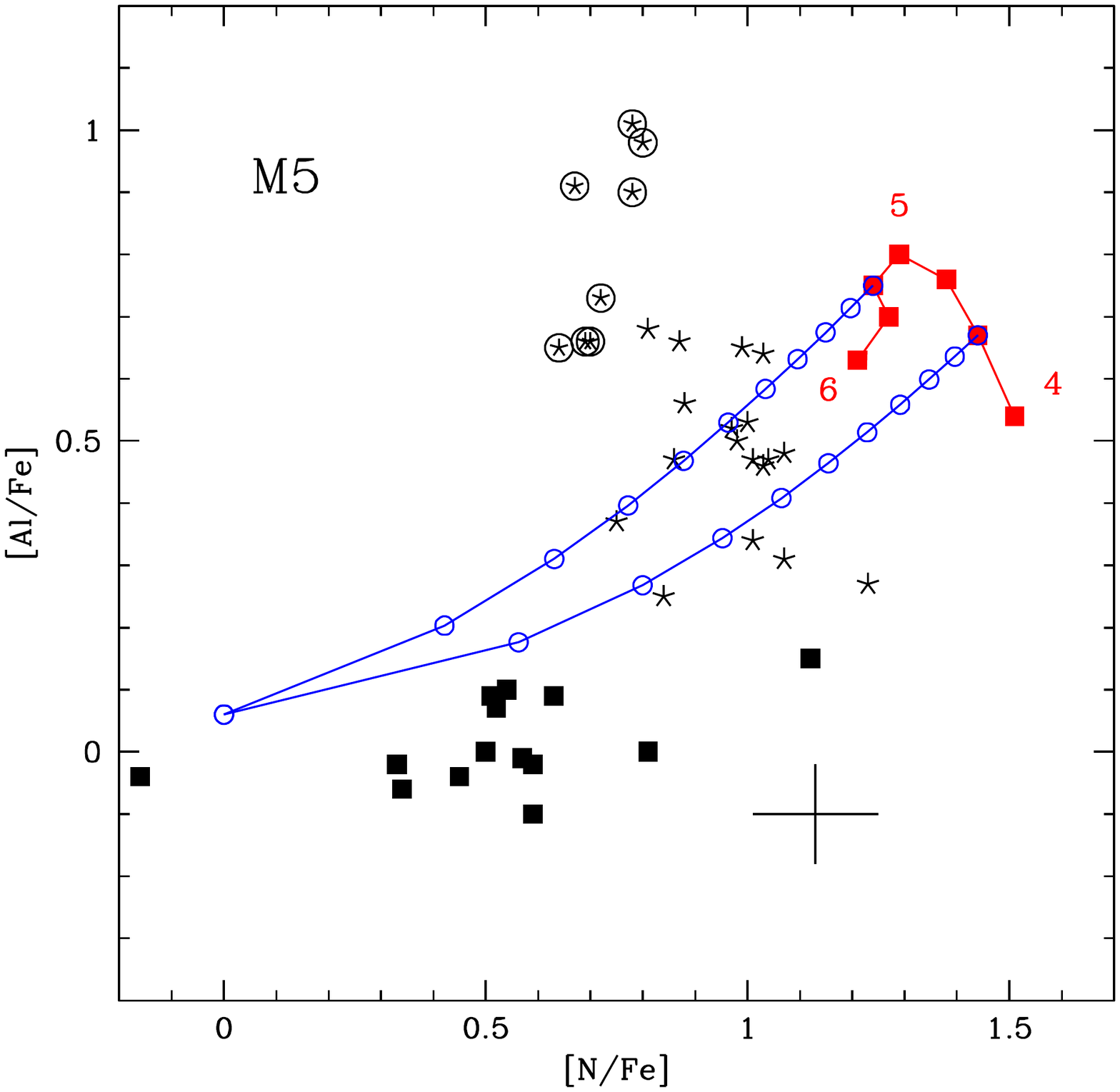}}
\end{minipage}
\vskip-10pt
\caption{The Mg-Al (left), O-Al (center) and N-Al (right) distribution of stars in M5. The observations are compared with AGB models of metallicity Z$ = 1.5\times 10^{-3}$. The meaning of the symbols is the same as in Figure \ref{M53}.}
\label{M5}
\end{figure*}

\subsection{M5: another cluster on the edge?}
The large number of stars observed (122) for M5 allows a solid interpretation of 
the different populations present in this cluster. The results of the comparison
of the distribution of the data points with the chemistry of AGB stars of the
same metallicity (Z$= 1.5\times 10^{-3}$) is shown in the three panels of Fig. \ref{M5}.

The most relevant difference in the comparison with the clusters analyzed
so far is that only a small Mg depletion is observed: ME15 claim that the
magnesium difference between FG and SG stars is below $0.1$ dex for M5.
Conversely, the Al spread is large, with a few stars having $[\rm{Al/Fe}] \sim 1$ \footnote{[Al/Fe]$_{SG}=1.08 \pm 0.04$ with rms=0.5}.
This is what we would expect if the gas from which SG stars formed was expelled by 
AGB stars. Within massive AGB stars with the metallicity of M5 ($[\rm Fe/H]=-1.25$) we find 
that Mg burning and Al production are activated, but the temperatures are not sufficiently 
hot to allow a very advanced nucleosynthesis, with a strong reduction of the surface 
magnesium.
This is the reason why the distribution of the stars in the Mg-Al is almost vertical.
Note that the small depletion of magnesium is not at odds with the large spread
in aluminium: as discussed previously, the initial Mg is much higher than Al,
thus the consumption of an even small amount of Mg is enough to produce
large quantities of Al.

The small spread in magnesium makes the O-Al plane\footnote{[Al/Fe]$_{SG}=0.9\pm 0.03$ with rms=0.38} the most suitable to infer the
presence of multiple population in M5 and, more generally, in higher metallicity GCs.
On this regard, the results from AGB modelling allows a nice fit of the distribution of stars
in the O-Al plane, particularly in the interpretation of the stars with largest Al,
thus with the most extreme chemical composition, which show a reduction of
oxygen $\delta [\rm{O/Fe}]\sim -0.6$ dex. We see in the central panel of Fig.~\ref{M5} 
that the dilution curve well reproduce the O-Al pattern traced by the observations. 
The almost uniform spread of the data across the dilution curve suggests a
continuous process of formation of the SG, with different degrees of dilution of
gas from AGB stars and pristine gas, in agreement with the preliminary findings by \citet{ventura16}.

Taking together, the O-Mg-Al data indicate that M5 stars have a peculiar metallicity:
massive AGB stars sharing this chemical composition experience HBB sufficiently strong
to consume oxygen and to produce aluminium, but the HBB temperatures do not allow
a significant depletion of magnesium. 

Similarly to the previous cases (M13 and M2), even in this cluster the N measurements by ME15 of the Al-rich/O-poor group of stars is once more against our immediate understanding. They are less enriched in nitrogen than the remaining SG stars; we believe that more reliable N measurements are needed to draw any conclusion on this regard.

The above results, particularly the presence of stars with an oxygen
content 4 times smaller than FG stars and aluminium mass fractions 10 times larger,
indicate that M5 harbours SG stars. This is not surprising, as high resolution
spectroscopy outlined an O-Na anticorrelation among RGB stars \citep{carretta09a}
and HB stars \citep{gratton13}; interestingly, the latter study 
shows that the stars with the anomalous chemical composition populate the blue
side of the HB, as expected based on their higher helium. 

The presence of stars with a variety of O abundances (see the middle panel of
Fig.~\ref{M5}), when compared with the chemistry of the AGB ejecta, leads us
to the conclusion that the SG in M5 formed with various degrees of dilution
of AGB gas with pristine matter, and that some stars with almost a pure AGB
chemical composition formed. If part of the SG formed from the AGB ejecta without
dilution, the duration of the process must have been longer than $\sim 30-40$Myr:
a shorter process, as discussed in section \ref{helium}, would imply the presence
of a group of stars extremely enriched in helium, which is ruled out based on
the studies focused on the morphology of the HB of M5, indicating that the
largest helium of SG stars in this cluster is $Y \sim 0.31$ \citep{dantona08}.
This is consistent with the helium in the ejecta of lower mass AGB stars,
which evolve on time scales of the order of $\sim 30-40$Myr.

Concerning the helium content of M5 stars, we find that the spread 
invoked by \citet{lee17} to explain the morphology if the RGB bump is too
small ($\delta \rm Y \sim 0.04$) to allow compatibility with the presence of 
stars with very low oxygen abundances.

\begin{figure*}
\begin{minipage}{0.33\textwidth}
\resizebox{1.\hsize}{!}{\includegraphics{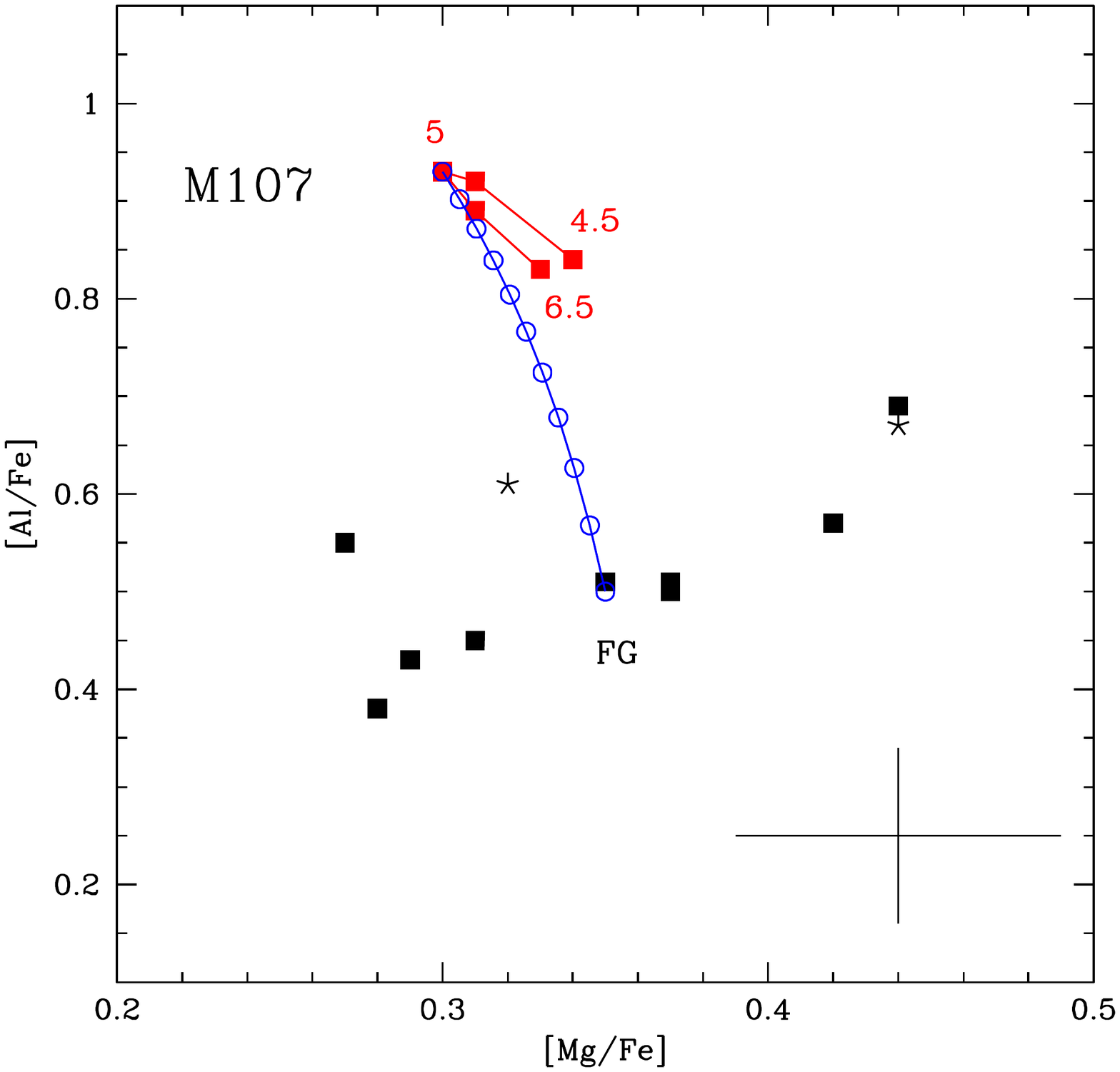}}
\end{minipage}
\begin{minipage}{0.33\textwidth}
\resizebox{1.\hsize}{!}{\includegraphics{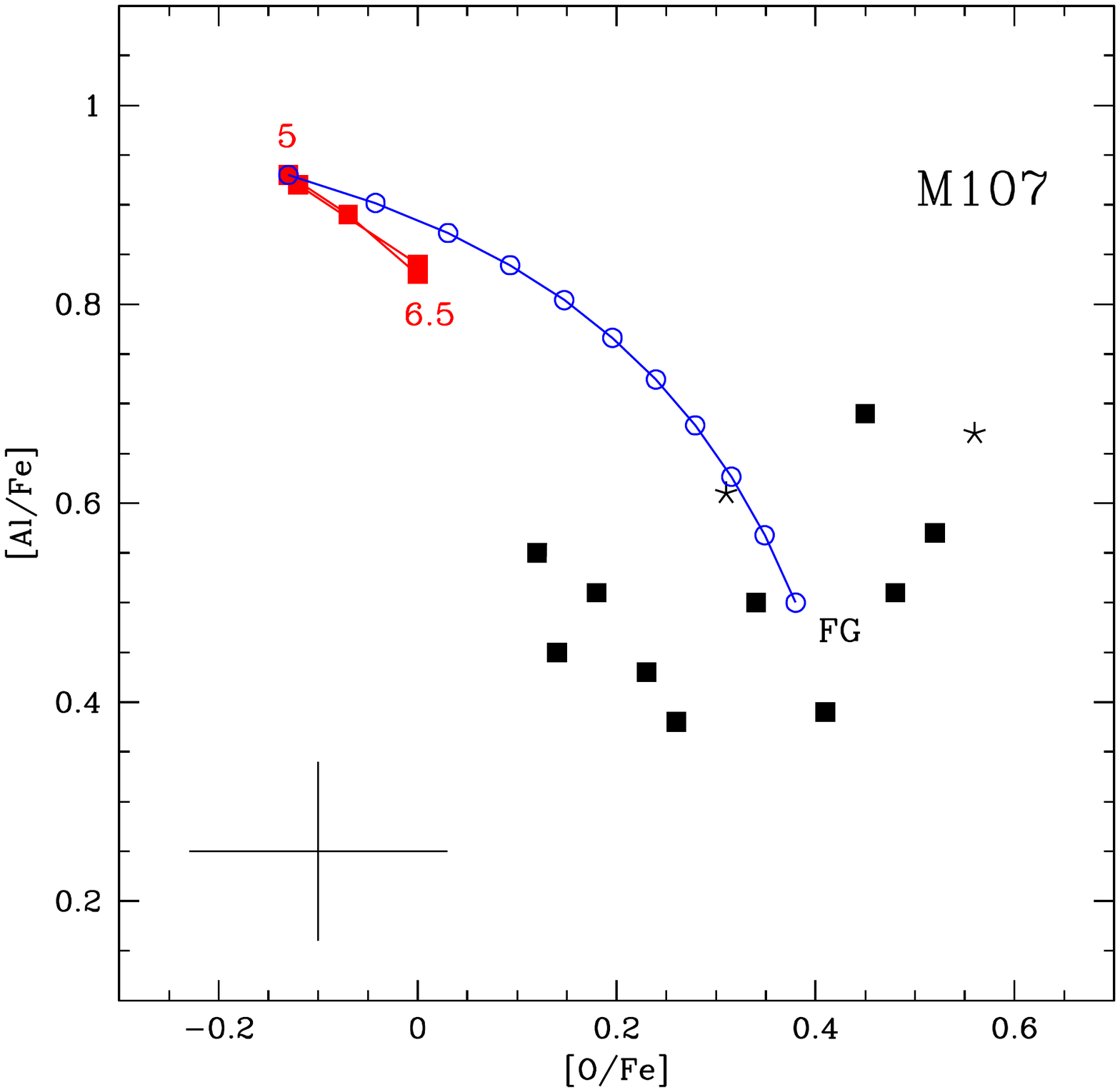}}
\end{minipage}
\begin{minipage}{0.33\textwidth}
\resizebox{1.\hsize}{!}{\includegraphics{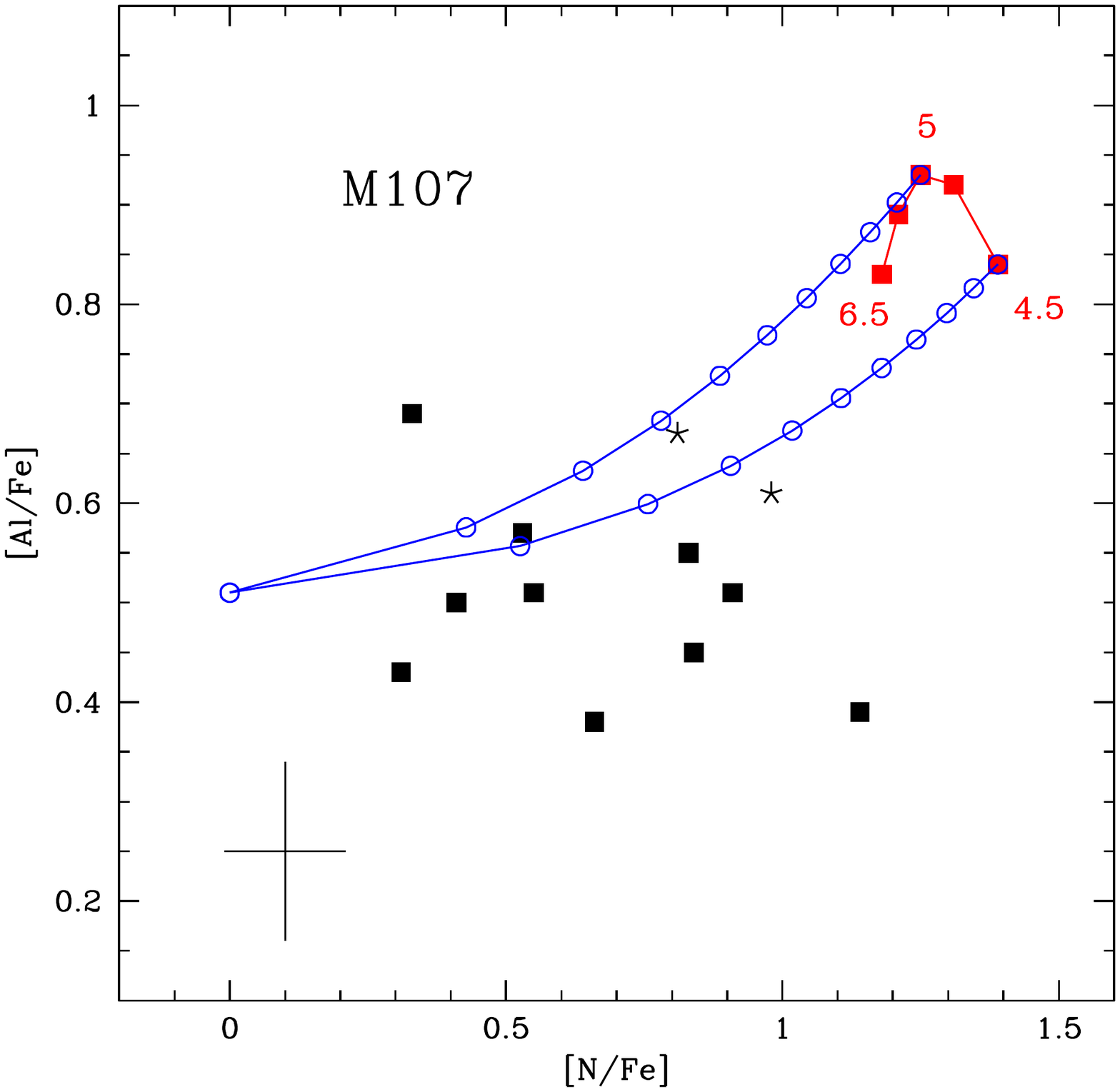}}
\end{minipage}
\vskip-10pt
\caption{The Mg-Al (left), O-Al (center) and N-Al (right) distribution of stars in M107. The observations are compared with AGB models of metallicity Z$ = 2.5\times 10^{-3}$. The meaning of the symbols is the same as in Figure \ref{M53}.}
\label{M107}
\end{figure*}

\subsection{M107: magnesium off the game}
\label{secM107}

This cluster is one of the two with the highest metallicity ($[\rm Fe/H]=-1$,
corresponding to Z$=2.5 \times 10^{-3}$) in the ME15 sample. The entire sample 
is made up of 42 stars; for 12 of them the measurements of CNO are available. 
We prefer to concentrate on the latter, smaller subsample for the analysis of 
this cluster, because ME15 found a trend of the Mg abundances with the 
effective temperature, which required some ad hoc corrections.
We compare the data with the massive AGB yields of the proper metallicity in
the 3 panels of Fig.~\ref{M107}.

No Mg-Al anticorrelation is found among the observed values (see left panel
of Fig.~\ref{M107}), in agreement with AGB models, according to which 
at these metallicities the HBB temperatures are too cool to allow 
any significant depletion of magnesium (see Fig.~\ref{Tbce}).

The Al-spread, if present, is very small: this is the first case within the
present work, where it is difficult to distinguish the different populations in 
the Mg-Al plane. Note that the AGB yields, run with the Al of the stars in the
FG of the cluster, are enriched in aluminum with respect to the FG, with 
$\delta [\rm{Al/Fe}] \sim +0.4$ dex; this is shown in the left panel of Fig.~\ref{M107}.

For the metallicity of M107 stars, based on the arguments
given in the previous section, we find that the depletion of oxygen is a much
more valuable indicator of the degree of contamination of SG stars, compared
to the Mg spread: as shown in the middle panel of Fig.~\ref{M107}, the AGB
yields are expected to show the imprinting of O-depletion, with 
$\delta [\rm{O/Fe}] \sim -0.5$ dex and, more generally, a O-Al trend, which is indeed
not observed. On the contrary, the data points cluster around the chemistry
of the FG of the cluster, a strong hint that either M107 is a FG-only 
cluster or it hosts SG stars formed with a large degree of dilution of AGB
gas with pristine matter.

Note that the present conclusions are different from what deduced by
\citet{ventura16}, that SG stars definitively exists in M107: that result
was based on the assumption that no Al-enhancement would be expected in
metal-rich clusters if the initial Al is higher than solar-scaled,
whereas the present results, self-consistently based on AGB models with the
same initial Al of FG stars, partially disregard that assumption.

The data shown in the N-Al plane (see right panel of Fig.~\ref{M107})
seems to confirm the conclusion drawn based on the distribution of the stars in the
O-Al plane. Any SG star should enriched 
both in N and Al, according to the dilution curve shown in the figure. On the 
contrary, we note that the data points do not show any significant spread in Al,
whereas the N-enrichment ranges from $\delta [\rm{N/Fe}] \sim +0.3$ to 
$\delta [\rm{N/Fe}] \sim +1$: the only possible explanation is that either the N-rich star
are FG stars, which were subject to deep mixing during the RGB phase, or
that they belong to the SG of the cluster, formed from AGB gas largely diluted with
pristine matter. The latter explanation likely holds for the two stars showing a
Al-excess, indicated with asterisks in the figure. 

In the case of M107 no deep analysis of the HB is present in the literature,
thus the conclusions given above cannot be tested against any result from photometry.
While the lack of any Mg and probably Al spread is in agreement with the 
expectations if AGB stars were the polluters, the lack of SG stars with the chemical
composition of pure AGB ejecta prevents a solid evaluation of the reliability of
the AGB stars of this metallicity.

\begin{figure*}
\begin{minipage}{0.38\textwidth}
\resizebox{1.\hsize}{!}{\includegraphics{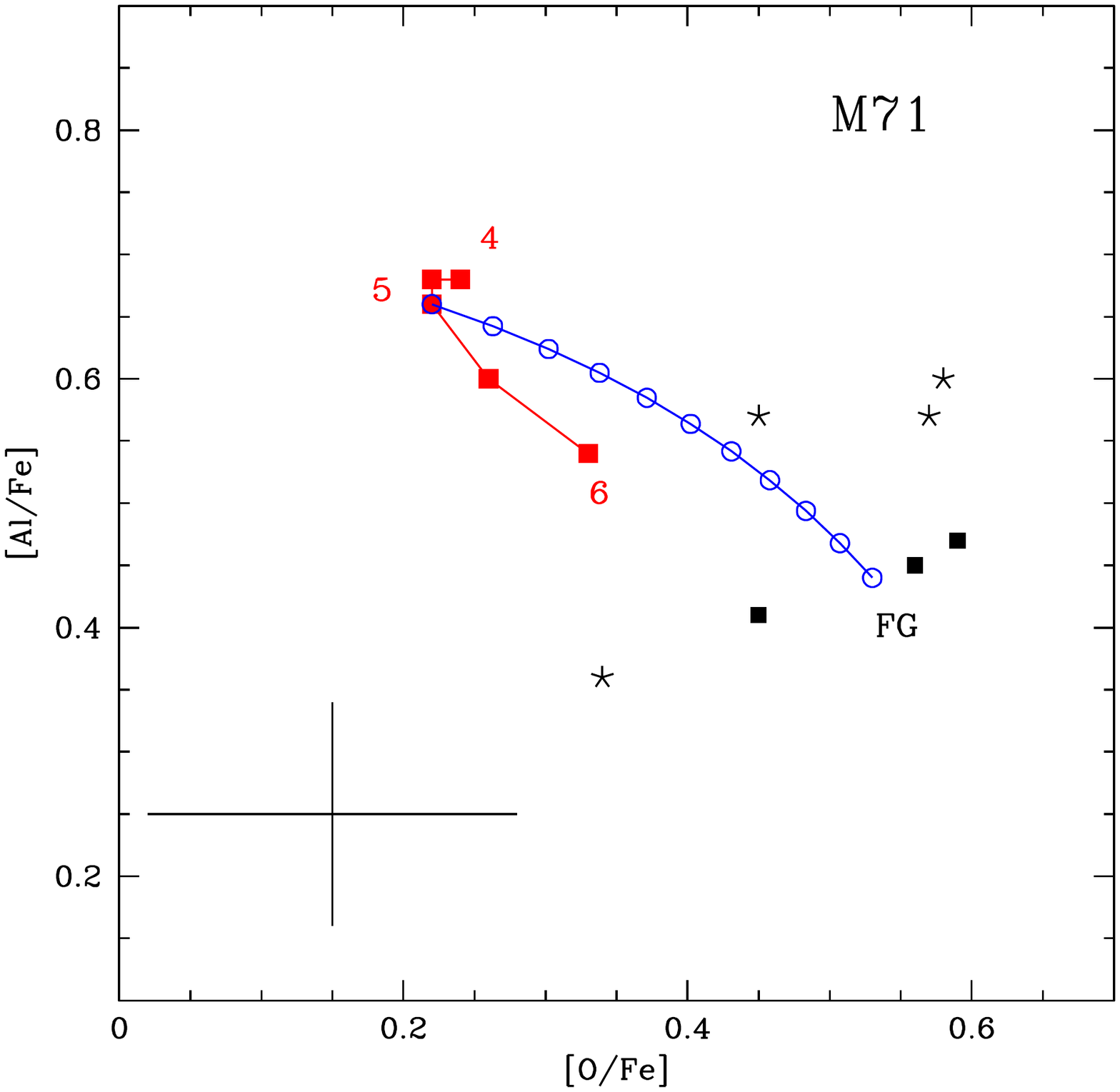}}
\end{minipage}
\begin{minipage}{0.38\textwidth}
\resizebox{1.\hsize}{!}{\includegraphics{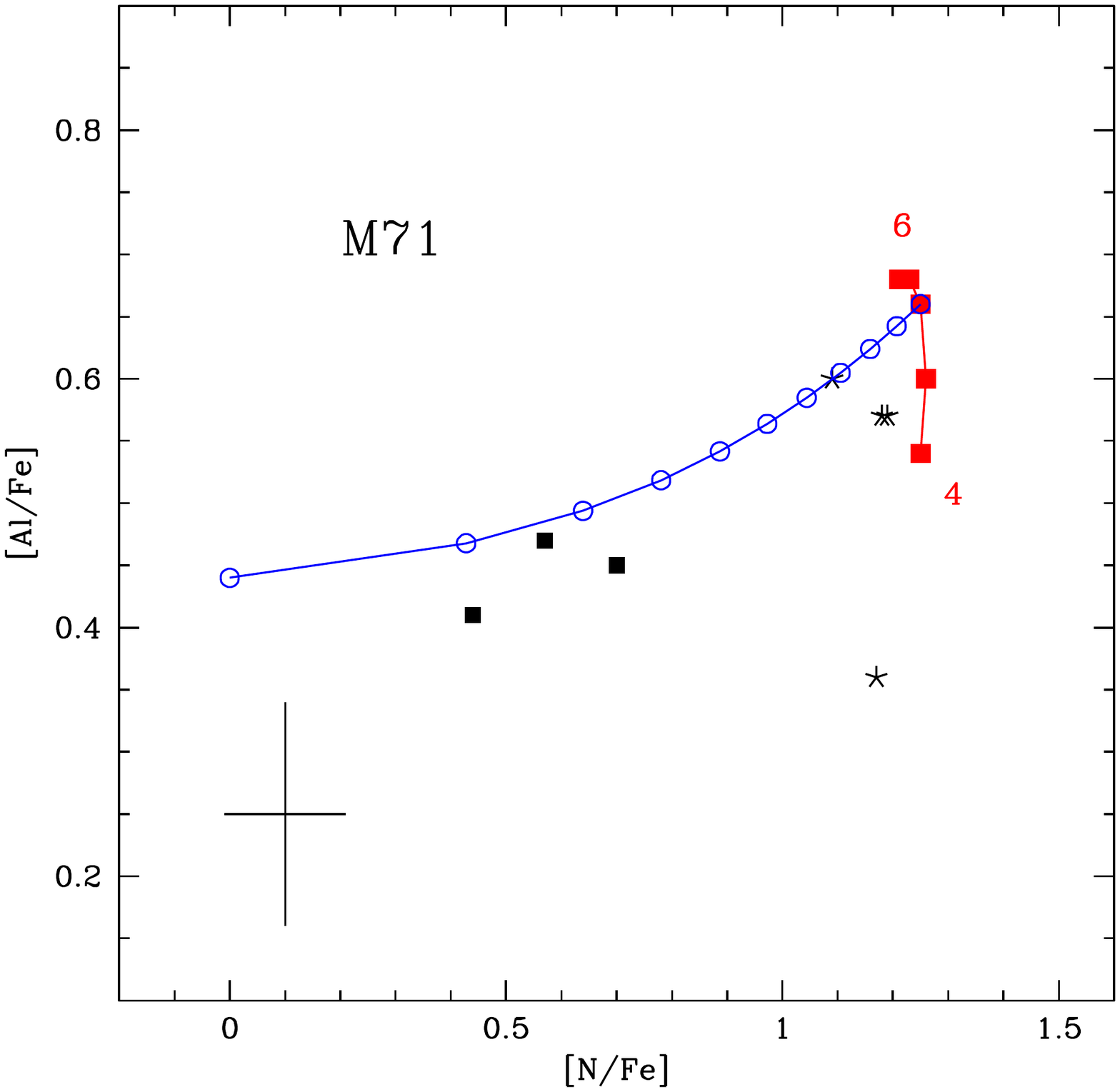}}
\end{minipage}
\vskip-10pt
\caption{The O-Al (left) and C-N (right) distribution of stars in M71. The observations are compared with AGB models of metallicity Z$ = 6\times 10^{-3}$. The meaning of the symbols is the same as in Figure \ref{M53}.}
\label{M71}
\end{figure*}

\subsection{M71}
\label{secM71}

The metallicity measured for M71 is $[\rm Fe/H]\sim-0.7$. This is the highest 
metallicity cluster in the sample we have considered. Theoretically, this represents 
a very useful case, to understand how self-enrichment works in high metallicity GCs,
and to confirm further how the chemistry of self-enrichment changes with the
metallicity of the cluster. Unfortunately only for 7 stars we have the determination of 
all the chemical abundances, including the CNO elements. This partly prevents a
solid understanding of how the SG formed in this cluster.

The data available show no sign of Mg-dispersion, in agreement with 
\citet{carretta09b} and with the expectations: this cluster is even more metal-rich than 
M107, which also shows no Mg dispersion.

We show the data points of M71 stars in Fig.~\ref{M71}; the yields from AGB stars of
the same metallicity (Z$=6\times 10^{-3}$) are also shown. The analysis of a possible pollution from AGB
stars is not straightforward in this case, because the expected Al enhancement
($\delta [\rm{Al/Fe}] \sim +0.2$) and oxygen depletion ($\delta [\rm{O/Fe}] \sim -0.3$ dex) are
only slightly higher than the errors associated to the data for the two elements.

The simultaneous analysis of the O-Al and N-Al planes allows us to better identify stars that show 
the imprinting of AGB pollution with a partial dilution with the pristine gas. This is the case 
of the stars marked with the asterisk.
In the O-Al plane we clearly identify 3 FG stars, indicated with solid squares.
The 4 stars left, indicated with asterisks, show-up an anomalous chemical composition, 
which might suggest that they belong to the SG of the cluster. One of the stars
(2M19535064+1849075) show an enhancement in aluminium and a depletion of 
oxygen compatible with AGB gas $50\%$ diluted with pristine matter. For the 2 stars
with slightly enriched in aluminium (2M19535764+1847570 and 2M19533986+1843530), the possibility that the belong to the FG
holds only if their oxygen is overestimated: alternatively, they are FG stars, with
the Al content overestimated by $\sim 0.1$ dex. While the error bars are sufficiently
large to make both cases possible, we are more confident about the first possibility,
because of the simultaneous Al and N enrichment of these stars (see right panel
Fig.~\ref{M71}).

In the N-Al plane these 3 stars show both the enhancement in nitrogen ($[\rm{N/Fe}]\sim 1.1$) and aluminium $([\rm{Al/Fe}]\sim0.5$), which correspond to
$\sim$50\% of pollution from AGB stars. One of the three stars discussed (2M19535064+1849075) is also depleted 
in oxygen ($[\rm{O/Fe}]\sim0.45$) in nice agreement with the dilution curve, while the other 
two stars show initial abundances of oxygen ($[\rm{O/Fe}]\sim0.6$) likely
overestimeated (this is not an issue though, if we take into account the errorbars).
On the contrary, one star (2M19534827+1848021) shows the evidence of oxygen depletion and nitrogen enhancement, 
while aluminium appear to be almost not touched ($[\rm{Al/Fe}]\sim0.35$).

The possibility that SG stars in M71 formed from the AGB ejecta was first invoked
by \citet{boesgaard05} to explain the abundances of $\alpha-$elements and neutron capture
elements. In a more recent study \citet{alves08} detected CN-Na and Al-Na correlations
among RGB stars below the bump, a clue of the presence of SG stars. The present data,
within the context of the self-enrichment by AGB stars, confirm the presence of SG stars
and indicate that some dilution is required to model the most extreme chemical compositions.
On the other hand, the robustness of this conclusion is partly affected by the small
O and Al spread expected in metal-rich environments, which are unfortunately comparable
with the extent of the error bars. The observations of a much higher number of stars is
required to make this conclusion more reliable on the quantitative side.

\begin{figure*}
\begin{minipage}{0.48\textwidth}
\resizebox{1.\hsize}{!}{\includegraphics{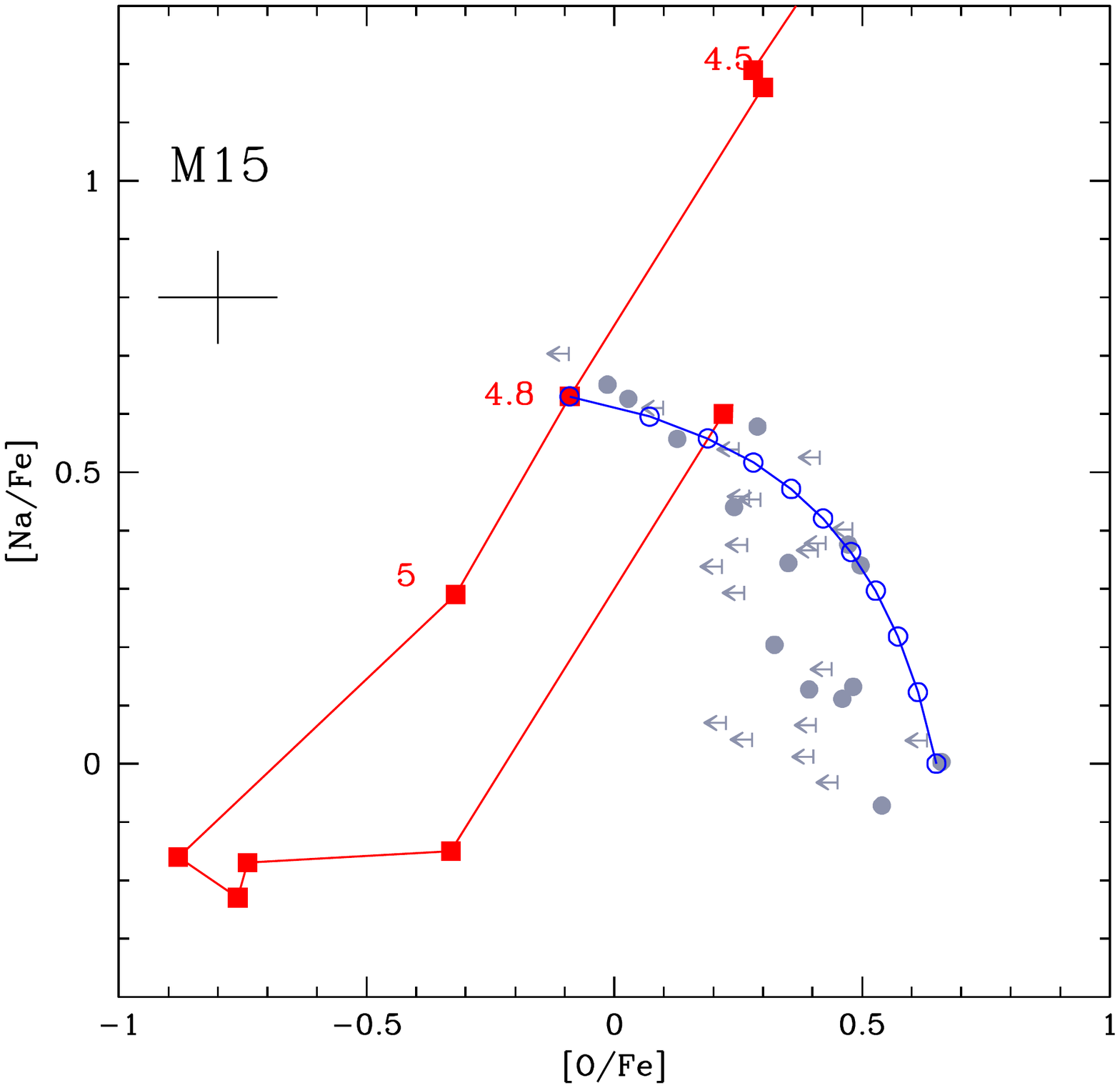}}
\end{minipage}
\begin{minipage}{0.48\textwidth}
\resizebox{1.\hsize}{!}{\includegraphics{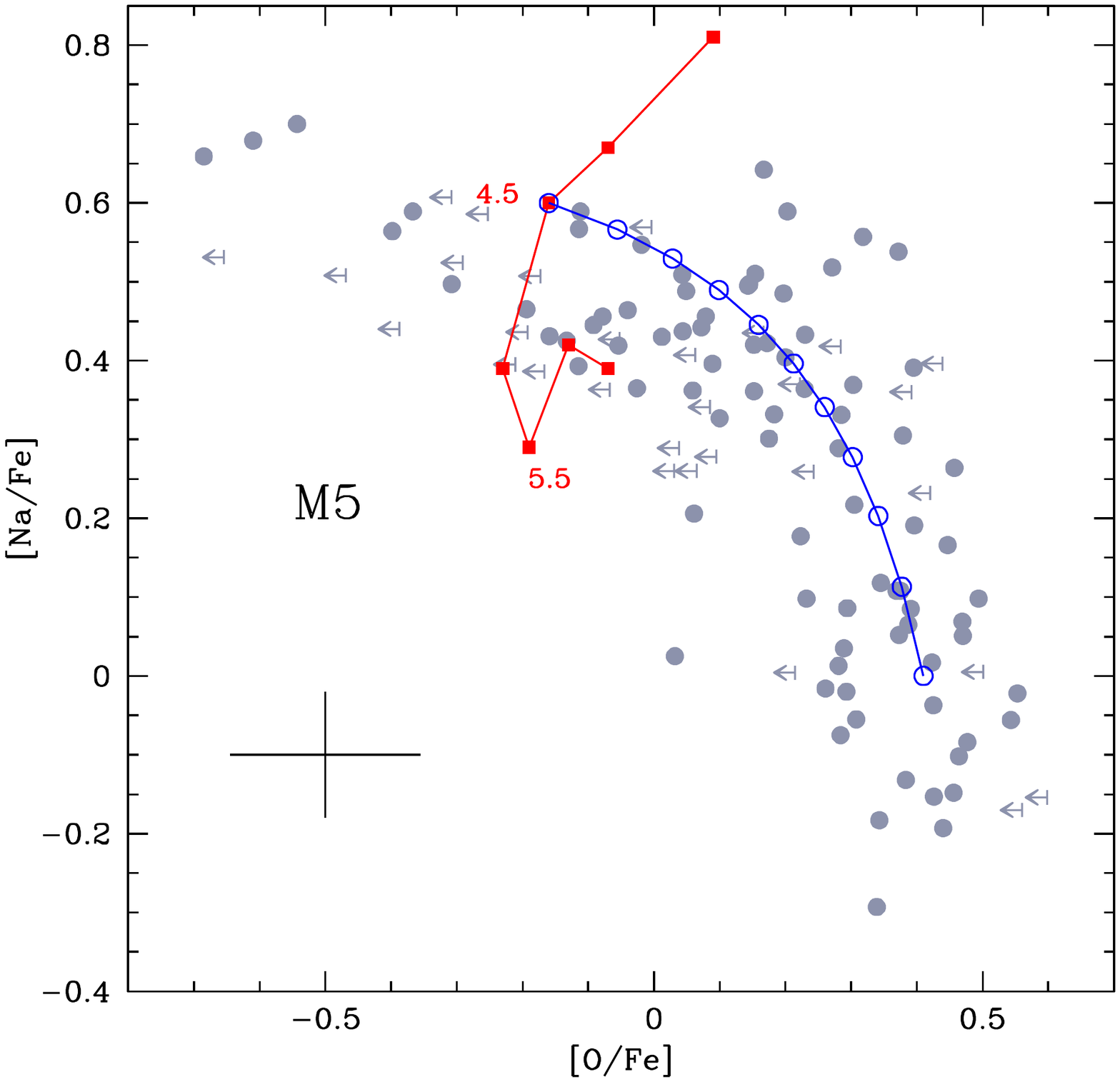}}
\end{minipage}
\vskip-70pt
\begin{minipage}{0.48\textwidth}
\resizebox{1.\hsize}{!}{\includegraphics{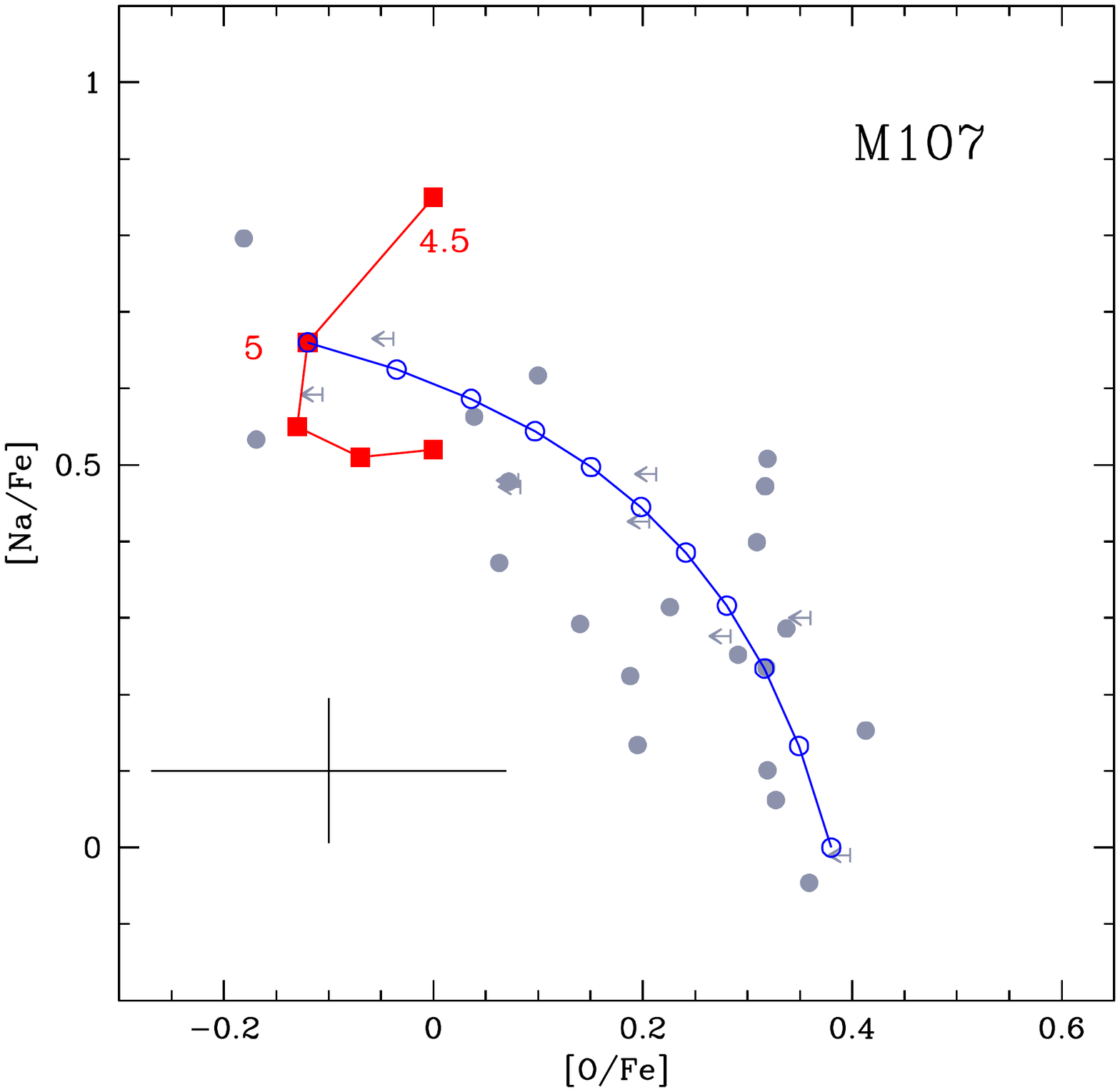}}
\end{minipage}
\begin{minipage}{0.48\textwidth}
\resizebox{1.\hsize}{!}{\includegraphics{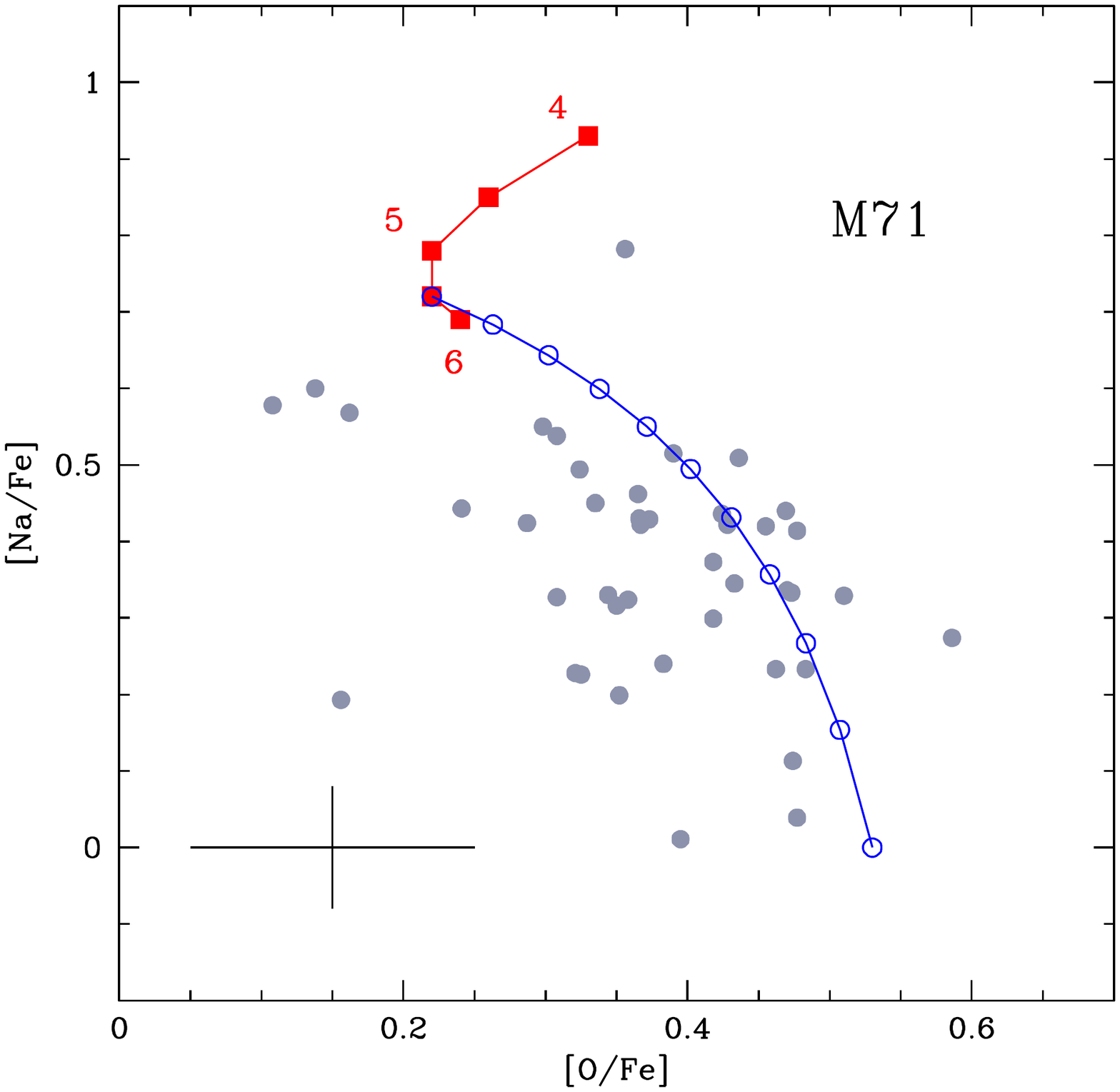}}
\end{minipage}
\vskip-50pt
\caption{Dataset from \citet{carretta09a, carretta09b} in the O-Na plane are shown in gray for M15 (top-left), M5 (top-right), M107 (bottom-left) and M71 (bottom-right). Upper limits in O abundances are shown as
arrows, detections are indicated as circles. AGB yields (red squares) and their corresponding dilution curves (blue open circles) are also shown according to the metallicity of each cluster.}
\label{carretta}
\end{figure*}

\section{The "classic" O-Na anticorrelation in Globular Cluster stars}
\label{ONacarretta}

Before entering the discussion of the implications of the results presented so far, we
want to further test the hypothesis that pollution from massive AGB stars provided the
contaminating material from which SG stars formed in GCs, by comparing the yields of
these stars with the O-Na pattern traced by GC stars. 

Unfortunately, the ME15 APOGEE abundances dataset do not
include the Na measurements because the Na spectral lines present in the H-band
are too weak (even for the most metal-rich GCs like M107 and M71; see ME15 for
more details). Thus, we are forced to consider complementary optical data (see
below) that are not fully homogeneous with those used so far.
On the other hand, while the main conclusions will be drawn based on APOGEE data solely,
we believe important to present this further test, because the O-Na anticorrelation
has traditionally being considered as the 
key signature of the GCs chemical patterns, as it has been observed
in practically all GCs so far studied in the literature \citep[e.g.][]{carretta06, gratton12}.

Understanding the behaviour of Na in massive AGB stars is not trivial, because
the surface abundance of this species stems from the equilibrium between the production
and the destruction channels. The former is mainly determined by $^{22}$Ne proton capture
activity, whereas the latter is made up of two proton capture reactions by Na nuclei,
the dominant channel being the Na$(p,\alpha)^{20}$Ne reaction \citep{hale02, hale04}.

The synthesis of Na is favoured until temperatures $T \sim 100-110$MK, whereas the
destruction prevails in hotter environments. In low-metallicity massive AGB stars, Na
increases during the initial AGB evolution, owing to $^{22}$Ne burning, then decreases
in the later phases, when the destruction channel becomes dominating \citep{ventura06}.

What renders tricky the interpretation of the results is that there is not clear trend
of the behaviour of Na with mass and metallicity: while AGB yields generally produce
Na-rich gas, the production factor with respect to the Na initially present in
the star depends on the duration of the phase during which the temperature of the base
of the envelope exceeded the threshold given above: the longer was this phase, the
smaller will be the average Na in the gas lost.

If we consider the clusters on which this study is based, the compilations by 
\citet{carretta09a, carretta09b} include data for M15, M5, M107 and M71. Fig~\ref{carretta} shows the data from \citet{carretta09a, carretta09b}\footnote{For stars in common between \citet{carretta09a} and \citet{carretta09b}, we show the results from the latter where the [O/Fe] and [Na/Fe] are obtained with a higher level of accuracy.}, overimposed
to the yields of AGB stars of the appropriate metallicity for each cluster. An offset among the [O/Fe] abundances is
found between the two samples (APOGEE vs. Carretta et al.). Because of the
different methods used in the chemical analysis (e.g., EWs vs spectral
synthesis, estimation of the stellar parameters, model atmospheres, spectral
range, solar reference abundances, etc.), this is somewhat expected. Besides
that, the APOGEE O abundance is obtained  from the near-infrared OH molecular
lines that are likely more affected by 3D/NLTE effects than the optical [OI] 630
nm line; see, e.g., \citet{dobrov15} who found that O-abundances from OH
lines could be $\sim$0.2 to 0.3 larger than the optical [OI] line in metal-poor
([Fe/H] $<$ -2.5) stars. This should be kept in mind when comparing [O/Fe] from
optical works, as \citet{carretta09a, carretta09b}, with APOGEE.

We start from M107 and M71, the two highest metallicity clusters in this group.
The chemistry of the yields of stars of different mass are pretty similar in these
cases \citep{ventura13}, because the metallicity is sufficiently large that the 
destruction channel of Na is never fully activated during the AGB phase, not even 
in the most massive stars. This makes the comparison between the data and the yields
very simple, as also the trend of the dilution curve, giving the results expected
when the gas from AGB stars is mixed with pristine matter in the cluster.

For both M107 and M71 we find that the trends defined by the data from \citet{carretta09a,
carretta09b} on the O-Na plane are nicely reproduced by the yields of AGB stars. In
particular, the latter can account for the observed spreads in oxygen
($\delta [\rm O/Fe] \sim -0.55$ and $\delta [\rm O/Fe] \sim -0.35$, respectively, for M107 and
M71) and Na ($\delta [\rm Na/Fe] \sim +0.7$ in both clusters).

The Figure \ref{carretta} (as compared to the findings based on APOGEE data only) allows a deeper analysis of the
possible presence of multiple populations in these clusters. In Sections \ref{secM107}
(M107) and \ref{secM71} (M71) we found that the few stars observed by APOGEE belong to
the FG, while from the O-Na data (with a much better statistics; see Fig. \ref{carretta}) we
can conclude that both clusters harbour a signicant fraction of SG stars; part
of which (those with the lowest O and highest Na) formed from gas with a small
degree of dilution with pristine matter.

Turning to M5, the O-Na data, shown in the top, right panel of Fig.~\ref{carretta}, trace a 
well defined anticorrelation pattern, the spreads in oxygen and Na being, respectively
$\delta [\rm O/Fe] \sim -1$ and $\delta [\rm Na/Fe] \sim +0.6$. The dilution pattern between the
yields of AGB models of the same metallicity as M5 stars and pristine matter can
account for most of the observed stars, with the exception of a few objects exhibiting
an oxygen depletion of a factor $\sim 10$ with respect to the FG, a factor of 2 in excess 
of the theoretical expectations. We propose that the stars with the unusually low 
oxygen content are SG stars of the cluster, which underwent deep mixing during the RGB
ascending, owing to the lower entropy barrier left behind by the first dredge-up, as
a consequence of the higher initial helium. As shown by \cite{dantona07}, such
a deep mixing in SG stars would provoke a significant decrease in the surface oxygen,
leaving the Na content practically unchanged.

The explanation of the distribution of stars in the O-Na plane is consistent with
the discussion on the APOGEE data of M5 stars, addressed in section 3.4, that this cluster
harbours SG stars formed with a variety of dilution factors of AGB gas with pristine
gas, including also stars with a chemical composition very similar to the AGB ejecta.

The interpretation of the O-Na observations of M15 stars is more cumbersome, because
the Na yields at these low metallicities are extremely sensitive to the mass of
the star (see the bottom, left panel of Fig.~\ref{carretta}). This is related to the above 
discussed competition between the production
and destruction channels of Na, because the higher mass AGB stars of this metallicity
reach at the base of the envelope temperatures sufficiently large (see Fig.~1) that the
destruction channel prevails for a significant part of the AGB evolution, thus
producing Na-poor ejecta. The trend of the Na yields with mass is further
complicated by the strong winds suffered by the most massive AGB stars, which favour the
loss of the external mantle before a significant destruction of the Na 
accumulated in the surface regions during the early AGB evolution may occur. 

As shown in Fig.~\ref{carretta}, the Na yields span an interval in excess of 1 dex, 
ranging from $\sim 6~M_{\odot}$ stars, for which $\delta [\rm Na/Fe] \sim -0.2$, to
$\sim 4.5~M_{\odot}$ stars, for which $\delta [\rm Na/Fe] \sim +1$. The interpretation of
the observations in the context of the self-enrichment by AGB stars in this case
would require the details of the timing of the formation of SG stars, i.e. when
the formation of SG started and when the gas from AGB began to mix with pristine matter.
This effort is clearly beyond the scope and the possibilities of the present
analysis.

\section{Discussion}

The sample by ME15 offers a valuable opportunity to understand the
formation of multiple population in GCs, given the wide
range of metallicities, covering the interval $-2.4 < [\rm Fe/H] < -0.7$.
The simultaneous knowledge of the abundances of various chemical elements 
allows a full, though more complex analysis, not limited to the 
interpretation of a single observational plane.

Moving across the different clusters, we find that the relative importance
of the data on a given chemical species depends on the metallicity.
The silicon spread is a key indicator of what was going on during the
formation of metal poor GCs, the Mg-Al trend was crucial to reconstruct the
star formation history in intermediate-metallicity clusters, whereas the
analysis of more metal-rich GCs relies essentially on the extent of the O-Al 
anticorrelation. 

Nitrogen deserves a separate discussion. During the RGB ascending the surface N of
stars undergoes significant variations, owing to convective mixing of the envelope
with stellar regions exposed to nuclear activity. This prevents a straightforward
use of the N data to deduce the initial chemical composition of the star.
However, the ejecta of metal-poor AGB stars
are expected to be largely enriched in nitrogen, up to a factor $\sim 100$
or more, compared to the initial abundance. Giants formed with so large 
quantities of nitrogen are not expected to undergo any change in the surface
N, because any deep mixing would reach internal, CNO processed material,
where the equilibrium N is not overbundant with respect to the surface content.
Therefore, reliable measurements of the surface N in low-metallicity ($[\rm Fe/H] < -1$)
GCs would definitively help in understanding whether star-formation from polluted gas
occurred, providing also indications on the degree of dilution with pristine matter.

In the present analysis we describe how the formation of the SG depends on
the metallicity of the cluster, checking for compatibility among the data 
and the predictions of the self-enrichment scenario by massive AGB stars.
The ME15 is particularly suitable to this aim, because among the various
actors proposed so far to explain the formation of multiple populations
in GCs, the ejecta of massive AGB stars, as discussed in section \ref{hbb},
are those most sensitive to the chemical composition of the stars.

In the comparison between the expected and the observed variations in the
light elements abundances, we have to keep into account that the abundances
of SG stars with the most extreme chemistry are determined by two factors:
a) the nature of the polluters, which reflects into the chemical composition
of the ejecta; b) the degree of dilution of the gas ejected with pristine 
matter of the cluster, sharing the same chemical composition of FG stars. 

The maximum spread of each element for the individual GCs would 
correspond to the quantities predicted by the AGB ejecta of the same metallicity
only in case that SG stars with no dilution formed, which is expected to
occur only in a limited number of cases. When dilution occurs, the
chemistry of SG stars will be less extreme than the ejecta: what is relevant
in confirming or disregarding any pollution scenario is that the spread
for the various chemical elements in all the clusters is smaller than
expected based on a pure contamination. This is indeed the case, with the only exception of the
magnesium spread in M13 as discussed in section \ref{secM13}.

The chemical composition of the ejecta from massive AGB stars presented
here outlined a remarkable capability of following the trend with
metallicity traced by data of stars in the GCs analysed by ME15.

\begin{enumerate}

\item{
Strong HBB at the base of the envelope of metal-poor AGBs is capable
of producing the silicon spread observed in the lowest metallicity clusters. As stated
previously, given that silicon is one of the most abundant metals within the stars,
an enhancement $\delta [\rm{Si/Fe}] \sim +0.2$ dex
is the clue that the gas from which SG stars formed was exposed to
a very advanced nucleosynthesis.}

\item{ 
The temperatures at the bottom of the external mantle reached by 
AGB stars of metallicity $[\rm Fe/H] = -2$ are sufficiently large to
start an advanced Mg-Al nucleosynthesis, but not hot enough to
produce any significant silicon enhancement. The analisysis of the data show a clear anticorrelation only in the lowest metallicity cluster.}

\item{
In the intermediate metallicity GCs we still see Mg-Al
trend, but the magnesium spread is shorter than observed 
in more metal-poor GCs. This has a clear explanation from the results of AGB 
modelling, because a less advanced HBB nucleosynthesis is expected
on higher metallicity AGB stars.}

\item{
$\delta [\rm{Al/Fe}]$ is observed to vanish in the highest metallicity clusters, suggesting a progressively lower Al self-enrichment process for the medium where SG form.
This find a confirmation in the AGB scenario: the strength of HBB becomes weaker and weaker as the
metallicity increases, procuding ejecta less rich in Al.
}
\end{enumerate}

A more robust test of the self-enrichment by AGB stars demands more statistically
significant data set of metal-rich clusters. The present sample includes only two metal-rich clusters, M107 and M71, with 
a small number of stars, and both characterized by a large dilution with pristine 
matter. The lack of any Mg and Si spread and a reduced O excursion in this kind of clusters
would provide a more solid clue that AGB stars were the key polluters of intra-cluster
medium of GCs, from which SG stars formed.

For all the clusters examined, we used the data of the stars with the most
extreme chemical composition to infer whether direct formation of SG stars from
the AGB ejecta occurred or, alternatively, which is the minimum dilution with pristine 
matter required. This is crucial to determine the largest helium enhancement
expected for the most contaminated stars and offers a valuable opportunity to test
the present conclusions with indirect estimates of the helium spread, based on the 
morphology of the HB. For the clusters for which these estimates were available, it was possible to verify a satisfactory consistency.

The APOGEE data used in this work are of paramount importance, because they confirm that
the material from which SG stars formed was exposed to a very advanced nucleosynthesis, 
with the activation of Mg burning and, in the most metal-poor clusters, the synthesis of 
silicon. This is of great help in the way towards the understanding of the possible GC
polluters, because the possibility that either fast rotating massive stars 
(Krause et al. 2013) or massive binaries (De Mink et al. 2009) played a role in this context
can be ruled out. Indeed in neither of the two cases the temperatures of the nuclear
burning regions are sufficiently hot to allows such an advanced nucleosynthesis.

The only possibility left besides the massive AGB scenario are super massive main sequence
stars. As shown by Denissenkov \& Hartwick (2014), the core temperature during H-burning
reach $\sim 70$ MK, thus triggering the activation of Mg burning and the production
of aluminium. A straight comparison between the predictions of the AGB hypothesis and
the chemical patterns expected when pollution from super massive stars is considered
is cumbersome, because while it was shown that in the interiors of this class of objects
the ignition of Mg burning occurs, the trend with metallicity, if any, needs further
investigation. While in the AGB case the sensitivity of the HBB nucleosynthesis to the
metallicity is a direct result of stellar evolution modelling, it is not clear how
the chemical composition would affect the core H-burning of super massive stars. At the
moment, this problem is still open.

\section{Conclusions}
We discuss the chemical composition of giant stars in 9 Galactic globular
clusters, for which the abundances of C, N, O, Mg, Al and Si are available.
The wide range of metallicities ($-2.4 < [\rm Fe/H] < -0.7$) 
and the simultaneous knowledge of the elements mentioned above 
offer a valuable opportunity to test the origin of multiple populations in these complex stellar systems.

We concentrate on the possibility that the polluters of the intra-cluster
medium, providing the gas from which SG stars formed, were massive AGB 
stars ($4~\rm M_{\odot} \leq \rm M \leq 8~\rm M_{\odot}$), which experienced HBB at the 
base of their external envelope. To this aim, we calculated ad hoc AGB models, 
with the same metallicity and the same mixture of the stars belonging to the 
FG of each cluster.

In agreement with previous studies on GCs in the Milky Way, we find that
all the clusters show the presence of SG stars, as 
deduced on the star-to-star chemical differences, which define abundance 
patterns in the different observational planes.

We find that the chemical patterns observed (Mg, Al, Si and O in  particular) are in good agreement with the
theoretical expectations: a) on the qualitative side, the spread 
observed for the different species follows the results from AGB 
evolution theories, according to which the strength of HBB, the phenomenon
by which the contaminated gas is produced, becomes
weaker the higher is the metallicity of the stars; b) on a 
quantitative point of view, the spread observed for the various
chemical species are in agreement with the chemistry of the AGB
ejecta. The only exception to this is the Mg spread of M13 stars. 

When available, we check consistency with results from photometry in the literature,
particularly with the helium spread deduced on the basis of the morphology of the HB; 
this poses important constraints on the degree of dilution of the gas from AGB stars with 
pristine matter in the cluster, from which the SG formed.

A more solid confirmation of the general conclusions drawn in the present work requires 
data for a larger sample of stars, particularly for metal-rich
clusters: this is crucial to confirm that the extent of the spread observed
between the chemical composition of FG and SG stars tend to vanish as the metallicity
increases, as expected based on the results of AGB modelling. In addition, the lack of a clear N-Al
correlation in the ME15 data, which has been previously confirmed in the
literature and it is expected also according to the AGB models, suggests that
the determination of N abundances in the H-band for GC giants should be
revisited. Such data may be provided by the on-going APOGEE-2 survey, which is almost triplicating the number of GC stars observed in the H-band and continuously improving the precision/reliability of the derived abundances.

\section*{Acknowledgments}
FDA, DAGH, TM, and OZ acknowledge support provided by the Spanish Ministry of Economy and
Competitiveness (MINECO) under grant AYA-2014-58082-P. DAGH was also funded by
the Ram{\'o}n y Cajal fellowship number RYC-2013-14182.
SzM has been supported by the Premium Postdoctoral
Research Program of the Hungarian Academy of Sciences, and by the Hungarian
NKFI Grants K-119517 of the Hungarian National Research, Development and Innovation Office.

\end{document}